\documentclass[twocolumn,onecolappendix]{aastex6}

\fullcollaborationName{}

\usepackage{newtxtext}
\usepackage{newtxmath}
\usepackage{array}
\usepackage{amsopn}
\usepackage{graphicx}
\usepackage{multirow}
\usepackage{enumitem}

\newcommand \etal {\textit{et al. }}
\newcommand*\rfrac[2]{{}^{#1}\!/_{#2}}

\begin{document}
\graphicspath{{/home/avivofir/SSD/AllPlanetsPolynomials/}}


\title[A spectral approach to TTVs]{A spectral approach to transit timing variations}


\author{Aviv Ofir \altaffilmark{1,2}, Ji-Wei Xie\altaffilmark{3,4}, Chao-Feng Jiang\altaffilmark{3,4}, Re'em Sari\altaffilmark{5}}
\and
\author{Oded Aharonson\altaffilmark{1}}


\altaffiltext{1}{Department of Earth and Planetary Sciences, Weizmann Institute of Science, Rehovot, 76100, Israel}
\altaffiltext{2}{email: avivofir@weizmann.ac.il}
\altaffiltext{3}{School of Astronomy and Space Science, Nanjing University, Nanjing 210093, China}
\altaffiltext{4}{Key Laboratory of Modern Astronomy and Astrophysics in Ministry of Education, Nanjing University, Nanjing 210093, China}
\altaffiltext{5}{Racah Institute of Physics, the Hebrew University, 91904,Jerusalem, Israel}

\begin{abstract}

The high planetary multiplicity revealed by \textit{Kepler} implies that Transit Time Variations (TTVs) are intrinsically common. The usual procedure for detecting these TTVs is biased to long-period, deep transit planets whereas most transiting planets have short periods and shallow transits. Here we introduce the Spectral Approach to TTVs technique that allows expanding the TTVs catalog towards lower TTV amplitude, shorter orbital period, and shallower transit depth. In the Spectral Approach we assume that a sinusoidal TTV exists in the data and then calculate the improvement to $\chi^2$ this model allows over that of linear ephemeris model. This enables detection of TTVs even in cases where the transits are too shallow so individual transits cannot be timed. The Spectral Approach is more sensitive due to the reduced number of free parameters in its model. Using the Spectral Approach, we: (a) detect 131 new periodic TTVs in \textit{Kepler} data (an increase of $\sim2/3$ over a previous TTV catalog);  (b) Constrain the TTV periods of 34 long-period TTVs and reduce amplitude errors of known TTVs; (c) Identify cases of multi-periodic TTVs, for which absolute planetary mass determination may be possible. We further extend our analysis by using perturbation theory assuming small TTV amplitude at the detection stage, which greatly speeds up our detection (to a level of few seconds per star).
Our extended TTVs sample shows no deficit of short period or low amplitude transits, in contrast to previous surveys in which the detection schemes were significantly biased against such systems.
\end{abstract}




\section{Introduction}

A transiting exoplanet's Keplerian orbits impart precisely evenly spaced transit events. Any deviation from such regularity indicates that other forces are at play. Importantly, if there are other massive bodies in the system they would interact with the transiting planet and perturb its orbit, imparting deviations from strict periodicity known as transit timing variations (TTVs). This important effect was predicted (Holman \etal 2005, Agol \etal 2005) and later observed (Kepler-9 system, Holman \etal 2010, Dreizler \& Ofir 2014). TTVs are stronger and easier to detect in systems near mean motion resonances (MMRs) and generally appear as sine-like deviations. Inverting the observed TTVs back for the parameters of the physical system that produced them is difficult (Lithwick \etal 2012) primarily since the important parameters (e.g., planetary mass and eccentricity) are degenerate when higher-order effects are not significant ({\it i.e.} departures from exact sine-shaped TTVs).

In fortuitous cases where higher-order effects are observed in multi-transiting systems, the TTVs may yield the absolute masses of the planets ({\it e.g.}: Kepler-9, Kepler-11, Kepler-87, Holman \etal 2010, Lissauer \etal 2011, Ofir \etal 2014, and others) without the use of high precision radial velocity (RV) measurements, allowing the determination of planetary masses even beyond current RV capabilities in some cases ({\it e.g.} Kipping \etal 2014). TTVs can be useful without higher-order effects, especially in multi-transiting systems, since anti-correlated TTVs constitute dynamical confirmation of the candidates as true planets in the same system, undergoing angular momentum exchange ({\it e.g.} Steffen \etal 2013, Xie 2013, Xie 2014) - albeit with weak mass constraints. Even TTVs that are observed in only one transiting planet in a system are still useful since they may reveal the presence of additional planets in the system that may not be transiting at all (Nesvorn{\'y} \etal 2012, 2013). Finally, the fact that TTVs are observed mostly for planets near MMRs allows using TTVs to address questions related planet formation and migration - and the means by which planets are captured in MMRs ({\it e.g.} Xie \etal 2014, Mills \etal 2016).
 
For these reasons, a significant amount of work has gone into identifying TTVs, particularly in the \textit{Kepler} dataset  ({\it e.g.} Ford \etal 2011, Xie 2013, Mazeh \etal 2013, Holczer \etal 2016). In this work, we develop and apply a new technique -- the Spectral Approach -- to detect TTVs. We focus on detecting low-amplitude TTVs (defined in \S~\ref{Domain}), with the assumption that high-amplitude TTVs were already identified (at least in the \textit{Kepler} data). We further increase the sensitivity of TTVs detection - both to the fundamental signal and to its higher-order components. To demonstrate this improvement we compare our work to the recent Holczer \etal (2016) results, hereafter H16, using both their "long-term" and "short-term" TTVs combined (TTV periods of above and below 100~d, respectively), and treating multi-periodic TTV as separate signals.

Below we will present the spectral approach in \S~\ref{GridSearch} and its more computationally-efficient perturbative approximation in \S~\ref{Perturbative}. We then describe the details of applying the spectral approach to \textit{Kepler} data in \S \ref{ApplicationToKepler}, the resultant TTVs catalog in \S~\ref{Results}, and conclude in \S~\ref{Conclusions}.

\section{A spectral approach to TTVs} 
\label{GridSearch}

\subsection{Motivation and description}
\label{Motivation}
This paper is concerned with low-amplitude TTVs, so we assume that a candidate transiting planet signal was already identified using simple linear ephemeris. When described using the Mandel-Agol 2002 formalism (hereafter MA02), the transit model requires explicitly just two parameters: the normalized planetary radius $r$, and the normalized distance between the star and planet $d_i$ to calculate the observed normalized flux at the $i$th point along the orbit. At a given time $d_i$ is a function of four orbital parameters in the case of circular orbits: $P, T_{\mathrm{mid}}, a, b$ - for the planetary orbital period, time of mid-transit, normalized semi-major axis and normalized impact parameter -- and two additional parameters for eccentric orbits (where all normalized parameters are relative to the stellar radius). Limb darkening parameters are also needed but are usually not fitted (except in the highest signal to noise ratio (SNR) cases) and depend on the selected limb darkening law.

In the usual procedure, the TTV search begins by allowing the individual times of mid-transit to deviate from their linear ephemeris. The individual timings are then searched for excess scatter or periodic signals. We note that in this case there are a total of $N_{\mathrm{tr}}+4$ fitted parameters, where $N_{\mathrm{tr}}$ is the number of individual transit events in the data and four are the other circular-orbit parameters. Since $N_{\mathrm{tr}}$ can be several hundreds or more in the case of short-period planets observed throughout the four years of \textit{Kepler}'s normal operation, the large number of fitted parameters reduces the sensitivity of this technique.

To first order, planetary TTVs close to first-order resonances are theoretically expected to have sine-like shapes with a super-period significantly longer than the orbital period of the transiting planets: $P_{\mathrm{Sup}}={P_{\rm out}}/{(j\Delta)}$ where $\Delta=\frac{P_{out}}{P_{in}}\cdot\frac{j-1}{j}-1$ is the normalized distance to resonance (Lithwick \etal 2012). Indeed many of the detected TTVs ({\it e.g.} H16) are actually observed to be mainly sine-like, sometimes exhibiting other frequencies due to, for example, resonances of higher order or terms of higher order of a given resonance ({\it e.g.} Deck \etal 2014, Agol \& Deck 2016), but all are still well-described by sine functions. We therefore do not attempt to measure the individual times of mid-transit and then fit these timings, but rather assume that a sinusoidal TTVs exists in the data and then calculate the improvement of the model under this assumption relative to the linear model. We refer to this procedure as a Spectral Approach to TTVs.

We wish to scan all possible sine-like TTVs. This requires a three-dimensional search grid of TTV frequency $f$, TTV amplitude $A$ and TTV phase $\phi$ (to avoid ambiguity the last will later be translated to the time $T_0$ at which $\phi=0$). On one hand, the frequency search axis can be well defined: the minimum frequency $f_{\rm min}=\rfrac{1}{2s}$ where $s$ is the span of the data. The maximum frequency is set by the Nyquist frequency $f_{\rm max}=\rfrac{1}{2P}$. The critical frequency spacing scale is as usual for sine fitting $\Delta f_{\mathrm{crit}}=1/s$ and one may want to oversample this critical frequency spacing by a factor of a few to avoid missing local peaks. On the other hand, there is no natural upper limit to $A$ (other than $A<P$ which is not very informative) and no natural resolution to $A$ or $\phi$ - leaving the search grid undefined and the processing slow (the MA02 model function need to be invoked for each transit event separately, on each test position on the search grid). The procedure therefore appears plausible but numerically inefficient.

Signal detection can be significantly accelerated, however, if one limits the search to low- or medium- amplitude TTVs (see \S~\ref{Domain} for definition). This improvement is made by using perturbation theory, and is described in \S~\ref{Perturbative}, and will be henceforward called the Perturbative Approximation (PA) to the Spectral Approach to TTVs, as opposed to the full fit. We stress that PA is used for detection only. Indeed, and as done in \S~\ref{FinalFit}, a full non-perturbative model should be used for the determination of the final TTV signal parameters and their associated errors.

\subsection{Benefits of Spectral Approach}

The spectral approach to TTVs offers several advantages. These include: (i) It encapsulates all the TTV information in 3 parameters -- the sinusoid's period, amplitude and phase -- regardless of $N_{\mathrm{tr}}$.
(ii) Since the entire light curve is fitted with a single TTV model, the TTV detection sensitivity scales closely with the precision of the linear $T_{\mathrm{mid}}$. The reason is that of the four linear ephemeris parameters, $T_{\mathrm{mid}}$ is the only one affected by adding small TTVs. 
This results in elimination of the bias to long periods since the absolute precision on $T_{\mathrm{mid}}$ of the linear ephemeris (and thus the TTV detection sensitivity) actually improves with decreasing orbital period.
(iii) Short-period signals (of few days period or shorter) have few points in-transit in each individual event, making transit time fitting nearly impossible. The extreme case is KOI 1843.03 (Ofir \& Dreizler 2013) which has an orbital period of $\sim4.25$~hr and total transit duration not much longer than a single \textit{Kepler} long cadence. Ultra-short period planet candidates are also typically too shallow to be detected at all as single transit events. Searching for TTVs using the spectral approach presents no such limitations.
(iv) The resultant TTV spectra are compatible with later stages of system interpretation: theoretical considerations can predict the TTV frequencies and their amplitude ratios ({\it e.g.} Lithwick \etal 2012, Deck \etal 2014, Agol \& Deck 2016). Therefore this information is useful even when no single transit event is above the noise.

The classical approach to TTV detection is biased to long-period planets and deep transits (Steffen 2016) whereas most transiting planets have short periods and shallow transits. Indeed, in the the H16 catalog of \textit{Kepler} TTVs the median orbital period for planets that exhibit TTVs is 29.8~d, while the median period of all \textit{Kepler} candidates is less than a third of that number -- about 9.5~d. This bias is therefore relevant to a large number of objects, and the proposed Spectral Approach can all but remove it.

We note that some TTVs are not sine-shaped: TTVs which originate from non-gravitational processes ({\it e.g.} instrumental effects, starspots) may have different morphologies, and their description as sinusoids or sum of a few sinusoids may be inadequate.

\subsection{Theoretical comparison with the classical technique}

The probability density function (PDF) of the timing of an individual transit event is a delta function at the true time, which is then broadened by observational noise. In high SNR data, the PDF of transit timing fits is close to a gaussian centered on the true transit time, and with a width that depends on the SNR. As SNR decreases, the PDF of the transit time fits reduces in amplitude, broadens, and no longer resembles the true PDF. Instead, there is an almost uniform probability of finding the transit at any time in the search interval, with only a slight increase at the true transit time. Thus fitting these individual transit PDFs with individual gaussians, as in past work, results in poor fits that cannot approximate the true PDF when the SNR is low. In other words, when the SNR is low the PDF is so wide and shallow that many of the individual transits are detected far from the true signal. Thus they do not contribute to the coherent addition that allows searching for a periodic variation. Using a single global fit avoids this problem.

\begin{figure}
	\includegraphics[width=\columnwidth]{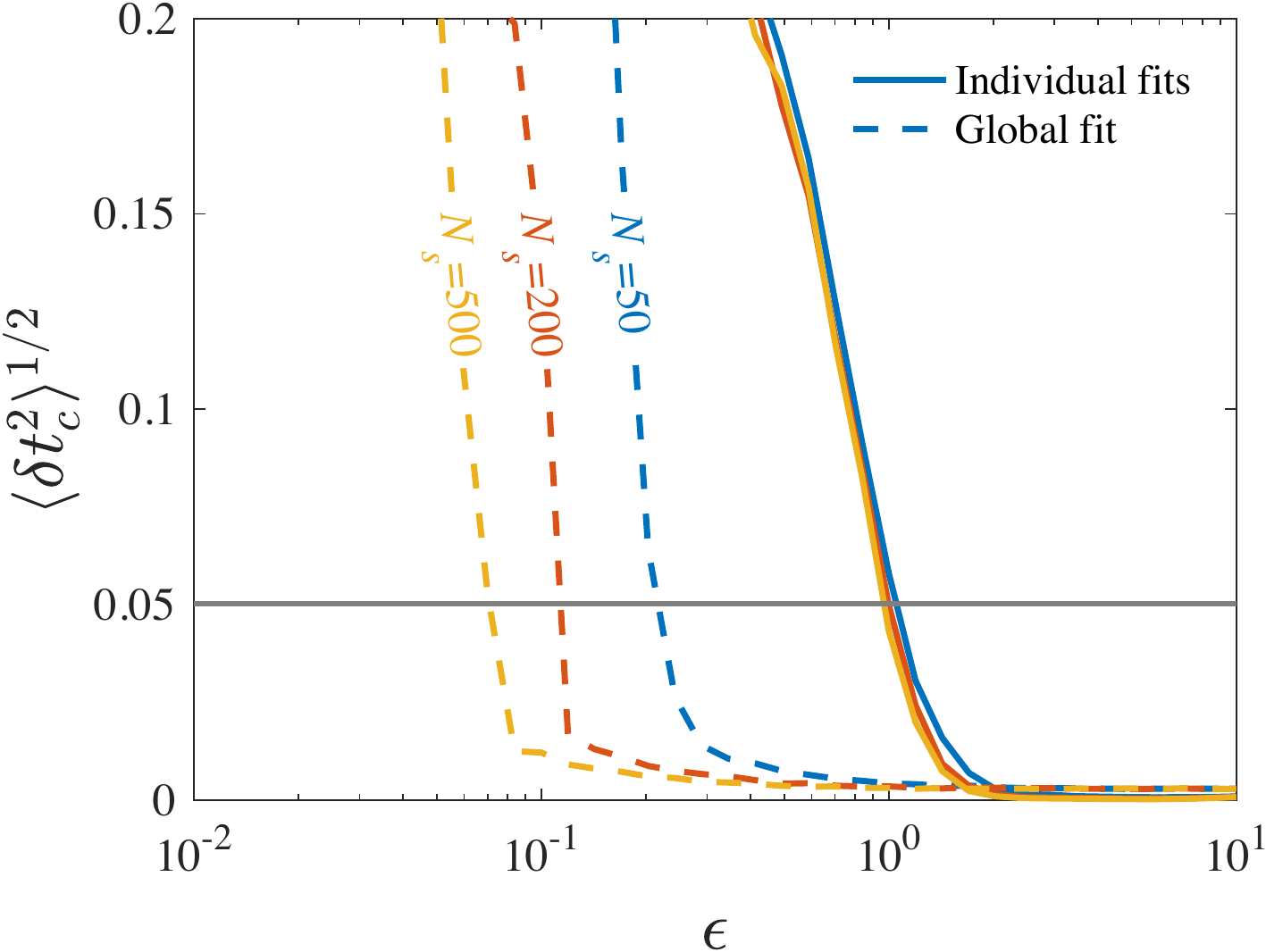}
	\caption{Illustration of improvement in timing fits that use global fitting of a periodic gaussian signal of amplitude $\epsilon$ embedded in unit noise.  Simulation using $N_p$=200 points, in $N_s$ segments shown in color, run over $N_r=200$ trials. The gray line is the width assumed for the simulated gaussian transit; the unit of time is chosen such that on average the segments have unit duration on either side of the simulated transit.}
	\label{gaussgame1}
\end{figure}

In order to illustrate this effect, we generated $N_s$ segments of duration unity on either side of a gaussian-shaped transit-like signal of height $\epsilon$, each segment with $N_p$ points including normally distributed noise of unit amplitude.  The simulated mid-transit time was chosen as a random (uniform) location within the central half of the domain. We then determined the best-fitting mid-transit time in two ways. We first fit a periodic gaussian shape to all the segments simultaneously. We also took the average of individual times measured by fitting each segment separately. The difference in time between the best fitting and the actual mid-transit location is $\delta t_c$. We repeated this experiment $N_r$ times, and averaged the results to produce smooth curves. Figure \ref{gaussgame1} shows clearly that the a global fit outperforms the average of individual fits, {\it i.e.}, the RMS of the timing error, $\left<\delta t_c^2\right>^{1/2}$, remains small up to lower values of the signal amplitude, $\epsilon$, relative to the noise. The mean timing error $\left<\delta t_c^2\right>^{1/2}$ rises sharply and exceeds the transit duration at signal amplitudes much lower (an order of magnitude for the representative parameters chosen) for the global fit than the individual fits. This again demonstrates the general effect that should be unsurprising: when the signal to noise is low enough to mask the individual transit such that its typical timing error exceeds the event duration, averaging many such fits does not improve the detection. However a global fit remains sensitive to lower SNR. 

\section{Perturbative approximation to the spectral approach}
\label{Perturbative}

\subsection{Basic perturbation theory}


To some general data set of $N$ points $\{t_i,F_i\}$ and $F$-errors $\sigma_i$ we fit a model $m_i=m(x_i)$. We wish to examine the sensitivity of the model to perturbation $g_i$ in the \textit{independent} parameter $t_i$ assuming the perturbation's functional shape is known ({\it e.g.}: linear trend, periodic variation, etc.). This perturbing function is known and normalized, only its scaling parameter $S$ is sought after. We therefore define the effective independent parameter:
\begin{equation}
t'_i \equiv t_i + Sg_i
\end{equation}
The perturbations are assumed to be small in the sense defined in \S~\ref{Domain}, and in this limit a linear approximation is possible:
\begin{equation}
\begin{array}{l}
m_i(t'_i) \approx m_i + Sg_im'_i, \\
m'_i\equiv\frac{dm}{dt}(t_i), \quad  \quad m_i\equiv m(t_i).\\
\end{array}
\label{Linearization}
\end{equation}
Since we are interested in TTVs, the independent parameter $t_i$ is the time $F_i$ is the normalized flux, and $m_i$ is the Mandel-Agol (2002) transit model. The derivative of the transit model $m'_i$ is computed with respect to time, but application to sparse data (such as \textit{Kepler}'s long-cadence light curves) should be done with care (see \S \ref{ApplicatioToTTVs}). Thus the amplitude $S$ has the desired meaning of TTV amplitude, and $m_i(t'_i)$ is the unknown TTV-inclusive model for which we solve perturbatively. We use least-squares to fit the new model:
\begin{equation}
\chi^2=\sum_i^N{\frac{\big(F_i-m_i(t'_i)\big)^2}{\sigma_i^2}}.
\end{equation}
Differentiating with respect to $S$, 
we obtain the best-fitting amplitude for the perturbation $g_i$:
\begin{equation}
\label{PA_amp_eq}
S=\sum^N_i{\frac{(F_i-m_i)g_im'_i}{\sigma^2_i}} \Bigg/ \sum_i^N{\bigg(\frac{g_im'_i}{\sigma_i}\bigg)^2 }.
\end{equation}

It is useful to generalize this result to two-parameter optimization: now two perturbations $g_i$ and $h_i$ perturb the model function $m(x)$ so that the effective independent parameter is now 
\begin{equation}
t'_i \equiv t_i + S_g g_i + S_h h_i,
\label{TwoParameterZ}
\end{equation}
giving the set of linear equations $\bf{MX}=\bf{V}$ where:

\begin{equation}
	\bf{M}=\left(
	\begin{array}{cc}
		\sum_i^N{\left(\frac{g_im'_i}{\sigma_i}\right)^2} & \sum_i^N{\frac{g_ih_im'^2_i}{\sigma_i^2}} \\
		\sum_i^N{\frac{g_ih_im'^2_i}{\sigma_i^2}}         & \sum_i^N{\left(\frac{h_im'_i}{\sigma_i}\right)^2} \\
	\end{array}
	\right),\\
\end{equation}
\begin{equation}
	\bf{X}=\left(
	\begin{array}{c}
		S_g \\
		S_h \\
	\end{array}
	\right),
\end{equation}
\begin{equation}
	\bf{V}=\left(
	\begin{array}{c}
		\sum_i^N{\frac{\left(F_i-m_i\right)m'_ig_i}{\sigma_i^2}}\\
		\sum_i^N{\frac{\left(F_i-m_i\right)m'_ih_i}{\sigma_i^2}}\\
	\end{array}
	\right).
\end{equation}
These linear equations may be solved for $\bf{X}$, the amplitudes of the two perturbations.

\subsection{Application to TTVs}
\label{ApplicatioToTTVs}
We now proceed to apply the analytical method above to TTVs. In this context, the model function $m$ is the linear ephemeris MA02 model and we assume that a linear model was already fitted to the light curve. At this point each data point has an assigned time from mid-transit $t_i$ which corresponds to a normalized distance $d_i$ for the MA02 model. We note that finding the true global minimum $\chi_{\mathrm{linear}}^2$ is of special importance since the TTVs analysis depend on the exact shape of the linear signal. We took great care in performing this step, as described in \S~\ref{ApplicationToKepler}. Furthermore, \textit{Kepler}'s long cadence means that $\frac{dm}{dt}$ cannot be computed simply from the time series as $m_i(x)=({m(t_i)-m(t_{i-1})})/({t_i-t_{i-1}})$ since the details of the ingress/egress are all but erased from the individual transits. To correct this effect $m'$ can be estimated either from a densely re-sampled flux model, or from the folded model light curve (in time modulo the period and not in phase, in order to be consistent with the TTV units of time and Eq.~\ref{Linearization}). In the folded light curve the ingress/egress region is much better sampled than in individual events, enabling a more precise evaluation of ${dm}/{dt}$.

We then add a perturbation - a sine in time which has a frequency $f$, amplitude $A$ and phase zero at time $T_0$. The effective independent parameter is therefor:
\begin{equation}
\label{TwoComponentModel}
\begin{array}{l c l}
t'_i & = & t_i + A \cdot \sin\left[2\pi f (t_i-T_0) \right] \\
     & = & t_i + A\left[ \sin(2\pi f t_i) \cos(2 \pi T_0 f) \right. \\
     &   & \left. - \cos(2\pi f t_i ) \sin(2 \pi f T_0)\right] \\
     & = & t_i + S_1 \sin(2\pi f t_i ) + S_2 \cos(2\pi f t_i) \\
\end{array}
\end{equation}
which encapsulate $\{A,T_0\}$ in $\{S_1,S_2\}$ in the form described by Eq.~\ref{TwoParameterZ}. This linearization allows us to analytically solve for the best fitting amplitude and phase of a sine-shaped TTV, removing the problems identified above and leaving only the TTV frequency as a single searched parameter (which remains well defined with clear boundaries and resolution as explained in \S~\ref{Motivation}). The resultant perturbed model (Eq.~\ref{TwoComponentModel}) appears similar to a true TTV-shifted signal but it is only a mathematical construct and not a physical model ({\it e.g.} it may have flux values greater than unity, unless these are clipped, see \S~\ref{ModelClipping}).

\begin{figure*}
	\includegraphics[width=1\textwidth]{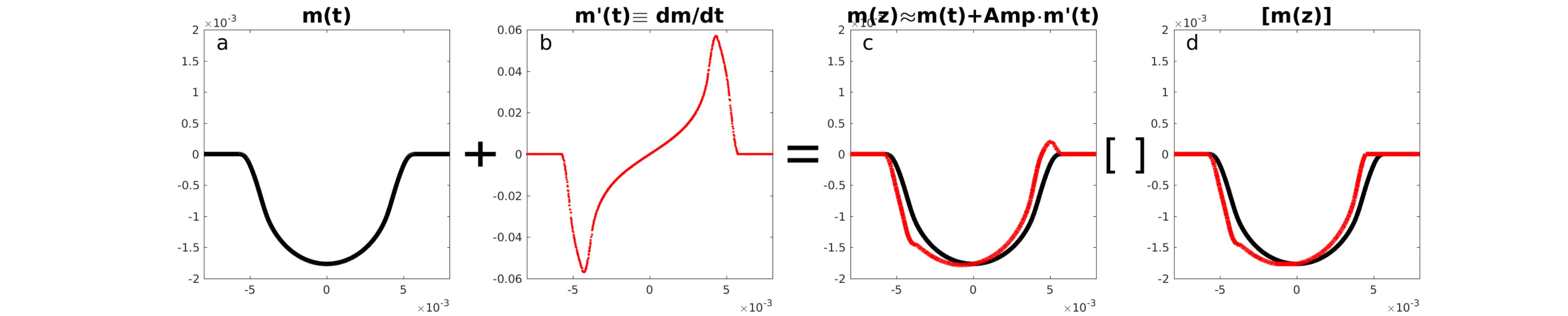}
	\caption{A visualization of the Perturbative Approximation (PA) to the Spectral Approach to TTVs: \textbf{a:} the phased linear-ephemeris model $m(t)$ vs. time since mid-transit. All subsequent black curved are identical to it. \textbf{b:} the model's time derivative $m'(t)$. \textbf{c:} the sum of the $m$ and $A \cdot m'$ (where $A$ is the instantaneous TTV amplitude, which itself is time-dependent $\mathrm{A}=S g_i$) in red. \textbf{d:} The square brackets stand for the two-sided clipping operation to the minimal and maximal $m(t)$ values of the above sum. The resultant perturbed model is very similar to a physically-calculated TTV. The model here is of KOI 641.01, and the simulated amplitude if of $0.01d$ -- close to- and between- the median and mode of High Confidence TTVs detected in this study.}
	\label{PA_visualization}
\end{figure*}

Figure \ref{PA_visualization} visualizes the process: the folded linear ephemeris model is cropped to just a region surrounding the linear-ephemeris transit, and the model derivative $dm/dt$ is calculated. The perturbed model is generated by adding $dm/dt$, scaled by some amplitude, to the linear model above -- to mimic a TTV-affected light curve. While the perturbed model usually appears just as a time-shifted model, some points
can obtain non-physical values in the perturbed model: above unity and below $\textrm{min}(m)$.
Model clipping (see \S~\ref{ModelClipping}) can be applied to counter this effect.

\subsection{Domain of validity}
\label{Domain}

PA approximates the light curve of TTV-affected planets assuming these TTVs are small. Indeed, the ability of the PA to find a better model than the linear model arises from the non-zero derivative of the model $m'_i$ in Eq.~\ref{PA_amp_eq}, which allows adjustment of the linear model to better fit the data. Transit light curves are roughly trapeze-shaped, so non-zero derivatives are significant mostly during ingress and egress. We can therefore divide the possible TTV amplitudes to three broad regimes of validity within the PA:

(i) Low-amplitude TTVs are those with amplitude lower than the ingress/egress duration. In this regime the TTVs are small enough for $m'_i$ to be able to approximate well the true TTV signal. We therefore expect here that the PA will both find the true TTV frequency and reproduce the TTV amplitude with optimal sensitivity.

(ii) Medium-amplitude TTVs have amplitudes larger than the ingress/egress duration, but smaller than the full transit duration. In this regime the PA-detected amplitude saturates as $(F_i-m_i)m'_i$ factor is nonzero but nearly independent of the TTV amplitude -- so the PA can only compensate for the ingress/egress duration part of the actual TTV amplitude. Here the correct TTV frequency would likely be identified but with a lower amplitude than the correct one.

(iii) High-amplitude TTVs have amplitude comparable to the transit duration or larger. In this case either $m'_i$ or $y_i-m_i$ in Eq.~\ref{PA_amp_eq} average to zero at all times, and so the perturbative approximation fails and it is unable to detect the TTVs. Note the full search without the linear approximation would still apply well.

These domains are simulated and shown in \S~\ref{SimulatedDetection}.

The PA approximation is valid for the majority of TTVs -- especially those yet undetected -- since the median TTV amplitude in the H16 catalog is $<23$~min while the median transit duration of the same objects $>300$~min. In other TTV catalogs, such as those of Xie (2013, 2014), the case is even clearer with median amplitude of  $<14$~min and median duration of $270$~min. While rare, high-amplitude TTVs do exist with amplitudes that may even exceed the duration of the transit itself. These are  better detected by other techniques ({\it e.g.} QATS (Carter \& Agol 2013), human eyes (Schmitt \etal 2014), or single event detection schemes (e.g. Osborn \etal 2016).

\subsection{Model clipping}
\label{ModelClipping}

The PA produces a perturbed model which is a mathematical construct (not a physical model).
This construct can be made nearly indistinguishable from a physical model by clipping the perturbed model on both the upper and lower ends, to zero flux decrement and to $\textrm{min}(m(x))$, respectively, before $\chi^2$ is calculated (see Figure~\ref{PA_visualization}). Model clipping is not used here to refine the PA-detected amplitude nor frequency of the signal, but improves the reliability of the $\chi^2$ values in discriminating true from spurious TTV signals.

\subsection{Correction to the Mandel-Agol (2002) model}

We use the standard MA02 model to calculate the size of the flux decrement during transit. This model is a set of a few mathematical formulae, each valid in its own section of the parameter space and with mathematically accurate transitions between these regions. While the computational implementation of these formulae has finite precision and thus some discontinuities in the transition between sections exist, the effect is small enough ($10^{-4}$ times the transit depth itself) to be largely ignored. 

\begin{figure}
	\includegraphics[width=0.5\textwidth]{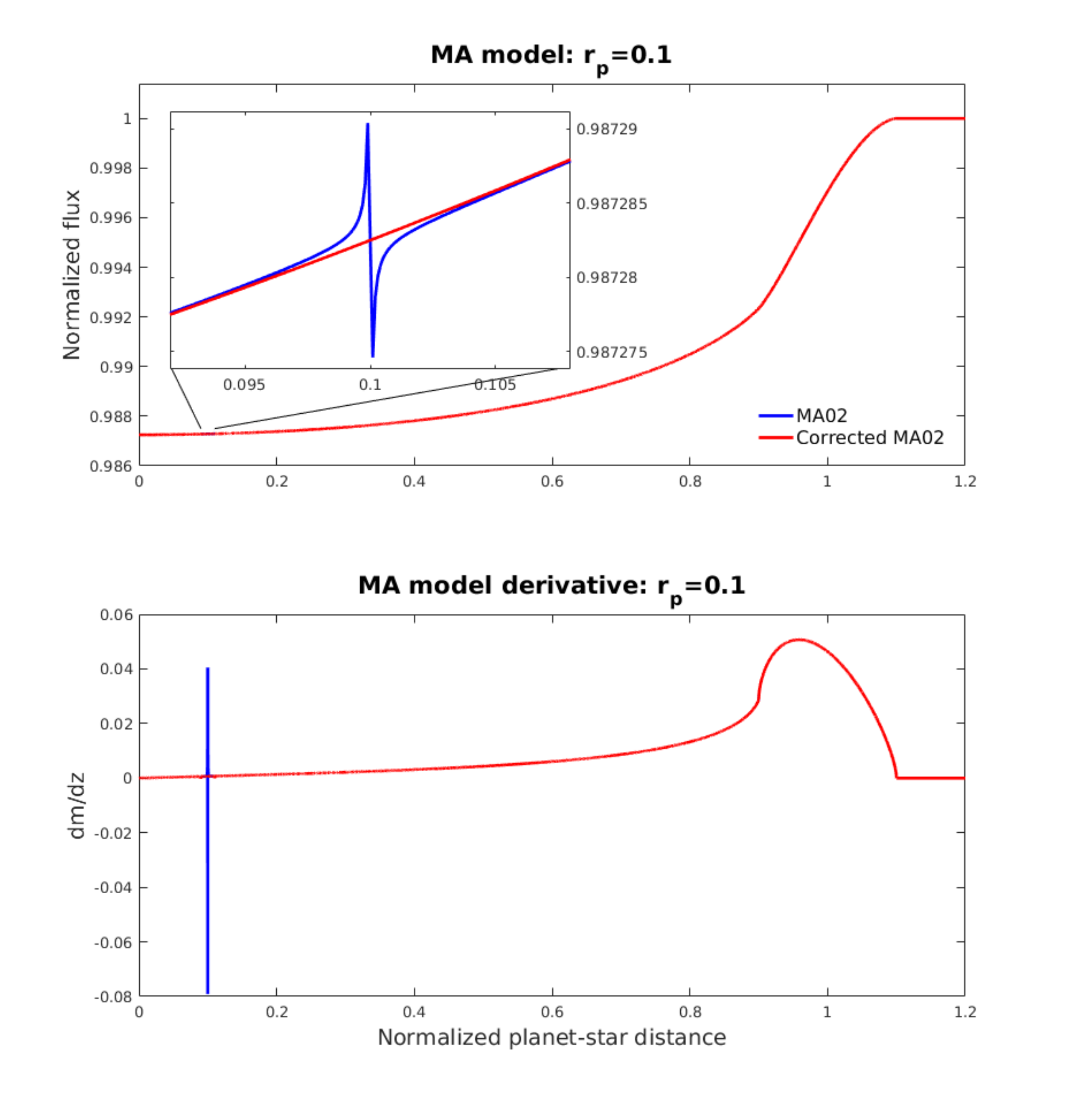}
	\caption{A scan though the MA02 model for a planet (with $r=0.1$) at a resolution of $dz=10^{-4}$. \textbf{Top panel:} The original and corrected model functions (indistinguishable on the main panel). The insert shows the region near the $z\approx r$ anomaly.
	\textbf{Bottom panel:} the derivative to the model amplifies the anomaly. At this resolution the $z\approx r$ anomaly has amplitude larger than the main $dm/dz$ signal, and the $z\approx 1-r$ is still not visible. At finer resolutions still (e.g. $10^{-5}$) both anomalies have much higher amplitude than the main signal.}
	\label{MA02_correction}
\end{figure}

However, in the PA the MA02 \textit{derivative} with respect to the planet-star separation $z$ is needed - and here the finite precision can no longer be neglected for points close to two problematic transitions at $z\approx r$ and $z\approx 1-r$. These numerical errors are small in amplitude but they also occur over a small range of $z$ values, causing their $dm/dz$ amplitude to rival that of the main signal (and actually diverge at the limit). This is demonstrated in Figure~\ref{MA02_correction}: we sample normalized distances of a $r=0.1$ planet in the MA02 model at a resolution of $dz\approx10^{-4}$. The small miscalculation of the model around the $z\approx r$ transition is observed with an amplitude $<10^{-5}$ (see insert on the top panel). At this resolution the anomaly near $z\approx 1-r$ is $\approx 100$ times smaller still (and thus not visible). Yet, the model's derivative has a large anomaly at $z\approx r$ (the one near $z\approx 1-r$ is still small). As the discontinuity is approached this becomes more severe: at higher resolution ({\it e.g.} $dz=10^{-5}$) the anomaly near $z\approx 1-r$ becomes dominant - reaching amplitudes several orders of magnitude higher than the main $dm/dz$ signal.

These numerical errors are inherent to the MA02 model and will persist even if numerical convergence criteria used in the MA02 are made stricter. We therefore propose an ad-hoc correction to the model. The $z\approx r$ anomaly has continuous slopes on either side (there is no physical change at $z=r$), while the $z\approx 1-r$ anomaly does reflect a physical change at $z=1-r$ (the II \& III contacts). We therefore correct the former by polynomial interpolation from both sides, and the latter by polynomial extrapolation from either side, separately. In both cases the corrected region is $\Delta z=2\cdot10^{-4}$ from the anomaly, and the polynomials are fitted using a region up to $\Delta z=10^{-3}$ from it. The polynomials are of $2^{nd}$ order, unless $r>0.5$ in which case they are of $3^{rd}$ order. While satisfactory, this procedure is imperfect: each transition from MA02 model to polynomial interpolation/extrapolation involves a "stitching" point that produces an outlier point in the model's derivative - but these are points (vs. finite regions in the uncorrected model) and with far reduced amplitude, regardless of the proximity to the transition points.

\subsection{Model sampling}
A further consequence of using the model's derivative concerns the model oversampling needed due to \textit{Kepler}'s finite integration time. The commonly used criterion for computing the needed oversampling is given by Eq.~40 of Kipping \etal (2010), which calculates the inaccuracy caused by sampling the instantaneous model at cadence $\mathcal{I}$ relative to integrating it over intervals with cadence $\mathcal{I}$ (the latter is of course closer to a real measurement). If the signal has a depth of $\delta$ and ingress/egress duration of $\tau$ that inaccuracy is found to be  $\sigma(F_{\textrm{Resampling}})\propto \delta \mathcal{I}/\tau$ (The $\sigma$ sign is used since the inaccuracy can be viewed as a type of modeling noise). However, this computation is not valid in the case of PA since now the sampling noise must be compared not with the model flux but with the model's flux derivative. The slope which Kipping \etal 2010 sampled was $\delta/\tau$. Analogously, the slope that needs to be well-sampled in PA is approximately $\delta/\tau^2$. We therefore update Eq.~40 of Kipping 2010 for use in PA by dividing it with yet another ingress/egress duration, or:
\begin{equation}
\sigma(F_{\textrm{Resampling}})=\frac{\delta}{\tau^2}\frac{\mathcal{I}}{8N^2},
\end{equation}
where $N$ is the number of subsamples computed. In practice, we impose a lower limit of $N=10$ to ensure adequate subsampling.

\subsection{Speed}
Linearly fitting two of the three search dimensions allows a significant speedup in calculating TTVs. This part of the calculation for a typical system now takes 10 seconds or less on a single-threaded CPU (a far larger amount of time is required to fit the linear model of MA02) as zero calls to the MA02 model function are needed. 

\subsection{Detection of simulated signals}
\label{SimulatedDetection}
We simulated \textit{Kepler}-like data that included a transiting planet with sinusoidal TTVs with a TTV period shorter than the time baseline of the simulated data, and tried to detect those TTVs using PA. We considered a detection successful when the frequency of the most significant peak on the TTV spectrum was close to the simulated one (\textit{i.e.} $|\Delta f|<(\mathrm{time\,\,baseline})^{-1}$), and calculated the detection efficiency of the algorithm while scanning along various parameters. Notably, we also recorded the amplitude of the best-detected peak, {\it i.e.} the $\chi^2$ improvement (as a positive number) over the best-fit linear model: $\Delta\chi^2\equiv\chi^2_{linear}-\chi^2_{PA}$.

\begin{figure}
	\includegraphics[width=0.5\textwidth]{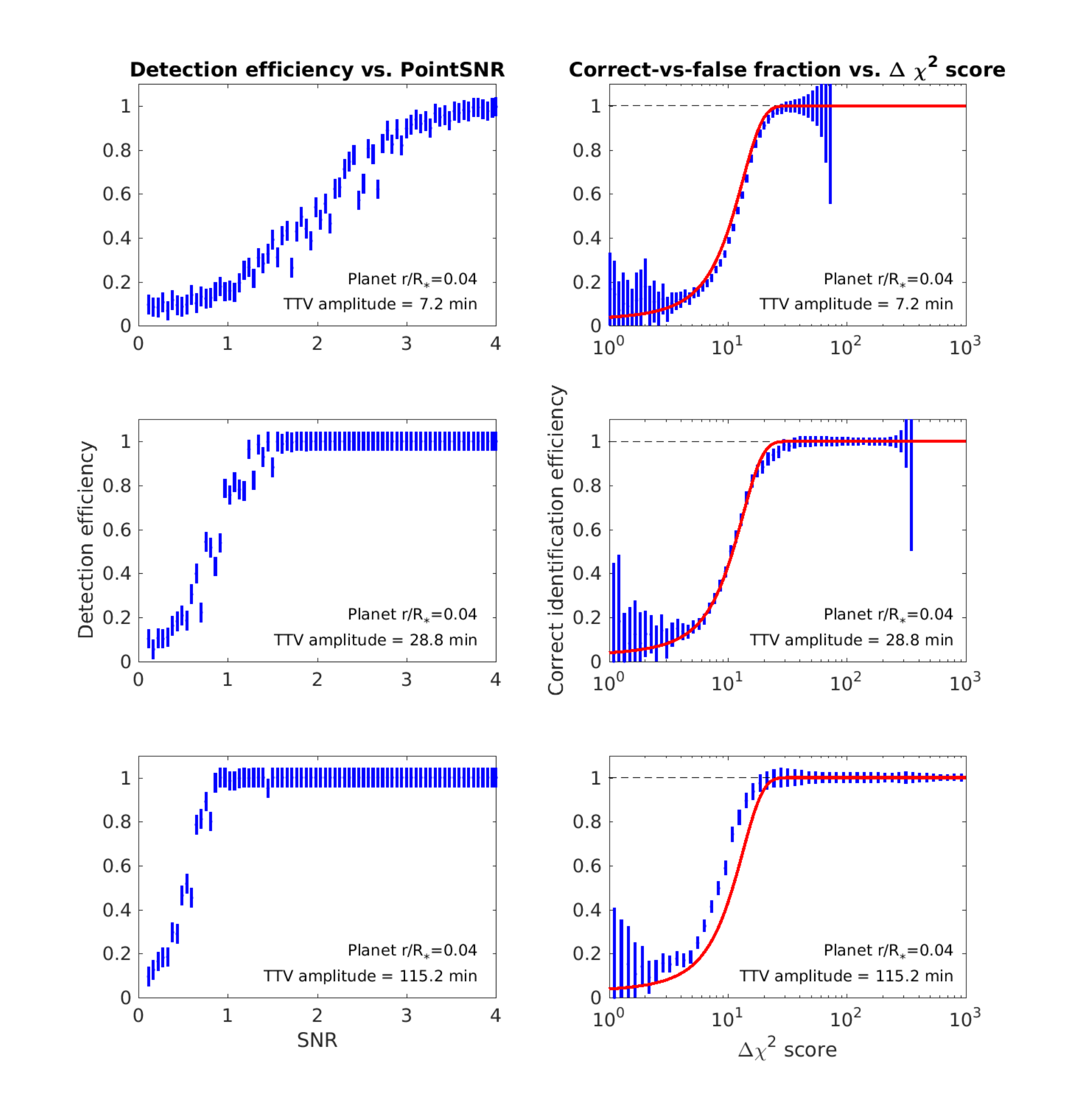}
	\caption{Transit detection efficiency of simulated light curves. The three rows show different TTV amplitudes of otherwise identical systems. \textbf{Left column:} detection efficiency at various single-transit SNR ratios .	\textbf{Right column:} the same data as the left column now sorted according the best-fit $\Delta\chi^2$. The red solid line is an empirical function (the error function), identical among the three panels, emphasizing the consistent behavior.}
	\label{DetectionEfficiencyFig}
\end{figure}

In Figure~\ref{DetectionEfficiencyFig} we show the results of three such parameter scans -- each along the transit SNR axis but at different TTV amplitude. Here, the SNR is defined per transit event, {\it i.e.}: the depth of the transit divided by the per-point uncertainty and the square root of the average number of points in a transit event. As expected, at a given transit SNR (left column of panels) detection becomes easier when the TTV amplitude increases. Scans of other parameters show similar and expected trends. Detection efficiency is higher when the planet is larger or when the noise or planetary orbital period are smaller. These different dependencies can make it challenging to establish an exact TTV detection limit of a given system. On the other hand, the panels on right column of Figure~\ref{DetectionEfficiencyFig} show the same data sorted by the $\Delta\chi^2$ score - and the red line is an empirical function (using the error function: $f(x)=\frac{1}{2}[1+\textrm{erf}((x-x_0)/\sigma)]$ with $x_0=11,\,\,\sigma=8$). The three panels show similar behavior across significantly different parameter values. Importantly, nearly identical $f(x)$ is obtained when varying all other parameters of the problem. We conclude that the spectral approach and its PA approximation allow TTVs to become uniformly detectable -- always becoming detectable with high efficiency when $\Delta\chi^2 \gtrapprox 20$ (which is a plausible threshold: $4-5 \sigma$ detections are usually regarded as reliable). This uniformity allows the spectral approach to detect TTVs using a full dataset even if no single transit is detectable, as is often the case for small planets.

\begin{figure}
	\includegraphics[width=0.5\textwidth]{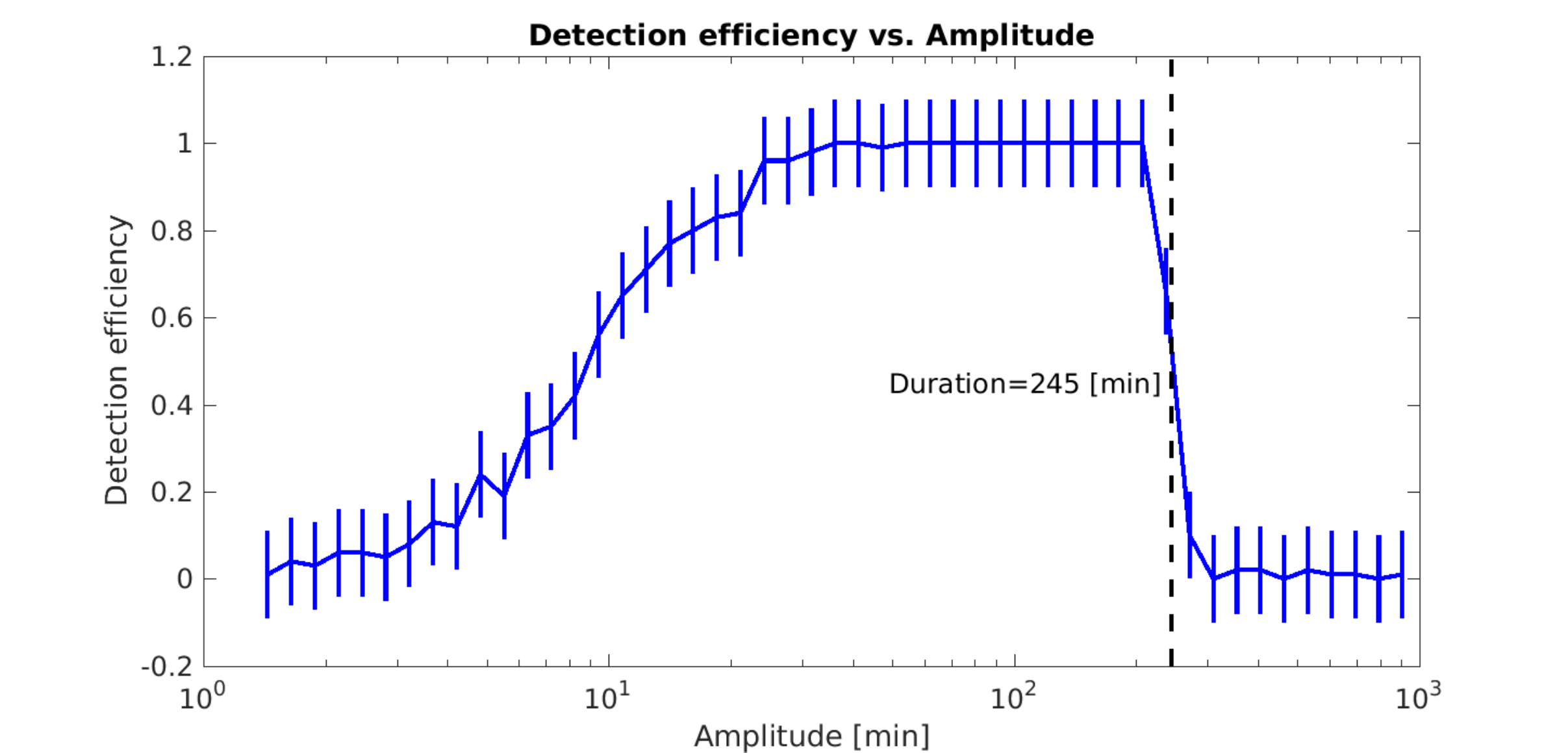}
	\caption{Transit detection efficiency of simulated light curves as a function of TTV amplitude. The simulated light curve has its noise sized to produce a low SNR of 2.5 on individual transit events if no TTVs are present - and TTVs would make the effective SNR even lower. Still, TTVs are detectable with high efficiency over a wide range / amplitudes - all the way to the expected limit of high-amplitude TTVs at TTV amplitude equal the transit duration (vertical dashed line).}
	\label{AmplitudeTestGrid}
\end{figure}

As expected, higher TTV amplitude does not always increase the detection efficiency: as explained in \S~\ref{Domain}, Figure~\ref{AmplitudeTestGrid} shows that high amplitude TTV have low detection efficiencies, if at all detected, using this technique. We note that the parameters chosen for the above test, especially the single-event SNR$<2.5$, is such that this planet would not even qualify for analysis by some catalogs (Mazeh \etal 2013) - but here PA obtains high detection efficiency over a wide range of values.

\subsection{The full spectral approach fit}
\label{FinalFit}

PA is a fast approximation to the full, but slower, spectral approach. Therefore, a full spectral approach model is computed based on the PA results for cases where significant TTVs are detected by PA. 
The full spectral approach model fits the non-periodic planet similarly to classical techniques of fitting the transit parameters simultaneously with the times of mid-transit ({\it e.g.} H16), except that the series of times of mid-transit is given by a sine function and not by a numerical vector. 
Grid-searching, as presented in \S\ref{GridSearch}, is sub-optimal. We therefore use a well-tested MCMC code ({\it e.g.} Ofir \etal 2014) to find the best-fit solution and the error ranges of eight parameters: four linear-ephemeris orbital parameters, normalized planet radius and three sine-TTV parameters. The results of the full fit usually agree well with the PA results and have somewhat better $\Delta \chi^2$ than PA (see also Figure~\ref{KOI_209_system}). Mismatches between the PA and the full fit are common when the TTV period is longer than the baseline of the data - causing the solutions range to be highly correlated (see Fig \ref{KOI_75.01}) - something that is not captured by the PA model. The adopted values are therefore the result of the full spectral approach fit.

\section{Application to \textit{Kepler} candidates}
\label{ApplicationToKepler}

\subsection{Data and pre-processing}
We used \textit{Kepler} Data Release 24 as the source data, and \textit{Kepler} Objects of Interest (KOIs) table downloaded from the NExSci archive\footnote{http://exoplanetarchive.ipac.caltech.edu/} on 25 December 2015 as the source of list of candidate signals, and processed 4706 object not dispositioned as "false positive". Many of the false alarm KOIs are eclipsing binaries (EBs), and eclipse timing variations in EBs may allow the detection of additional bodies in the systems and even circumbinary planets (e.g. Borkovits \etal 2016), and so are ostensibly closely related to the project at hand. However, high-precision modeling of EBs, and especially non-detached EBs, is significantly more complicated than modeling planetary transits. We therefore identify the likely EBs as KOIs that appear on the \textit{Kepler} Eclipsing Binaries Catalog (hereafter KEBC, Kirk \etal 2016) or have eclipse depth >10\% and are not otherwise cataloged as confirmed planets. We remove these EBs from the candidates list, and defer work on eclipse timing using the spectral approach to the future.

Having good linear ephemeris in advance is useful for optimizing the extraction of the transits. Transits of each KOI were extracted by fitting several polynomials (orders zero through five) to a small region bracketing each transit. Each such region extended three times the transit duration before and after the linear ephemeris expected time of mid-transit, and excluded the points in transit and two extra data points before and after it (since the transits were suspected to have some TTVs). Regions that did not include at least two data points before and two data points after the transit were not accepted, and these transits were ignored. Each fit was iteratively clipped to four sigma, and the common set of all remaining points in all six polynomial fits were compared such that the fit with the lowest Bayesian Information Statistic (BIS) was selected (where BIS$\equiv\chi^2+k\mathrm{ln}(n)$, $k$ is the number of model parameters and $n$ is the number of data points). The selected polynomial was then interpolated to the in-transit points to generate the background flux on each transit. Notably, variable stars with variability time scales similar to the duration of the transit itself (or shorter) will not be well-modeled using this procedure and the resultant background will be severely affected by this variability. PA results on such variable stars may not be reliable as is, and relevant information is given in our results below when such a case is identified.

Each normalized light curve was fitted with the MA02 model after accounting for \textit{Kepler}'s finite exposure time (Kipping 2010), using the MCMC code. We adjusted the five usual parameters $P, T_{\mathrm{mid}}, r, a, b$ for most KOIs, adding also the stellar limb darkening coefficients $u_1$, $u_2$ for planets with NExSci-reported model SNR >100. After this fit we rejected in-transit outliers and then calculated the $\chi^2$ of every transit event. Events that had $\chi^2$ larger by more than $4\sigma$ from the median (where $\sigma$ was assessed ignoring the event with the highest $\chi^2$) were rejected. If indeed either in-transit points or transit events were rejected then we re-fitted the remaining data.

We note that multi-planet systems where modeled such that before analyzing a given KOI all the other transit signals on the same host star were modeled-out using the NExSci linear ephemeris parameters. This has the effect that systems may be inaccurately modeled if more than one TTV-bearing planet exist and some of the transits are nearly overlapping in time. In such cases the TTVs of the \textit{other} planet (not the KOI of interest) are imperfectly modeled out by our pipeline, potentially hampering the analysis of the KOI of interest.

\subsection{Selection of significant TTVs}

In order to be selected as a significant TTV a candidate signal should pass a number of statistical tests: the boostrap analysis confidence test, and a set of tests based on cumulative $\Delta\chi^2$ (below).

\subsubsection{Boostrap analysis confidence test}
\label{Bootstrap}
For each KOI we performed a bootstrap analysis confidence test (Press \etal 1992): we generated $10^3$ artificial light curves, each by sampling with replacement the residuals to the linear model and then adding these re-sampled residuals back to the linear model. Each artificial light curve was then analyzed using PA and the peak of its TTV $\Delta\chi^2$ spectrum was recorded. The Confidence metric, equivalent to 1 minus the false alarm probability, is the fraction of those random artificial light curves that had a lower peak $\Delta\chi^2$ than the real data. Indeed, previously detected TTV-bearing KOIs clearly cluster around high Confidence metric ($\geq 0.999$) with most of the exceptions easily understood as either high-amplitude TTVs (for which PA is not suitable in the first place) or variable stars with special characteristics ({\it e.g.} at time scales similar to the transit duration, making the polynomial background estimation inadequate). We therefore use this Confidence metric as the main threshold for the selection of significant TTVs.

Note that this test checks for the reality of the most significant peak. Since the degeneracy in TTV inversion can be broken in cases where additional super-frequencies are detected, we use this test also as a rough guide to the significance of other peaks in the PA spectrum.

\subsubsection{$\Delta\chi^2$ test}

As shown in \S \ref{SimulatedDetection}, high efficiency detection is expected for $\Delta\chi^2\simeq20$ or larger. We therefore checked if any of the high-confidence signals had lower $\Delta\chi^2$; we found only one such case, with only a slightly lower $\Delta\chi^2$, confirming $\Delta\chi^2\simeq20$ is a good discriminant value.

\subsubsection{Cumulative $\Delta\chi^2$ definition and tests}
\label{CumChi2}

It can be difficult to judge if a $\Delta\chi^2$ for a given system is due to a real signal or just some outliers as $\Delta\chi^2$ is only summary total of the difference between the linear and the perturbed models. However, the two scenarios above differ in their rate of accumulating $\Delta\chi^2$ in time - a real signal would gradually accumulate $\Delta\chi^2$ whereas if due to an outlier event much of the $\Delta\chi^2$ would accumulate over a small number of data points. Below we detail a few statistical tests one can perform to identify false positive TTV signals when using the PA technique by distinguishing the cases with and without TTVs. The observed cumulative $\Delta\chi^2$ curve is:
\begin{equation}
\mathrm{Obs}_i =\sum_{j\le i}{\left(\frac{y_j-m_{T,j}}{\sigma_j}\right)^2 -
                     \left(\frac{y_j-m_{L,j}}{\sigma_j}\right)^2}
\end{equation}
where $\{m_L\}$ is the linear ephemeris model and $\{m_T\}$ is the best-fit TTV model for the data and errors $\{y, \sigma\}$, and where the best-fit model $m_T$ always has better (lower) $\chi^2$ than $m_L$. Moreover, if the best-fit model is correct ({\it i.e.} in the limit of $y\rightarrow{m_T}$) then there is specific shape to this $\Delta\chi^2$-gain curve which we call the "expected" curve:

\begin{equation}
\mathrm{Exp}_i =-\sum_{j\le i}{\left(\frac{m_{L,j}-m_{T,j}}{\sigma_j}\right)^2}
\end{equation}

\begin{figure*}
\begin{tabular}{ll}
    (a) & (b) \\
	\includegraphics[width=0.5\textwidth]{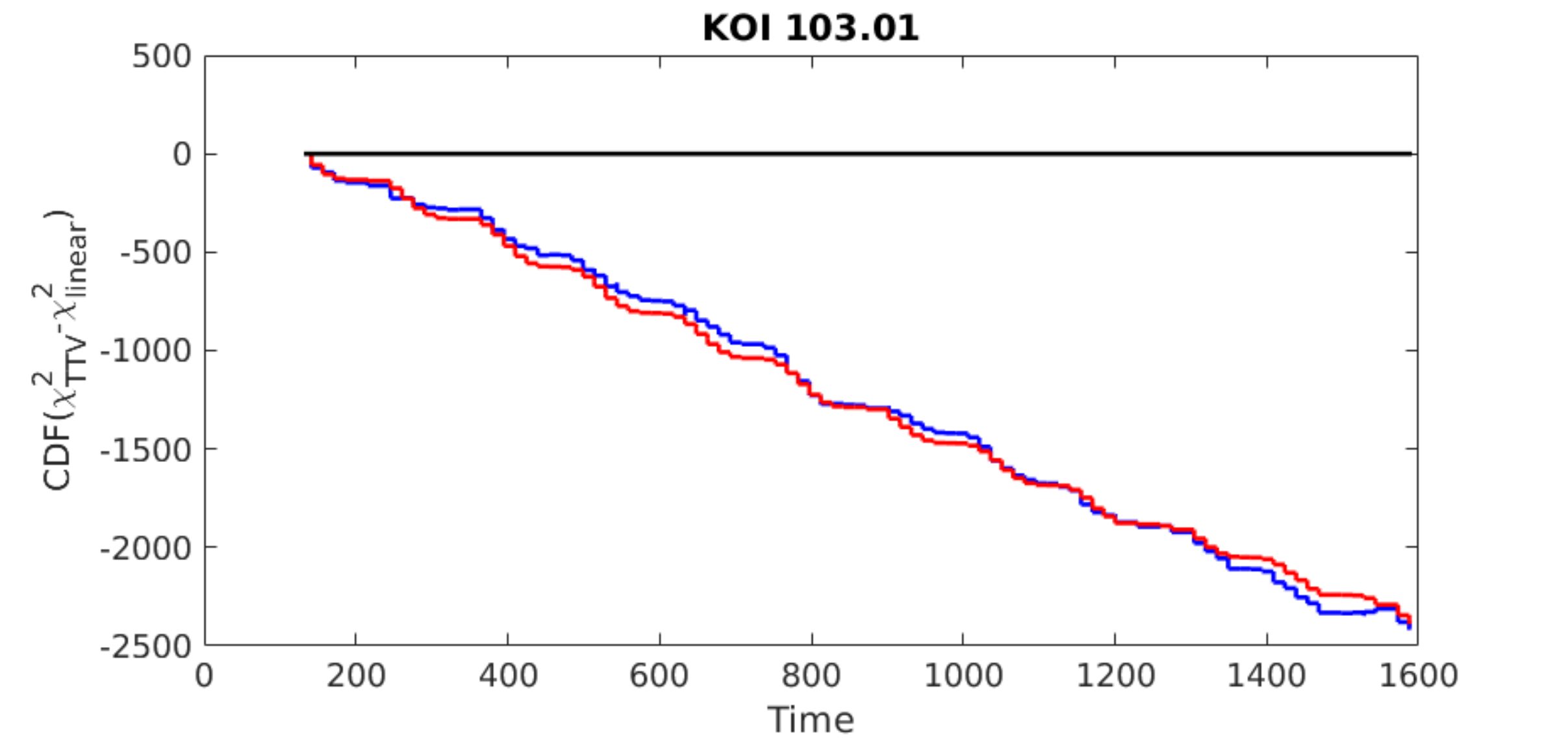} &
	\includegraphics[width=0.5\textwidth]{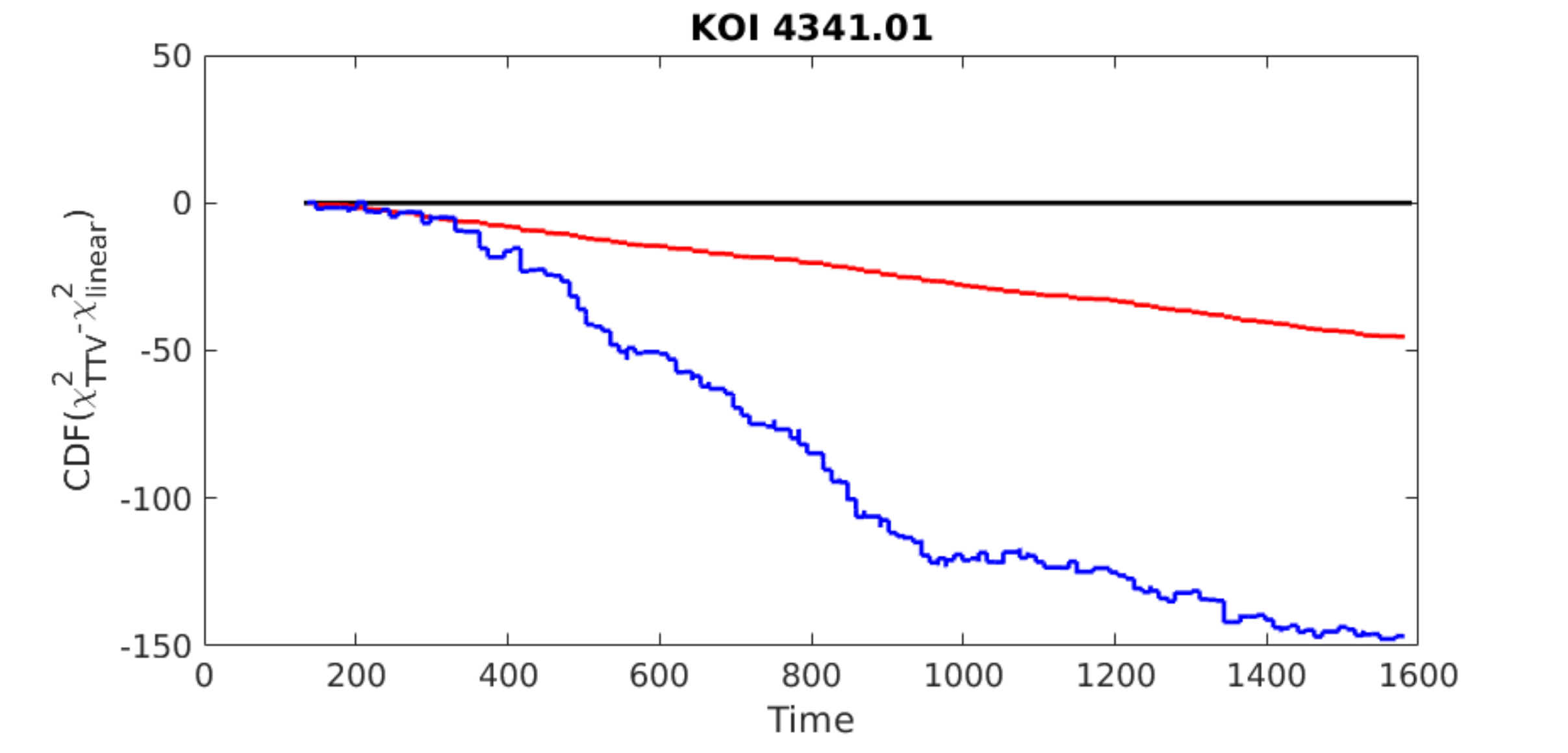} \\
    (c) & (d) \\
	\includegraphics[width=0.5\textwidth]{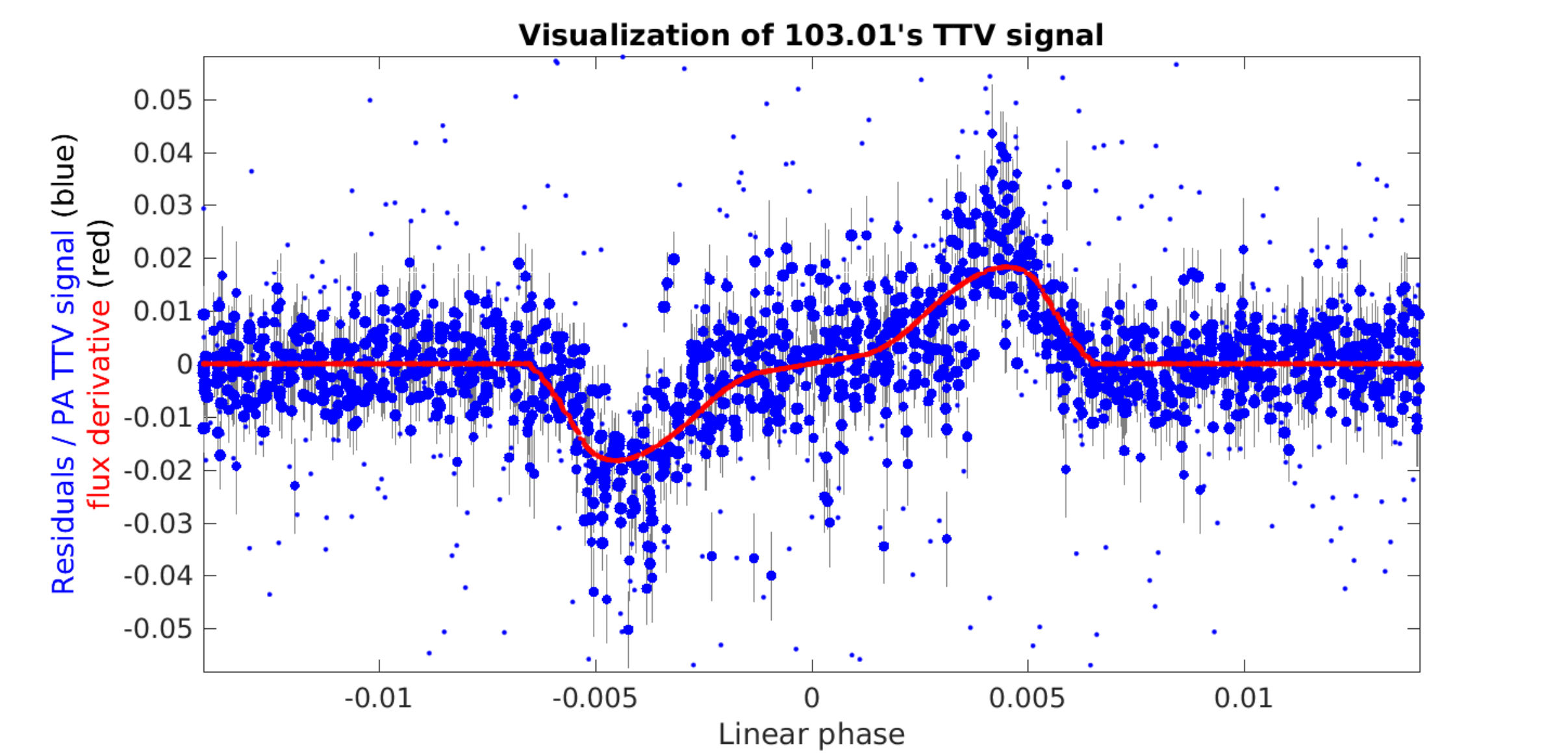} &
	\includegraphics[width=0.5\textwidth]{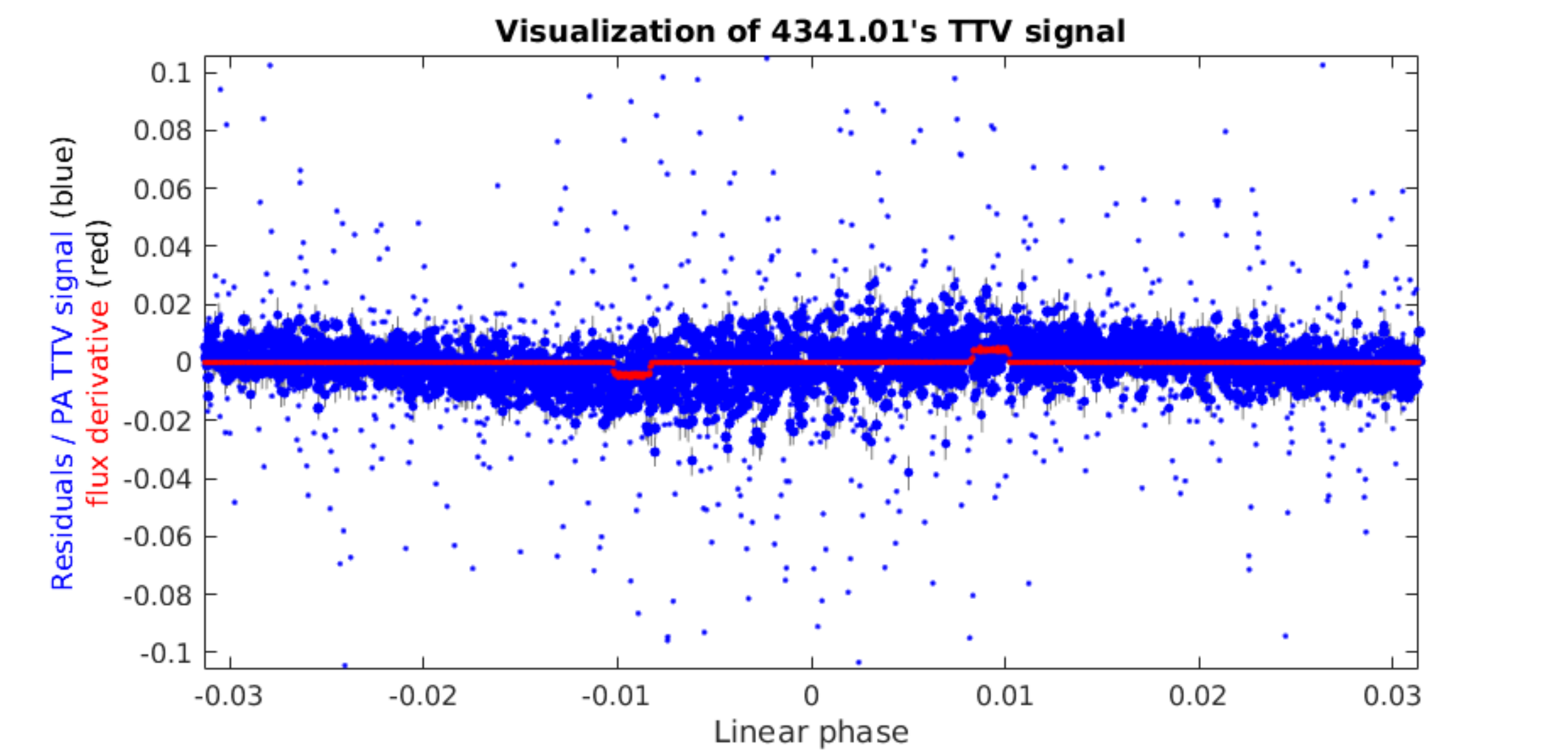} 
\end{tabular}
	\caption{\textbf{Top row:} The expected (red) and observed (blue) accumulation of $\Delta\chi^2$ relative to the linear ephemeris solution along the light curve of: (a) KOI 103.01 and (b) KOI 4341.01. Real TTV and a correct TTV model are manifested by slow accumulation of $\Delta\chi^2$ and shared common behavior of the two curves -- as is shown for KOI 103.01. On the other hand, on KOI 4341.01 the observed and expected $\Delta\chi^2$ curves are very different, creating a large area between them -- despite that the signal has high observed $\Delta\chi^2$ of $\approx150$ and high Confidence.
    \textbf{Bottom row:} A visualization of the (c) KOI 103.01 and (d) KOI 4341.01 TTV signals. For both planets their respective best fits are shown, but for KOI 4341.01 this best fit does not correspond to a real TTV signal. The linear ephemeris residuals divided by $S g_i$ are in blue and the linear ephemeris model's derivative is in red. The size of the points is scaled by $|g_i|$ and error bars are drawn only when $|g_i|>0.5$.}
	\label{Cumulative_DeltaChi2}
\end{figure*}

If the TTV model is correct, then $y\rightarrow m_T$ and $\mathrm{Obs}_i$ reduces to $\mathrm{Exp}_i$. On the other hand, if there are no TTVs in the data then $y\rightarrow{m_L}$ and $\mathrm{Obs}_i$ will differ from $\mathrm{Exp}_i$. Generally, a wrong TTV model would manifest itself by exhibiting significantly different observed/expected cumulative $\Delta\chi^2$ curves. In Figure~\ref{Cumulative_DeltaChi2} we demonstrate this metric on the well-known TTV bearing star KOI 103.01, and on another signal, apparently with high-confidence and high-$\Delta\chi^2$, that fails some of the cumulative $\chi^2$ tests.
The cumulative $\Delta\chi^2$ curves are information-rich and we use them to design the following tests for the reality of a PA-detected TTV signals:

\textbf{(i) Normalized area between curves:} the normalized area between the $Obs_i$ and $Exp_i$ curves can be defined as: $\sum_i{|\mathrm{Obs}_i-\mathrm{Exp}_i|}/\sum_i{\mathrm{Exp}_i}$. False positive detections are expected to have large normalized area between the two curves.

\textbf{(ii) maximum single-point $\Delta\chi^2$:} a real TTV signal would accumulate $\Delta\chi^2$ slowly over time, while an attempt of PA to minimize $\chi^2$ due to some outliers would accumulate most of the $\Delta\chi^2$ in a small region. By recording the highest single-point $\Delta\chi^2$ gain (absolute value) for each KOI one may detect irregular behaviour. We note that this test bears similarity to the KS test.

\textbf{(iii) RMS of difference:} the RMS of the difference between the $\mathrm{Obs}_i$ and $\mathrm{Exp}_i$ curves is computed. True TTV signals should have a low value for this RMS, so we required that all new TTV signals will have a smaller RMS value than the ones found on the H16 objects (that are not high amplitude).

\textbf{(iv) Correlation coefficient:} The correlation coefficient between the $\mathrm{Obs}_i$ and $\mathrm{Exp}_i$ curves is computed. True TTV signals should have high correlation between $\mathrm{Obs}_i$ and $\mathrm{Exp}_i$, so we required that the correlation coefficient for all new TTV signals will be larger than the ones found on the H16 objects (that are not high amplitude).

We computed these statistical measures above for all KOIs, and required that all new PA detections will be in the range spanned by the H16 high significance candidates. We found 8 KOIs that passed all other criteria but failed the normalized area test, and indeed all of them were found to be false positives by visual inspection (seven strongly pulsating stars and one EB). The remaining signals passed all the additional tests.

\subsubsection{Visualization of the TTV signal:}

By rearranging Eq.~\ref{Linearization} to express the model derivative, and applying it to the measurement set $F_i$, one obtains an observational estimation of the flux derivative. This can be used to visualize the PA-detected TTV signal, by comparing the observed shape to the expected one, {\it i.e.} a transit derivative.

\begin{equation}
  F'_i=\left(F_i-m_i\right)/(S g_i).
\end{equation}

The lower panels of Figure \ref{Cumulative_DeltaChi2} depicts this residual. Note that this is not the quantity that is fitted, but is only used for visualization. Since the oscillating function $g_i$ is in the denominator, some points diverge on the plot.

\section{Results of application to \textit{Kepler} candidates}
\label{Results}

\subsection{General statistics and comparison with H16}
\label{GenStat}

\begin{figure}
	\includegraphics[width=0.5\textwidth]{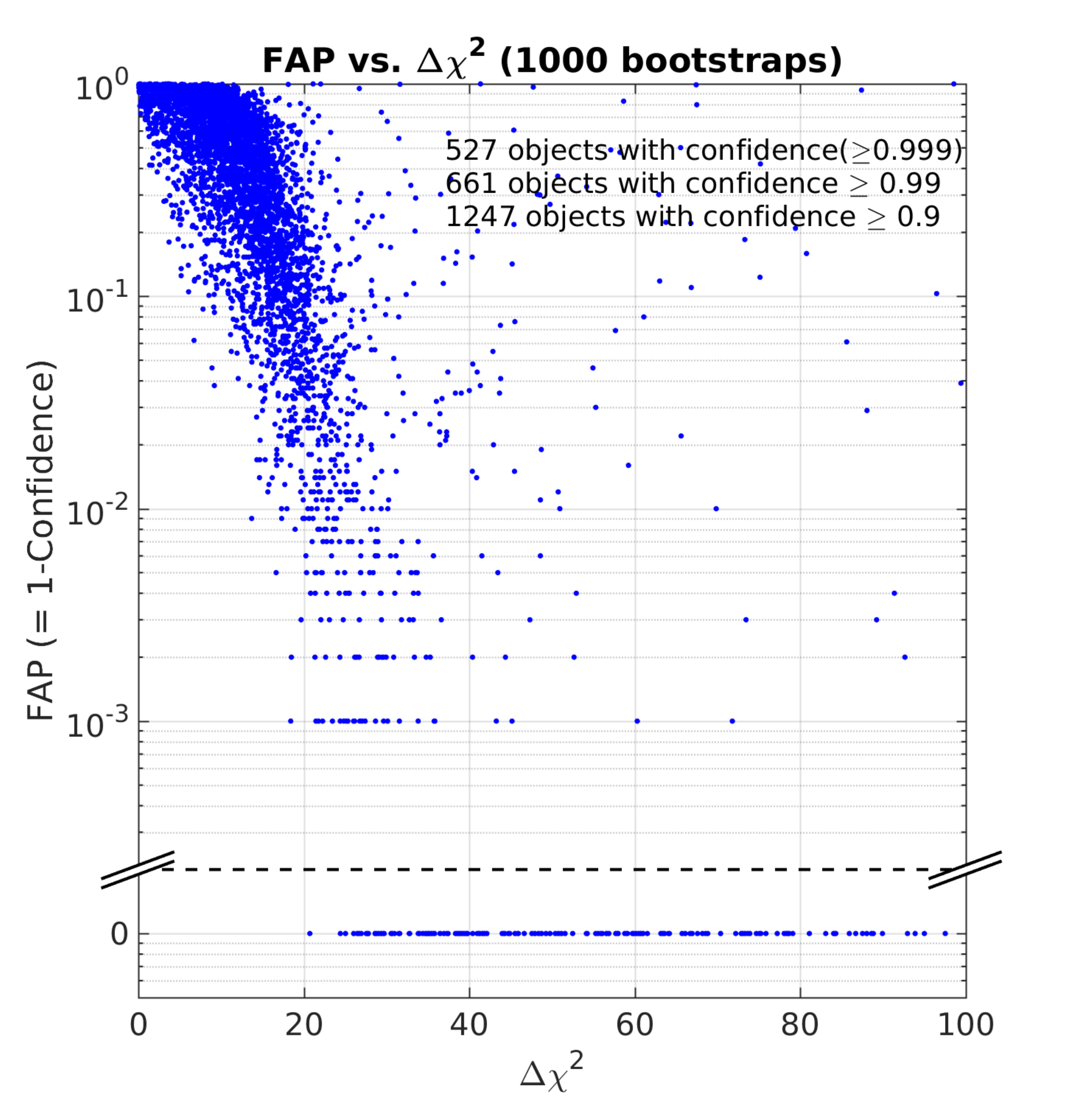}
	\caption{The distribution of FAP (false alarm probability) vs. $\Delta\chi^2$ for all analyzed KOIs (blue dots).
	Higher confidence object are appear on the bottom of the plot. The axes were chosen to highlight the distribution near the transition from low- to high- confidence objects -- most high-confidence TTVs have $\Delta\chi^2>100$, {\it i.e.} beyond the scale of this figure.}
	\label{Chi2_vs_Confidence}
\end{figure}

We analyzed 4605 KOIs that met the rather minimal requirements of not being marked as false positives by the NExSci archive, and including at least three usable transits after preprocessing. We found 527 objects with Confidence $\geq0.999$, all of them but one also with $\Delta\chi^2>20.7$ (see Figure~\ref{Chi2_vs_Confidence}), similar to the expected threshold of $\sim20$ from the tests on simulated data. False positive comprise 198 of these object, arising due to the following reasons: (i) they appear in the current KEBC; (ii) they have depth $>10\%$; (c) they have high-amplitude TTVs (making the PA-detected value likely wrong); (d) The TTV frequency was the maximal one (see discussion on \S~\ref{Recurring}) (e) they failed any of the cumulative $\Delta\chi^2$ test (\S~\ref{CumChi2}) (f) they had scatter-to-error ratio $>50$. Collectively these objects will be termed the EBs/FPs (eclipsing binaries and false positives) sample, while the sample of objects with Confidence $\geq0.999$ that are not EBs/FPs will be termed the High Confidence TTVs.

The list of 329 remaining KOIs with high-confidence TTVs was cross-referenced to the TTV catalogs of: H16, Xie (2013), Xie (2014), Hadden \& Lithwick (2014, 2016) (hereafter HL14, HL16), Jontof-Hutter \etal (2016) (hereafter JH16) and Van Eylen \& Albrecht (2015). In all cases we required that the reported TTV amplitude in these catalogs will be significant to at least $3\sigma$ as determined by each catalog, in order to be considered as a significant TTV detection by these catalogs, \textit{i.e.,} a "known object". After removing the known objects, 165 remaining KOIs with new periodic TTVs were identified, of which 34 are present in the H16 catalog with long-term and unconstrained periods (dubbed "polynomial TTVs"), Thus 131 TTV signals found here are considered new. Our results are given in Table \ref{ResultsTable}.

\begin{figure}
	\includegraphics[width=0.5\textwidth]{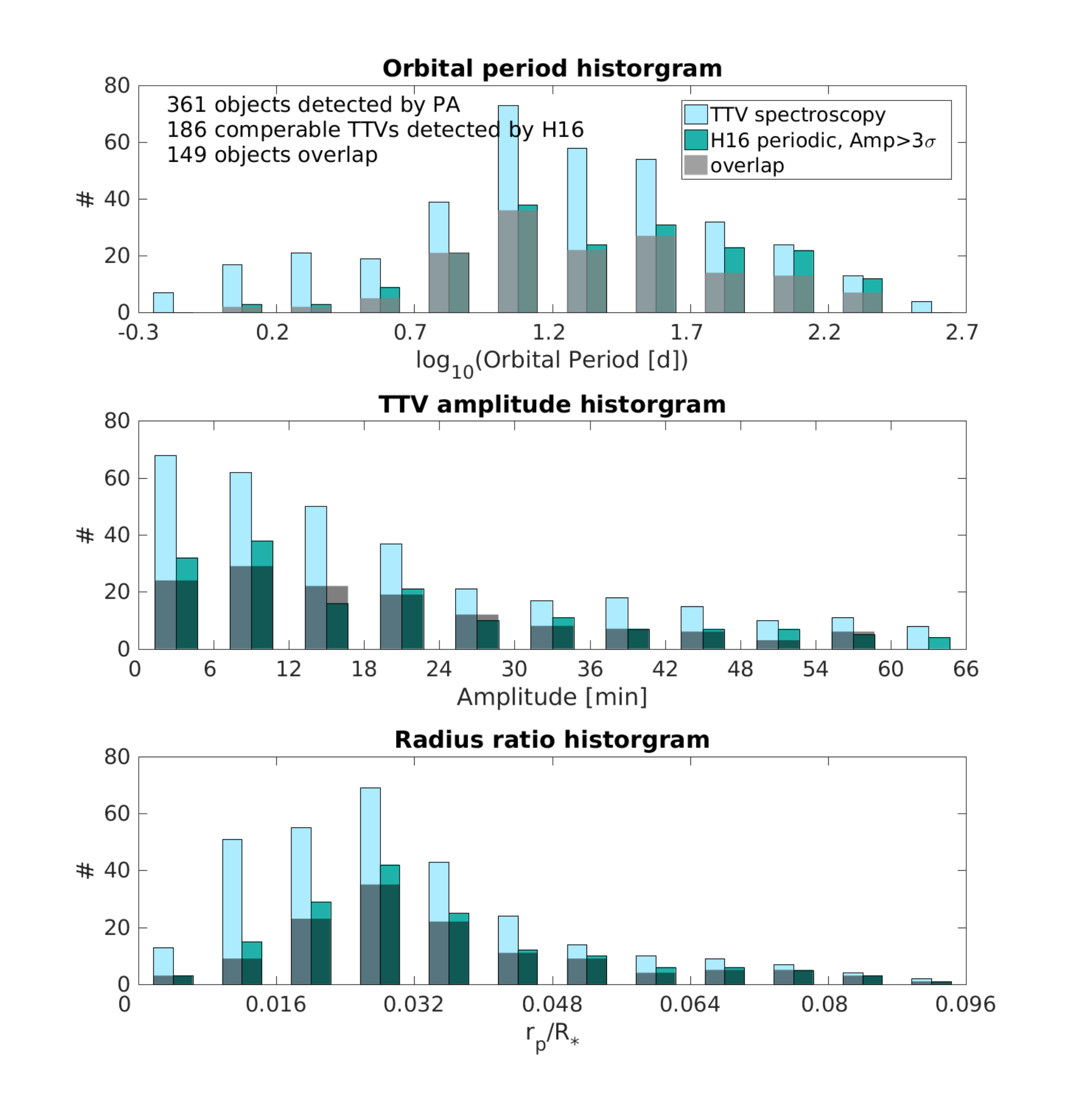}
	\caption{High Confidence TTVs detected by PA (light blue), the TTVs detected by H16 that have amplitude significant to more than $3\sigma$ and are not High Amplitude (cyan) and the overlap between the two sets (darker shade), focusing on the lower range of all parameters.
	\textbf{Upper panel:} Orbital periods distribution: PA overdetects TTVs on short-period planets.
	\textbf{Middle panel:} TTV amplitude distribution: PA overdetects TTVs of lower amplitude. Note that identical objects may have somewhat differently determined TTV amplitudes in each catalog, hence the slightly different overlap regions in similar bins.
	\textbf{Lower panel:} relative planets size distribution: PA overdetects TTVs on planets with smaller relative ratii. Note that since H16 does not quote this parameter, all $r_p/R_*$ were taken from the MAST database for both sets of KOIs.}
	\label{Compare_P_and_A_TTV_and_r}
\end{figure}

\begin{figure}
	\includegraphics[width=0.5\textwidth]{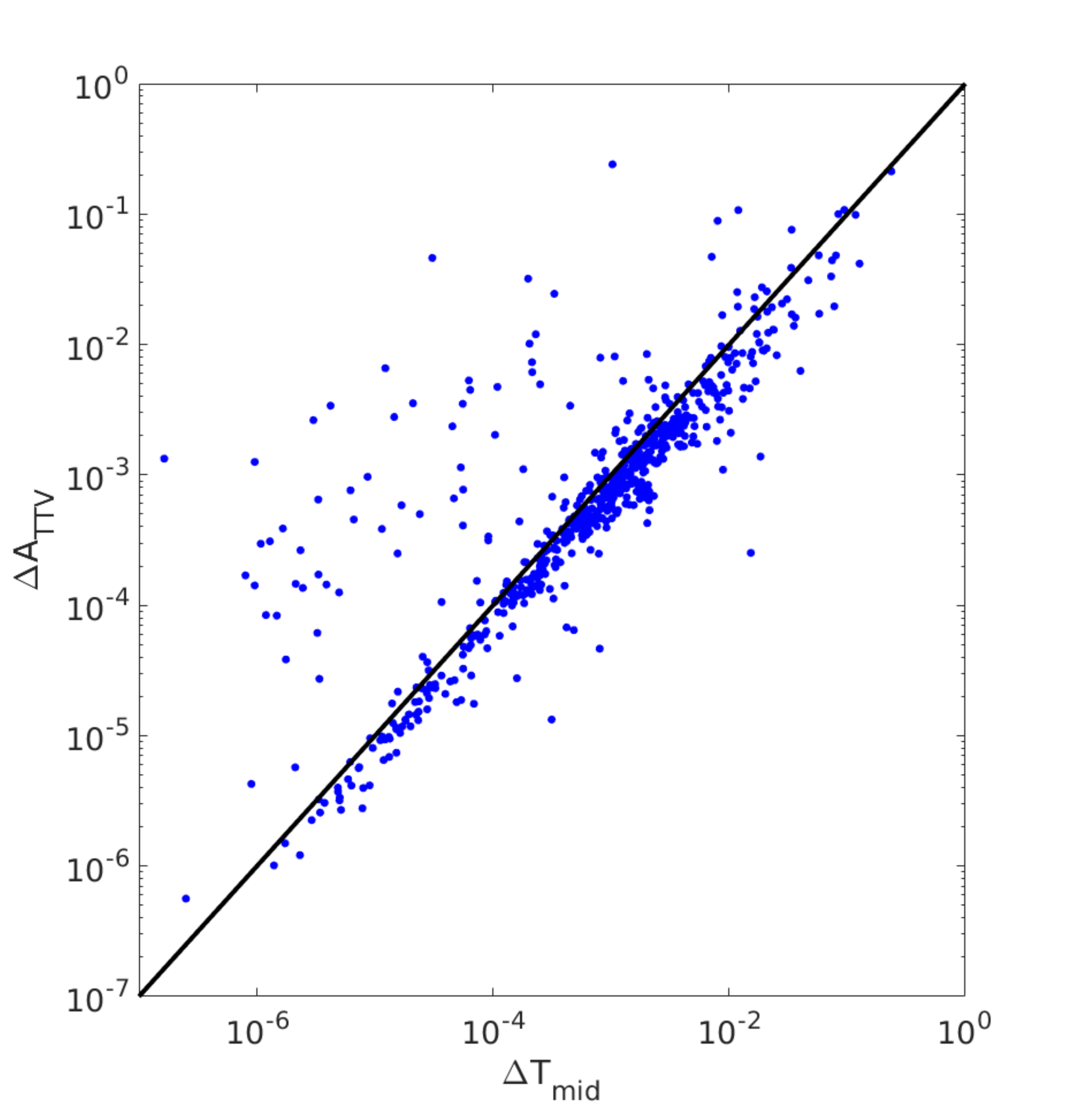}
	\caption{The $1\sigma$ error range for the TTV amplitude vs. the $1\sigma$ error range for the linear $T_{\mathrm{mid}}$ -- both taken from the non-linear fit. 
	The correlation is clear over multiple decades. The locus of the 1:1 ratio is shown as a thin line.}
	\label{Tmid_TTVamp_connection}
\end{figure}

The power of the spectral approach is seen by comparing the distribution of TTV-bearing KOIs detected using PA and the classical techniques ({\it e.g.} H16 in Figure \ref{Compare_P_and_A_TTV_and_r}). As predicted, we see that PA can detect TTVs for planets with shorter orbital periods, lower amplitude and smaller size than was previously possible, and the higher object count indicates that PA is also more sensitive. Numerically, as mentioned above the median orbital period for planets that exhibit TTVs in the H16 catalog is more than three times the $\sim9.5$~d median period of all \textit{Kepler} candidates, while the median period of the Spectral Approach new TTV detections is $\sim10.8$~d - eliminating the period bias. Similarly, the median transit depth of the H16 objects is $\sim1075$~ppm while the same number for of all \textit{Kepler} candidates and the Spectral Approach detections is $\sim428$~ppm and $\sim458$~ppm, respectively - eliminating the depth bias. Additionally, in Figure~\ref{Tmid_TTVamp_connection} we see that the precision in the amplitude of the detected TTVs (and thus, the sensitivity to them) scales exactly as the precision on the linear $T_{\mathrm{mid}}$ over more than five decades of amplitude.

During this work, we noticed a few issues with the H16 catalog to which we compare: (i) the TTV detection and final fits are separate steps, and their results are occasionally incompatible. In particular, 22 of the reportedly significant TTVs (p-values $<10^{-4}$, see H16) have low amplitude significance $\mathrm{A}/\sigma_{\mathrm{A}}<3$. These 22 TTV signals are questionable: we only detected high-confidence TTVs on two of these objects (KOIs 282.01, 1783.01), and only the first one is consistent with H16. We therefore term the remaining 20 signals 
as low-significance H16 KOIs, and do not include them in the following comparison with H16. (ii) The uncertainty of the linear ephemeris orbital period was found to be systematically too low in H16: while we obtained similar uncertainties for short-period KOIs, the H16 uncertainties were progressively smaller than ours towards longer period, typically about an order of magnitude smaller and approaching three orders of magnitude smaller at the long-period end. Examination of the individual signals confirms that these uncertainties are underestimated in H16. 
(iii) the H16 uncertainty on the linear-ephemeris time of mid-transit is almost always larger than our own. The difference is far smaller than in point (ii) above (median factor of 2.35) and thus is probably unrelated to it.

Many of the TTV-bearing KOIs we found were previously identified in the literature. Indeed, of the 144 objects in common with the H16 catalog, 132 have TTV period consistent within $3\sigma$ between the catalogs, demonstrating the compatibility of the spectral approach with the classical TTV identification techniques used in H16. It is  noteworthy that due to the lower number of free parameters the spectral approach allows improving the precision of the determination of the TTV parameters. The uncertainty on the TTV period was reduced by a median factor of $\sim10\%$ and the uncertainty on the TTV amplitude was reduced by a median factor of $\sim40\%$.

There are 30 stars with periodic and significant TTVs identified by H16 that were not detected by PA. We evaluated each of these stars manually, and found that about half of them reside in active stars with high frequency variability (scale of transit duration or shorter) making TTV detection a more subtle issue. Nearly all the rest were either low-significance to begin with (amplitude significant to less than $4\sigma-5\sigma$ according to H16) or that the same TTV frequency was detected by PA and H16 but different confidence levels were attributed to the signals. Indeed only in a two cases (KOIs 1581.02, 4519.01) no good explanation was identified for the PA missing robust H16 signals.

\subsection{Discussion}
\label{Recurring}
Below we list a number of recurring phenomena that can be useful in understanding the results given in Table \ref{ResultsTable}.

\textbf{The chosen cutoff levels are somewhat arbitrary:} 
We provide in Table \ref{ResultsTable} information on more than just the high-confidence targets (>0.999), to help identify the ones that may, with relatively little additional input data or detailed analysis, become high confidence.

\textbf{Best fit TTV frequency is the maximal one:} 
If an EB with two similar eclipses is misidentified as a transiting planet, any difference between the odd and even eclipses (that may actually stem from non-zero eccentricity, different surface brightness of the stars, etc.) may cause the PA fit to incorrectly add high-frequency TTV at twice the "planetary" orbital period. We found 75 High-Confidence TTV signal that are close to the maximal one, and these KOIs are suspected as EBs or otherwise false positives. Indeed, 60 of these were already in the EBs/FPs sample, and the rest (but one) appear in H16's suspected false positives list (their Table 1). 
In practice, the maximal frequency at which we searched for TTVs was limited by the orbital period $P$ of the transiting planet:
$f_{\mathrm{max}}={1}/{(2P)}$. Since the frequency resolution scales with
the time span of the data, we labeled systems with best-fit TTV frequency within $s^{-1}$ of $f_\mathrm{max}$ as suspect. We note that sometimes stellar pulsations that survived filtering also caused such high-frequency apparent TTVs.  To summarize, in cases where the best fit TTV frequency is consistent with the maximal one, it is a strong sign of a misidentified EB and those objects are labeled likely false positives. However they are given in Table \ref{ResultsTable} for completeness of high-confidence signals.

\begin{figure}
	\includegraphics[width=0.5\textwidth]{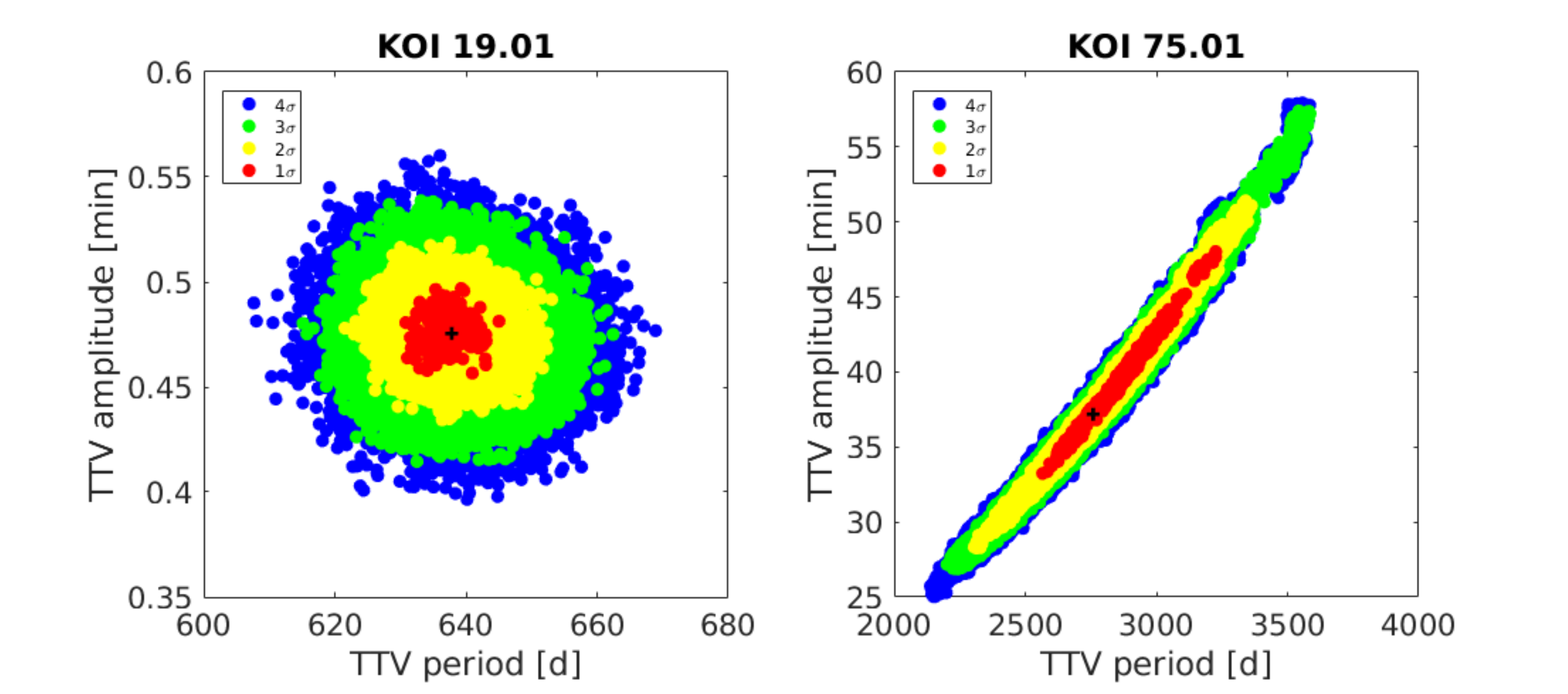}
	\caption{Posterior distributions of the spectral approach optimization for KOI 19.01 (Left) and KOI 75.01 (right) in the TTV period and TTV amplitude plane. Error ranges are the location of constant $\Delta\chi^2$ relative to the best-fit position, which is marked in a black '+'. Stars that have TTV periods smaller than the data time span show little or no correlation between the TTV period and its amplitude, while significant correlation exists for the longest TTV periods. For e.g., KOI 75.01 was determined by H16 to have polynomial TTVs but here we show that the TTV frequency can be well constrained (better than $3\sigma$) using the spectral approach - albeit with significant amplitude correlation. We note that the actual MCMC jump parameter is the TTV frequency, which is more uniformly distributed, but we presented here the TTV period which is more intuitive.}
	\label{KOI_75.01}
\end{figure}

\textbf{Very long TTV periods:} Usually the error ranges are symmetrical in frequency space. However, on very long periods $P_{\mathrm{TTV}}>s$ where $s$ is the span of the data, the period error range is highly correlated with TTV amplitude. This is expected at such long TTV periods as the data does not allow seeing even a single complete TTV period, making the observed amplitude either close to the total real amplitude (if $f_{\mathrm{TTV,true}}\leq s^{-1}$) or just a fraction of it (if $f_{\mathrm{TTV,true}}\ll s^{-1}$). Importantly, in such cases the best constraint is on some function of $f_{TTV}$ \textit{and} $A_{TTV}$ and not on each of them individually (see Figure \ref{KOI_75.01}). For this reason some objects in Table \ref{ResultsTable} appear to have low significance to either $f_{TTV}$ or $A_{TTV}$ however these objects all have very long TTV period and so the detection of some long-period TTVs on these objects is correct, but the exact amplitude and frequency of these TTVs are more poorly constrained.

\textbf{Stroboscopic frequency: } \textit{Kepler}'s finite exposure time may also produce a stroboscopic effect, exhibiting apparent TTVs on strictly periodic transiting planets, when the orbital planet's period of happen to be close to an integer multiple of the exposure time (Szab{\'o} \etal 2013, Mazeh \etal 2013). We therefore also provide in Table \ref{ResultsTable} the expected stroboscopic frequency and note that it was detected in practice multiple times.

\textbf{Unexplained significant TTV signals:} TTV signal with no apparent connection to any other known object in the system were detected multiple times. Such TTVs are very likely due to interaction with additional planets in the system that are not transiting. Such systems are  fertile ground to RV surveys to connect the inner and outer parts of multi-planet systems.

\textbf{Multi-periodic TTVs:} More than half (277 objedts) of all stars with high-confidence TTVs were found to have more than one significant TTV frequency. Additional TTV frequencies allow breaking the degeneracy in the TTV inversion back to absolute masses and are therefore very useful - provided they are real. Indeed, Pulsating/variable stars can create spurious TTV signals (see also the following paragraph). On the other hand, a high number of apparently significant TTV frequencies (we adopted the >5 threshold) is likely a sign of imperfect filtering of a variable star and not of multiple dynamical phenomena - and 91 stars exhibit it.

\textbf{Pulsating/variable stars:} These are more difficult to filter, and sometimes residuals of the variability signal remain, especially when the variability time is close to- or shorter than-  the transit duration itself. Such stars frequently exhibit TTVs but those are difficult to judge for reliability without individually-tailored filtering, spot- or pulsations- modeling, etc. To indicate this as well as cases that may have not been filtered and/or modeled well, we provide in table \ref{ResultsTable} the ratio of scatter around the linear model to the median error. As a general rule, ratios lower than two usually mean good filtering and modeling (unity being the white noise limit), and ratios larger than three should be reviewed on a case-by-case basis unmodeled phenomena likely exist. None of the objects with scatter-to-error ration $>50$ seemed to be actually reasonably modeled with linear ephemeris and thus these objects were removed from further analysis as strongly pulsating. We also visually inspected all high-confidence signals and commented on systems that appeared to be affected by such effects.

When computing the bootstrap analysis, we saved in addition to the best $\chi^2$ of each mock data, also its entire PA spectrum. This in turn allows us to build a smaller bootstrap test for each frequency individually: how often did test frequency $f$ had higher $\Delta\chi^2$ than our final $\Delta\chi^2$ cutoff? by counting these frequencies one can easily identify systems that likely include multi-frequency information (and thus possibly enable inversion for masses) even if this is not visible by eye. We note that poorly modeled systems that exhibit large ratio of scatter to the median error (also reported on Table \ref{ResultsTable}) are prone to exhibiting unrealistically large number of apparently-significant frequencies.


\subsection{Specific systems}

\begin{figure}
	\includegraphics[width=0.5\textwidth]{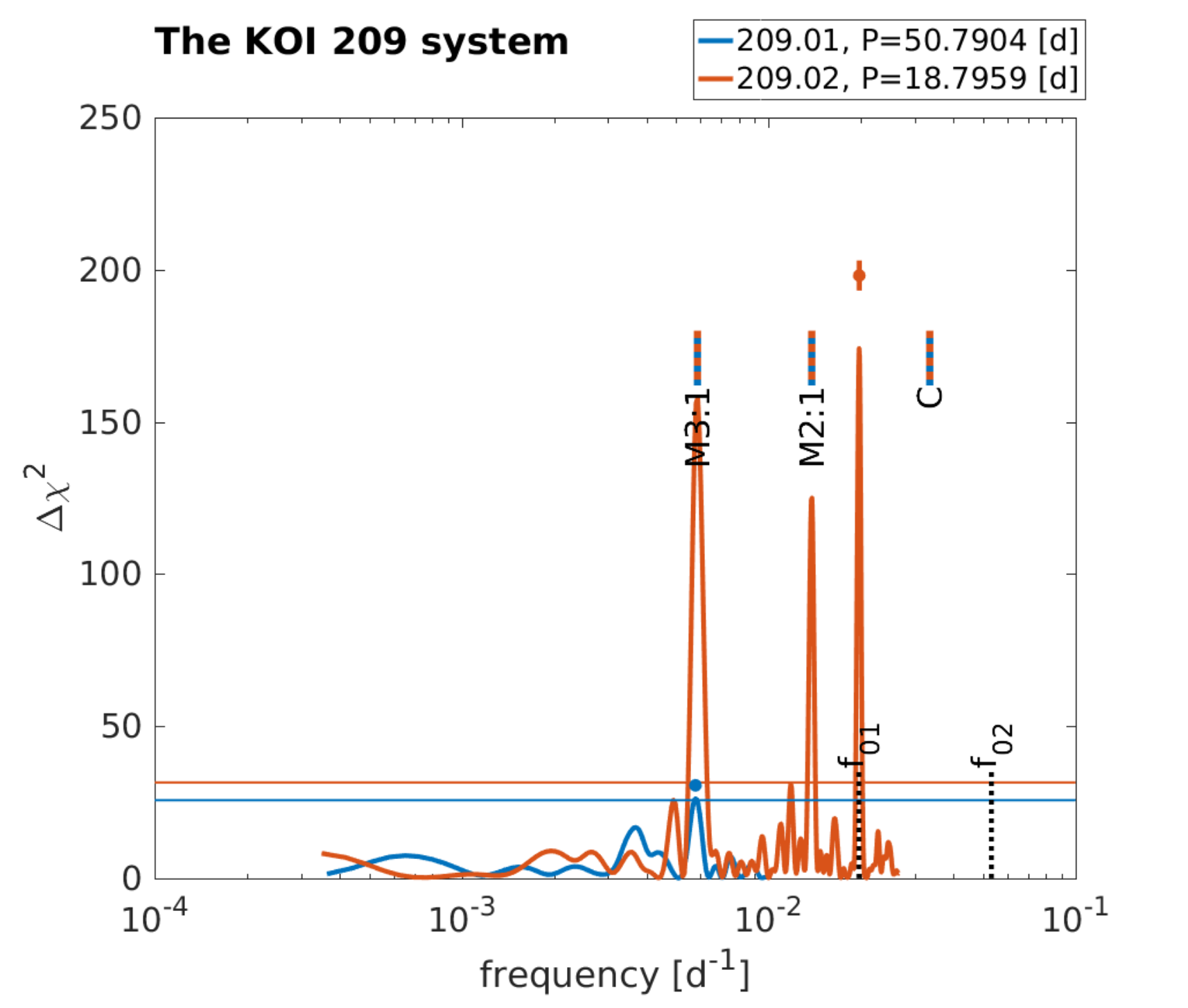}
	\caption{PA spectra of all KOIs in the KOI 209 system, each planet/signal is plotted in a different color. Relevant theoretically-expected super-frequencies between any two interacting planets that are discussed in the text are marked at the top in a colored dashed vertical line in the colors of the relevant pair and the letter "M", as well as an indication of the relevant MMR. Similarly, the chopping frequencies associated with each pair may also be indicated with a "C". The peak of the PA spectrum was the starting point for non-linear optimization, the results of which are given as a horizontal error bar in the appropriate color for each TTV signal that was subjected to non-linear optimization - usually directly above it and with somewhat better significance (higher $\Delta\chi^2$). The different orbital frequencies are indicated at the bottom with vertical black dotted lines marked with an "f" and the relevant signal's number. If a the stroboscopic frequency of a particular planet is in the scanned frequency range, it is similarly marked with a dashed line and the "S" label (not relevant for KOI 209). In the textbook-like case of KOI 209.02 it is a clearly simultaneously affected by two different nearby resonances and and the orbital period of 209.01.}
	\label{KOI_209_system}
\end{figure}

Here we discuss some of systems for which new information was gained by applying the above analysis. Our goal was to flag interesting systems and not to fully characterize them in depth (given the scope of this paper), and our main tool for analyzing the systems are plots as shown in Figure \ref{KOI_209_system}: in each such figure we superimpose the PA spectra of all transit signals in given system as well as all expected and TTV frequencies, which are of four types: (1) all possible super-frequencies expected from all MMRs with $j:j-N$ period-ratio up to $j=9$ (arbitrarily) and $N=j-1$ using equation from \S 5 of Deck \& Agol 2016.
(2) The orbital frequencies. (3) The so-called "chopping frequencies" which are the frequency of conjunctions between any planet pair: $f_\textrm{C}=f_\textrm{in}-f_\textrm{out}$. (4) The expected stroboscopic frequencies of individual planets. This rather dense figure allowed to quickly assign an observed PA peak with a possible physical meaning, even in high-multiplicity systems. Plots like Figure \ref{KOI_209_system} show only those expected frequencies which are found to be relevant to a given system and discussed in the text.


\begin{itemize}[noitemsep,topsep=1pt,parsep=1pt,partopsep=1pt,leftmargin=0pt]

\item{} \textbf{KOI 89 / Kepler-462:} (Figure \ref{KOI_89_system}) KOI 89.02 was detected by the PA to have significant TTVs with $f_{\mathrm{TTV}}\approx9\cdot 10^{-4} d^{-1}$. The non-linear fit revised this value to $f_{\mathrm{TTV}}=(5.39_{-0.74}^{+0.67}) \cdot10^{-4}$ - consistent with the predicted 5:2 MMR with KOI 89.01 at $f_{\mathrm{Sup}}=4.704\cdot 10^{-4} d^{-1}$, while previous analyses (e.g. H16) had a $>4\sigma$ discrepancy. This, together with the high-confidence of the TTV detection, confirms KOI 89.02, hitherto just a candidate, as a bona-fide planet. 
\begin{figure}
	\includegraphics[width=0.5\textwidth]{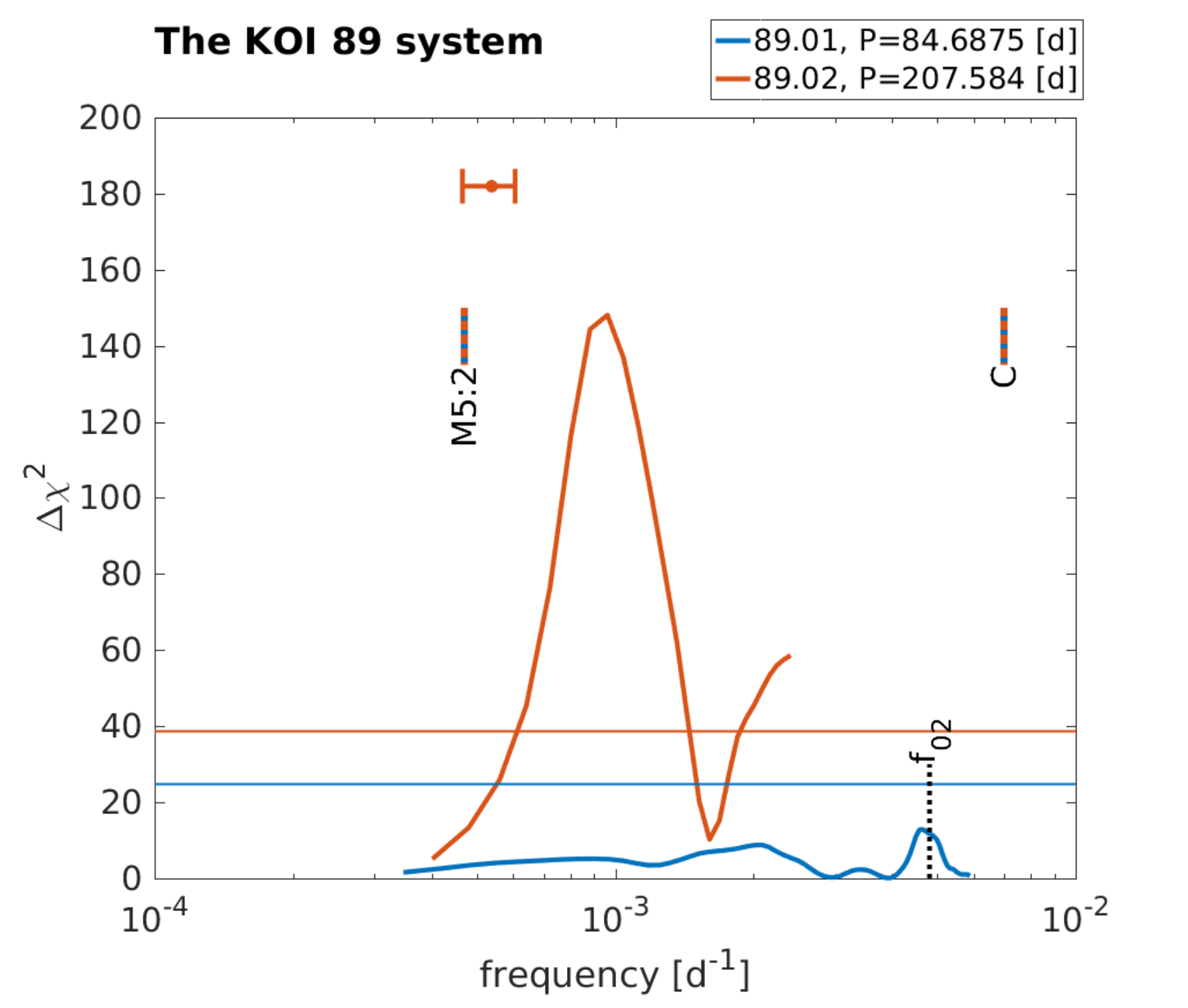}
	\caption{Similar to Figure~\ref{KOI_209_system}, for the KOI 89 system.}
	\label{KOI_89_system}
\end{figure}


\item \textbf{KOI 108 / Kepler-103:} (Figure \ref{KOI_108_system}) The PA approach detects two significant frequencies in KOI~108.02, and no TTVs in KOI~108.01.  The frequencies detected do not correspond to any known interaction frequency between the known planets, and hence may suggest the presence of additional objects in the system. These TTVs are consistent with those reported by H16, but are inconsistent with those first identified by Van Eylen \& Albrecht (2015), possibly owing to their use of only part of the data used here.

\begin{figure}
	\includegraphics[width=0.5\textwidth]{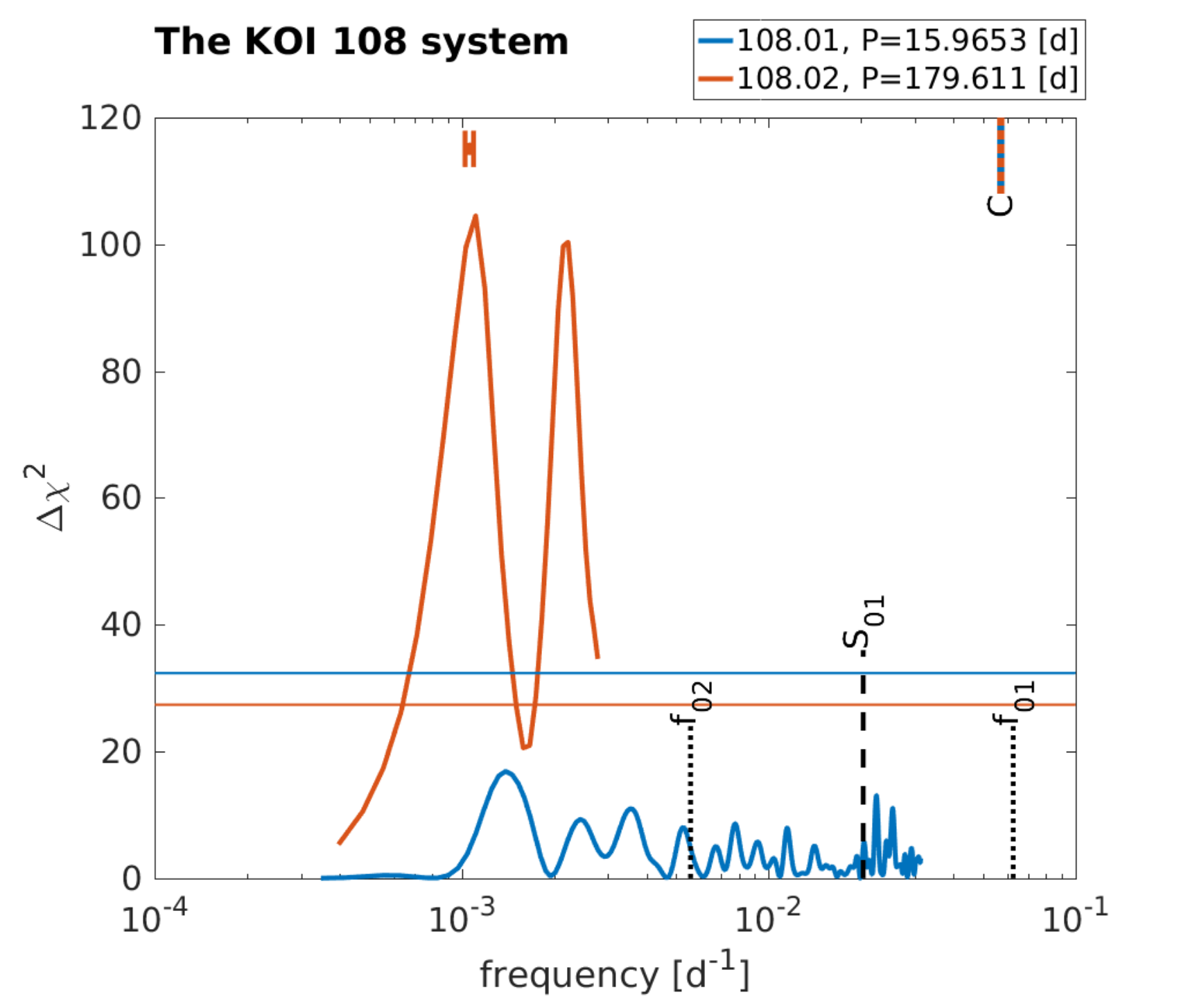} 
	\caption{Similar to Figure~\ref{KOI_209_system}, for the KOI 108 system.}
	\label{KOI_108_system}
\end{figure}




\item \textbf{KOI 185:} (Figure \ref{KOI_185.01}) This system presents the longest TTV period we were able to constrain (at $>3\sigma$). A TTV frequency of $f_{\mathrm{TTV}}=(1.75_{-0.45}^{+0.79}) \cdot10^{-4}$ (TTV period of $15.7 \pm 4.1$ years).  Note however that KOI-185.01 may not be due to a planet: with a $\sim3\%$ deep grazing transit on a $\sim0.8R_\odot$ star the occulting object may be too large for a planet.

\begin{figure}
	\includegraphics[width=0.5\textwidth]{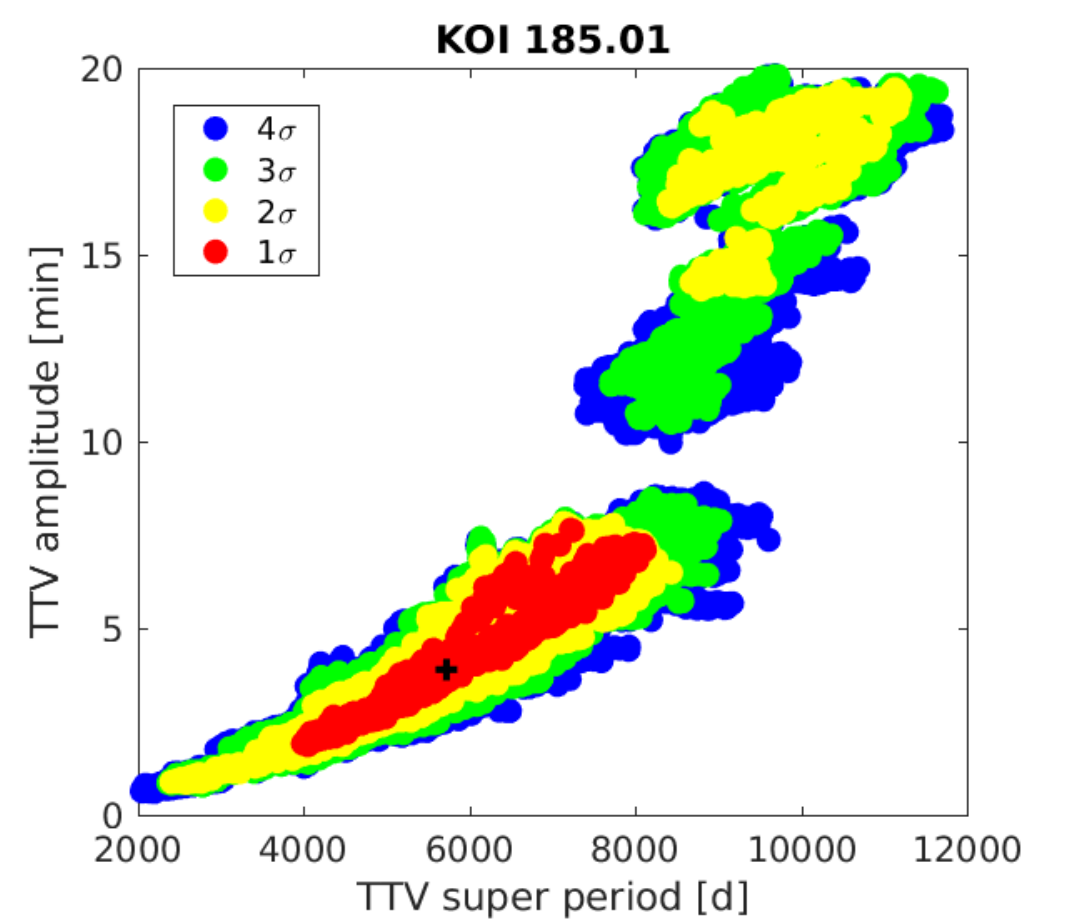}
	\caption{Similar to Figure \ref{KOI_75.01}, but for KOI 185.01 - the longest TTV period (but still constrained to better than $3\sigma$) TTV signal of about 18yrs.}
	\label{KOI_185.01}
\end{figure}

\item \textbf{KOI 209 / Kepler-117:} (Figure \ref{KOI_209_system}) This system is included for illustrative purposes. There are three distinct frequencies in the inner planet's TTVs (KOI 209.02): one matches the 2:1 MMR, another matches the 3:1 MMR, and the last matches the orbital period of the outer KOI 209.01 (Bruno \etal 2015) \footnote{and also Ofir \etal (2014) during the "The Space Photometry Revolution CoRoT Symposium 3" conference, Toulouse, France - see: https://corot3-kasc7.sciencesconf.org/33656}.


\item \textbf{KOI 262 / Kepler-50:} (Figure \ref{KOI_262_system}) 
Steffen \etal 2013 confirmed this two-planet system based on anti-correlated TTVs spanning $\approx700$~d. The data available today clearly shows that the two planets have TTVs of different frequencies: $f_{\mathrm{TTV,01}}=(9.74_{-0.11}^{+0.12}) \cdot10^{-4}$ and $f_{\mathrm{TTV,02}}=(15.42_{-0.14}^{+0.15}) \cdot10^{-4}$ Interpreting these TTV signals is not trivial: their period ratio is very close to the 6:5 MMR ($\Delta\simeq0.000131$), however, the expected super period is too long and we cannot resolve it using current data. The two different observed TTV periods could be a sign of separate interactions of each of the observed planets with yet another non-transiting planet in the system. For example, a planet on a $\sim11.7$ days orbital period could explain both observed TTV periods. We conclude that the confirmation above of the two planets by connecting the observed TTVs to mutual interaction between them is not correct, and speculate on the cause of the observed TTVs.

\begin{figure}
	\includegraphics[width=0.5\textwidth]{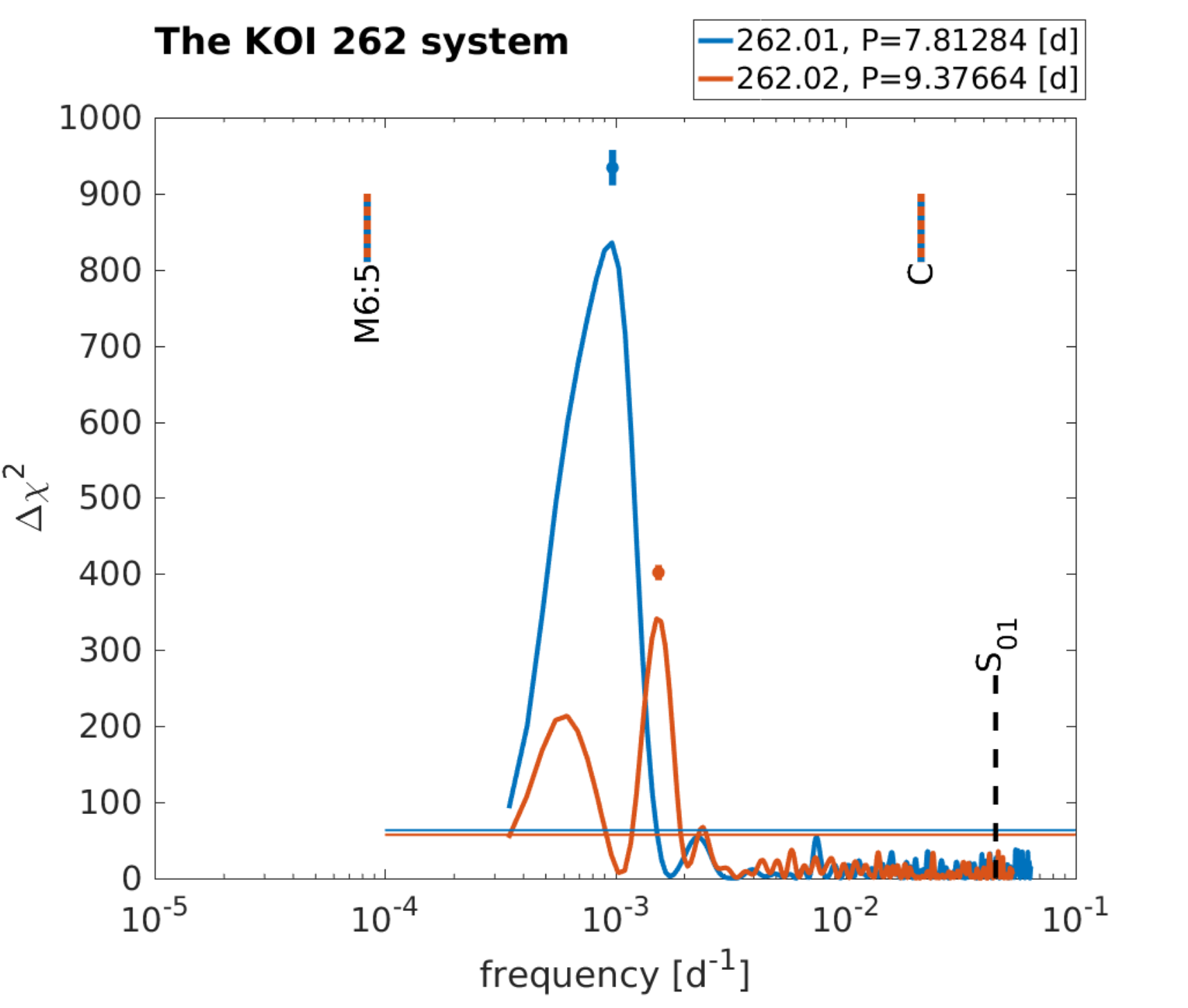}
	\caption{Similar to Figure~\ref{KOI_209_system}, for the KOI 262 system.}
	\label{KOI_262_system}
\end{figure}

\item \textbf{KOI 271 / Kepler-127:} (Figure \ref{KOI_271_system}) This system reveals dynamical interactions of KOI~271.02 with both KOI~271.01 and KOI~271.03.  The main peak of KOI~271.02 is consistent with the expect 5:3 resonant frequency among .01 and .02. Additionally, the 2:1 between .02 and .03 appears is the most significant peak of .03, and possibly contributing a shoulder to the spectral peak of .02. All three planets were hitherto only statistically validated.

\begin{figure}
	\includegraphics[width=0.5\textwidth]{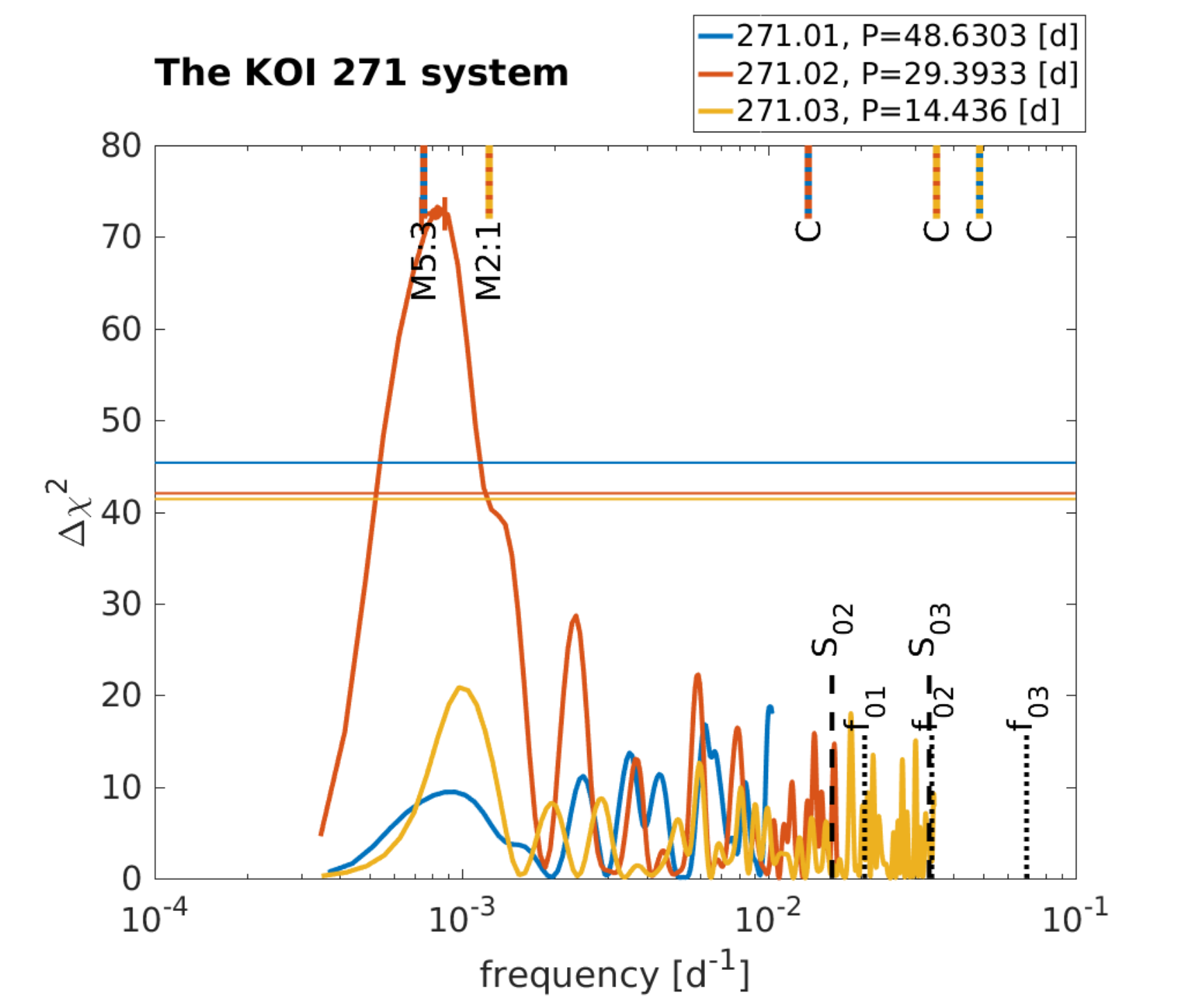}
	\caption{Similar to Figure~\ref{KOI_209_system}, for the KOI 271 system.}
	\label{KOI_271_system}
\end{figure}

\item \textbf{KOI 282 / Kepler-130:}  (Figure \ref{KOI_282_system}) KOIs 282.01 and 282.03 are close to 3:1 MMR. The known TTV signal on 282.01 at $f_{TTV}=(20.55\pm0.44)\cdot 10^{-4} d^{-1}$ is consistent with the predicted $f_{\mathrm{Sup}}=20.73\cdot 10^{-4} d^{-1}$ for that $2^{nd}$ order resonance, confirming both planets (which were only statistically validated thus far but had no dynamical confirmation).
\begin{figure}
	\includegraphics[width=0.5\textwidth]{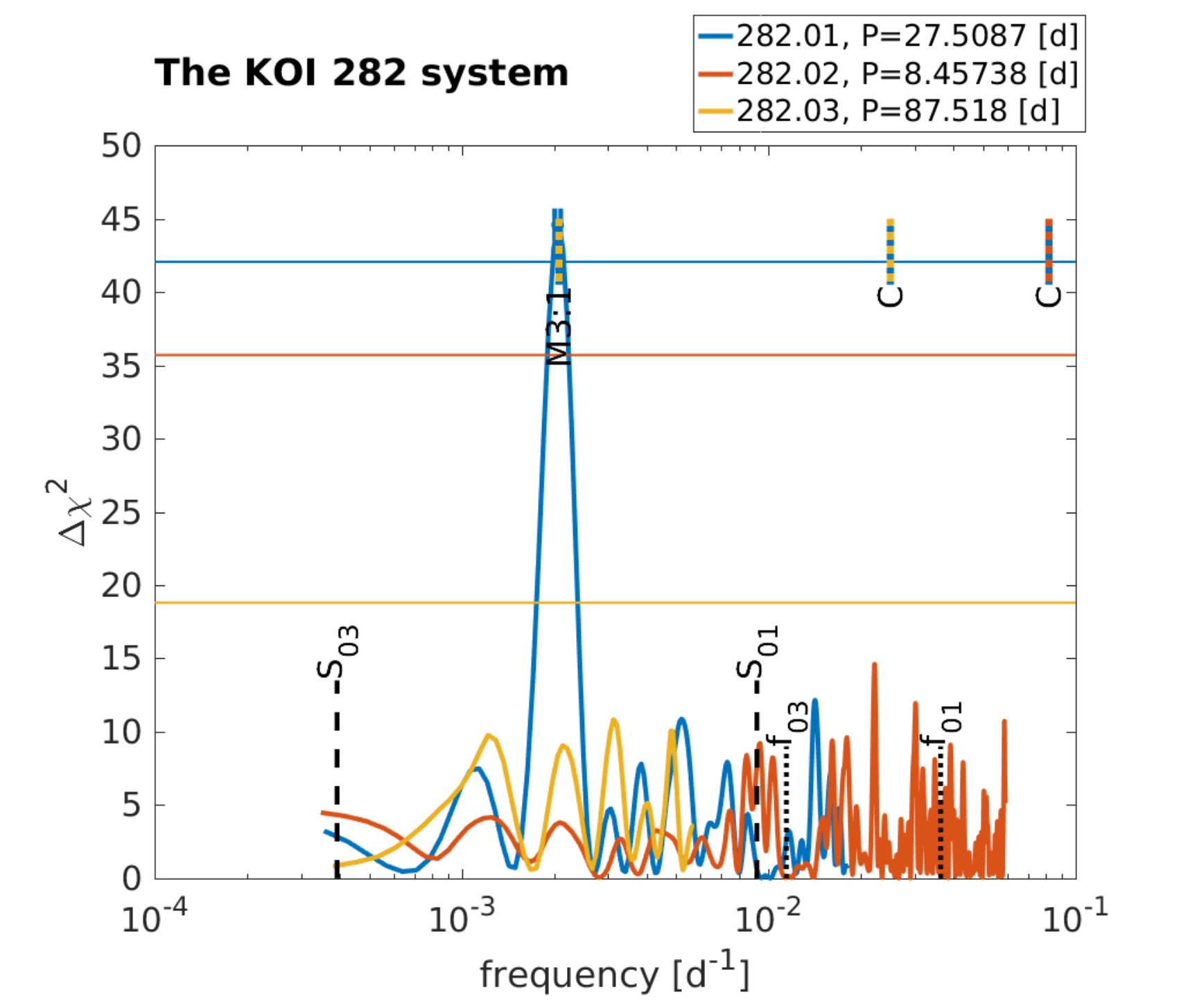}
	\caption{Similar to Figure~\ref{KOI_209_system}, for the KOI 282 system.}
	\label{KOI_282_system}
\end{figure}

\item \textbf{KOI 312 / Kepler-136:} (Figure \ref{KOI_312_system}) KOIs 312.01 and 312.02 are close to a 3:2 period commensurablity $\Delta=-0.056$, and both have significant TTVs, yet the TTVs are not close either the expected super-frequency or the expected chopping frequency. The system is not yet understood, and it is suspected there exists an additional
undetected perturbing planet.
\begin{figure}
	\includegraphics[width=0.5\textwidth]{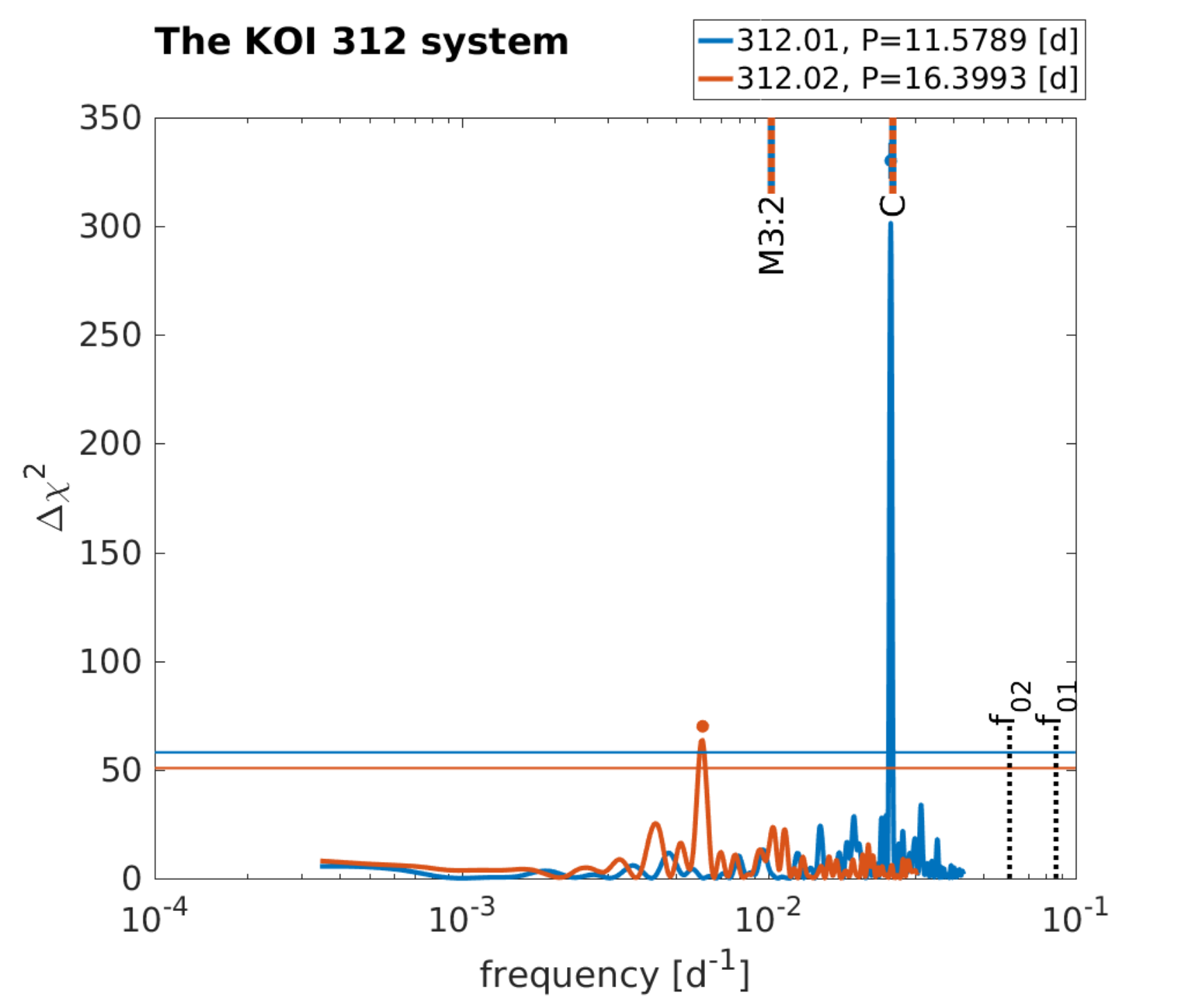}
	\caption{Similar to Figure~\ref{KOI_209_system}, for the KOI 312 system.}
	\label{KOI_312_system}
\end{figure}

\item \textbf{KOI 464 / Kepler-561:} (Figure \ref{KOI_464_system}: The two planets in the system are widely separated (periods of 5 and 58 days for KOIs 464.02 and 464.01, respectively) and thus likely not interacting. The outer transiting planet KOI 464.01 exhibits TTVs with two distinct and significant spectral peaks, that may be explained by one or more additional undetected planets.
\begin{figure}
	\includegraphics[width=0.5\textwidth]{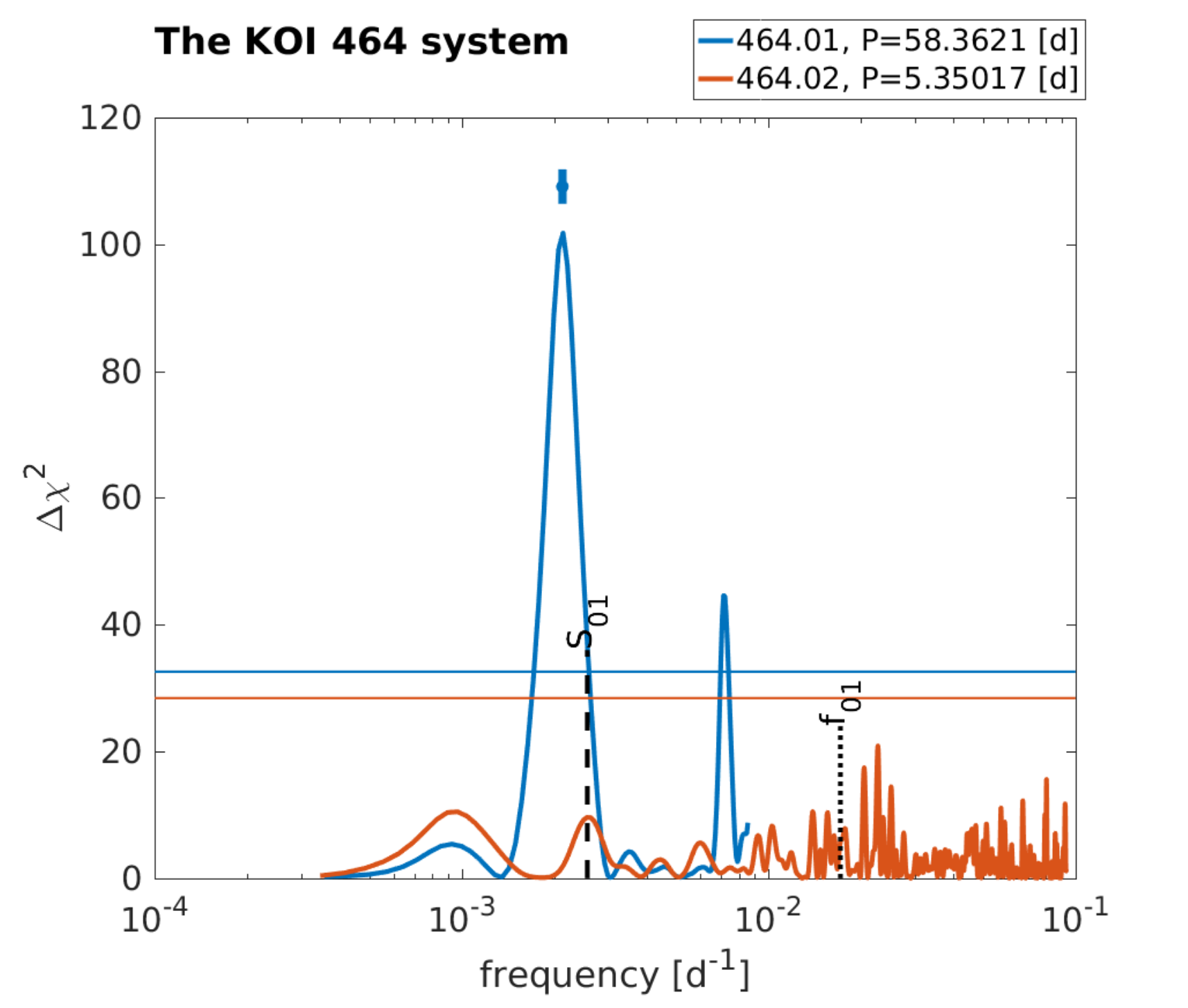}
	\caption{Similar to Figure~\ref{KOI_209_system}, for the KOI 464 system.}
	\label{KOI_464_system}
\end{figure}

\item \textbf{KOI 523 / Kepler-177:} (Figure \ref{KOI_523_system}): This system shows that in addition to the known (Xie 2014) TTV signal of KOIs 523.01 (at $f_{TTV,01}=(4.33_{0.70}^{1.08})\cdot 10^{-4} d^{-1}$) which is consistent with the super-frequency of the 4:3 MMR with 523.02, there exists an additional significant spectral peak at  $f_{TTV,2}\simeq62.8\cdot 10^{-4} d^{-1}$ in KOI 523.01 which may be related to the expected "chopping" frequency of $f_{Chop}=68.94 \cdot 10^{-4} d^{-1}$. Furthermore, hints of both of these peaks are seen in the PA spectrum of KOI 523.02. These new identifications may be used to improve on the current mass determination (Xie 2014).
\begin{figure}
	\includegraphics[width=0.5\textwidth]{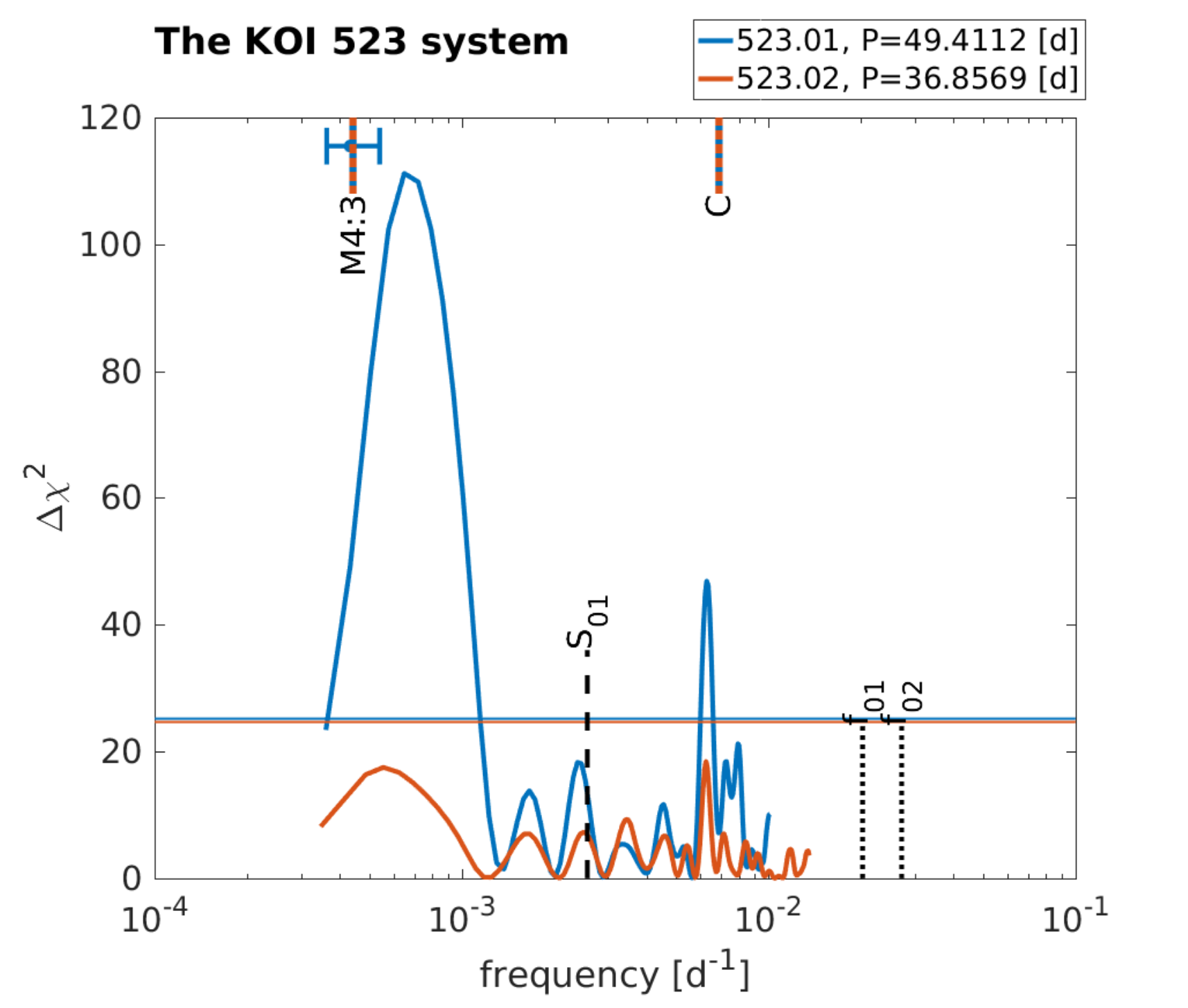}
	\caption{Similar to Figure~\ref{KOI_209_system}, for the KOI 523 system.}
	\label{KOI_523_system}
\end{figure}


\ 


\item \textbf{KOI 775 / Kepler-52:} (Figure \ref{KOI_775_system}): In addition to the previously known ({\it e.g.} H16, HL14) TTV peak of KOI 775.02 at a frequency close to the super-period associated with the 2:1 MMR, we detect a new peak in the spectrum of KOI 775.01 at nearly the same frequency ($f_{\mathrm{TTV}}=48.81 \cdot 10^{-4} d^{-1}$), though at a confidence of 0.998. This may allow improved mass determination for both planets.
\begin{figure}
	\includegraphics[width=0.5\textwidth]{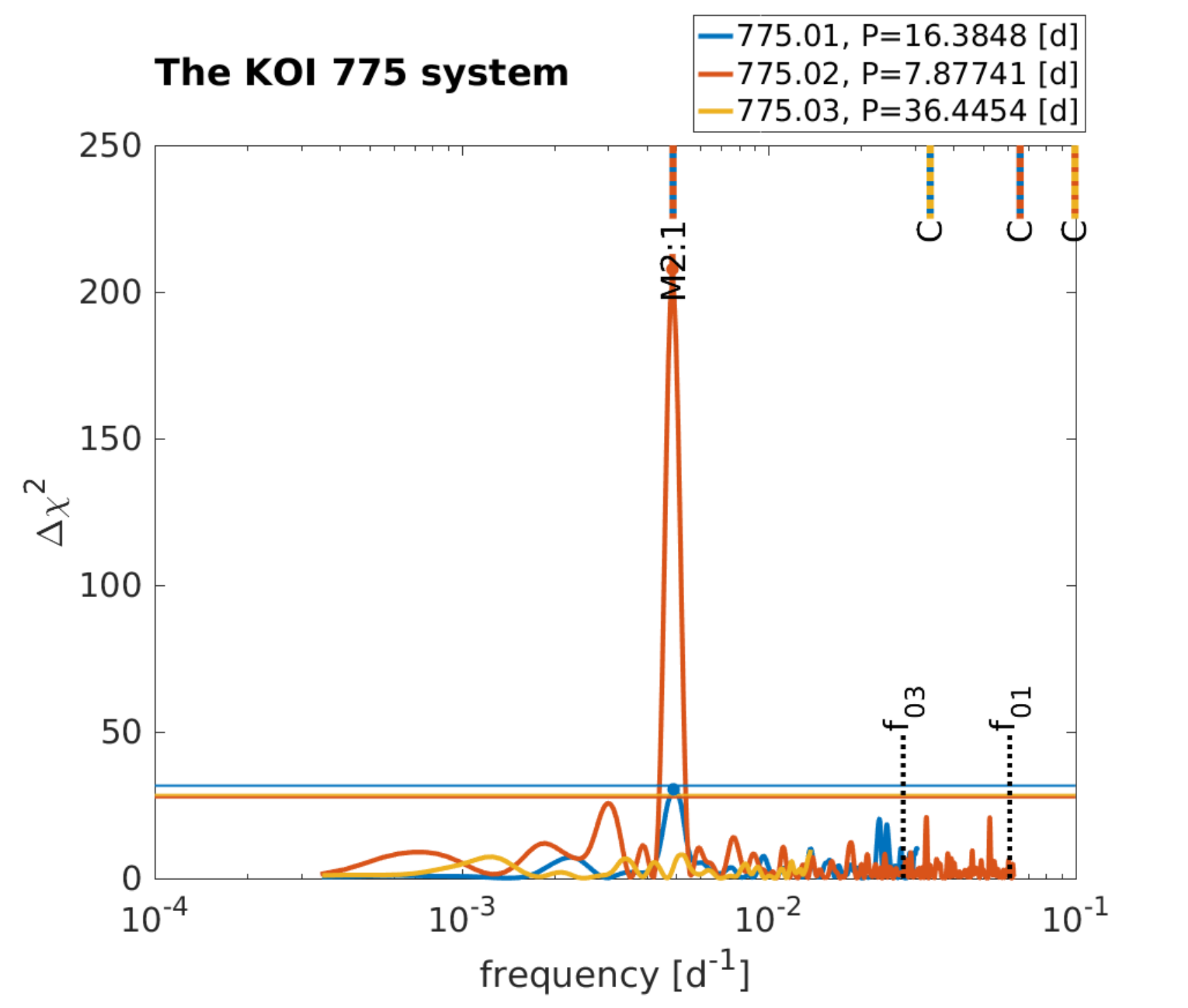}
	\caption{Similar to Figure~\ref{KOI_209_system}, for the KOI 775 system.}
	\label{KOI_775_system}
\end{figure}

\item \textbf{KOI 841 / Kepler-27:} (Figure \ref{KOI_841_system}): We detect two previously unidentified frequencies in the PA spectrum of KOI 841.02, in addition to the known primary frequency (Steffen \etal 2012, HL14). These peaks may allow improved constraints on the masses.
\begin{figure}
	\includegraphics[width=0.5\textwidth]{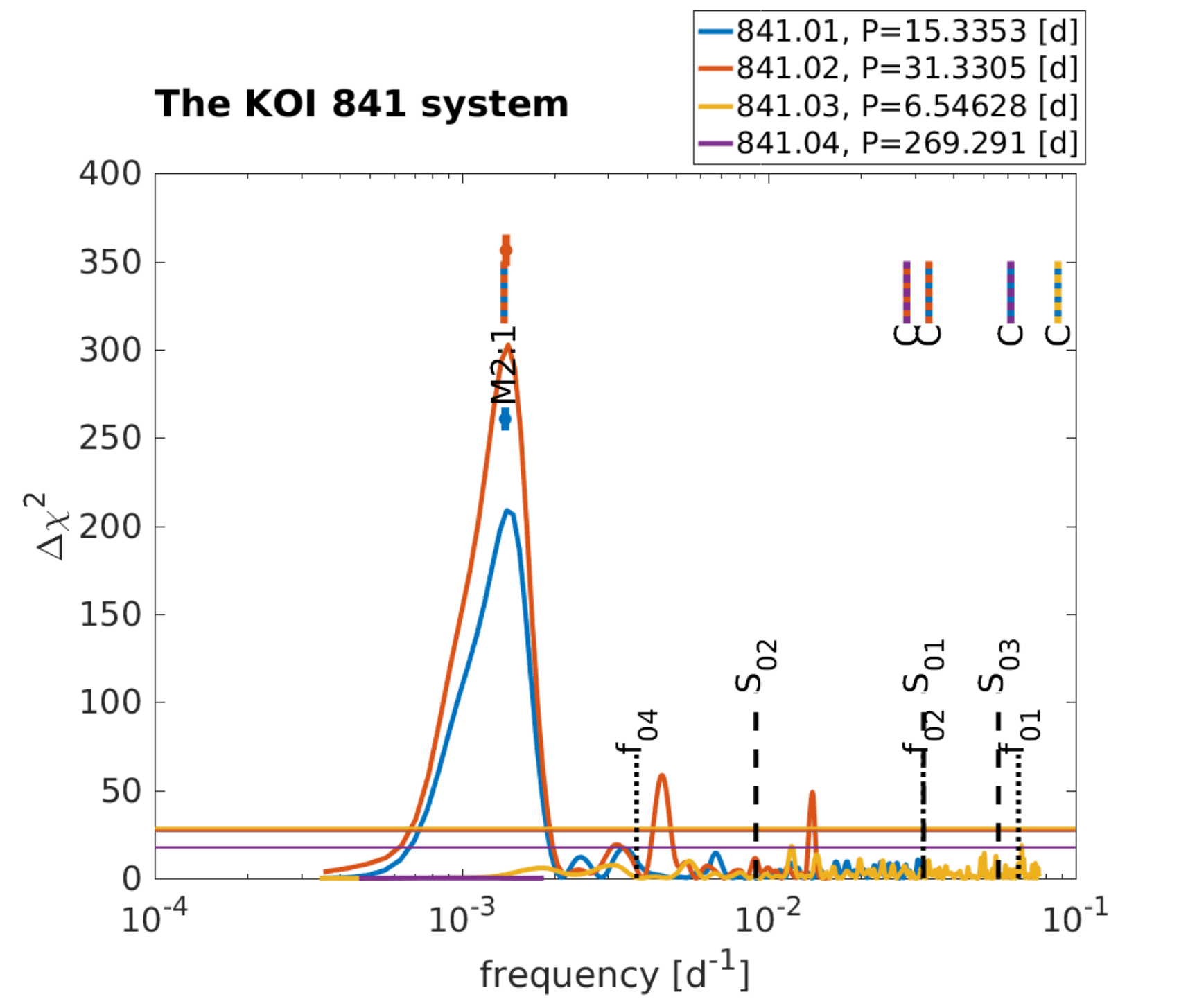}
	\caption{Similar to Figure~\ref{KOI_209_system}, for the KOI 841 system.}
	\label{KOI_841_system}
\end{figure}

\item \textbf{KOI 870 / Kepler-28:}
(Figure \ref{KOI_870_system}): KOI 870.01 and 870.02 are close to the 3:2 MMR and both exhibit significant TTVs at frequency consistent with the expected $f_{\mathrm{Sup}}=44.21 \cdot 10^{-4} d^{-1}$. Upper limits to the masses were given by Steffen \etal (2013), and low-significance ($m/\Delta m\leq3$) detection of masses by HL14, but only using the most significant TTV frequency. Here we detect secondary frequencies that are just below the high-significance threshold at bootstrap Confidences of 0.983 and 0.998 for KOIs 870.01 and 870.02 respectively, which may allow a better mass determination for both planets.
\begin{figure}
	\includegraphics[width=0.5\textwidth]{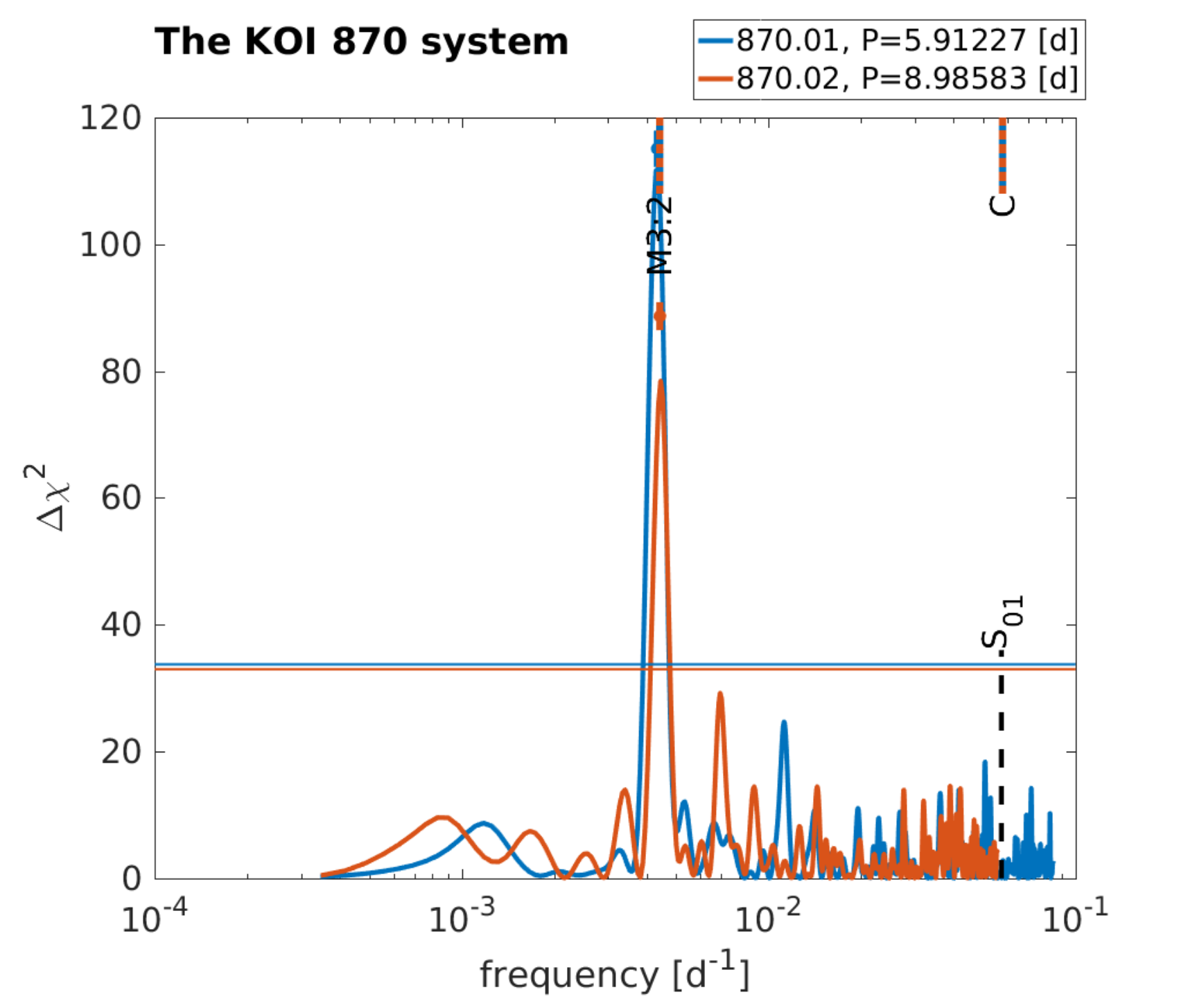}
	\caption{Similar to Figure~\ref{KOI_209_system}, for the KOI 870 system.}
	\label{KOI_870_system}
\end{figure}

\item \textbf{KOI 877 / Kepler-81:} 
(Figure \ref{KOI_877_system}): The TTVs on KOI 877.02 and KOI 877.01 are both consistent with the expected 2:1 MMR between them at $f_{\mathrm{Sup}}=18.1465 \cdot 10^{-4} d^{-1}$, although only the former is high-confidence while the latter has near-threshold confidence of 0.996. HL14 analysed the system and constrained the component's masses of KOI 877.01 and KOI 877.02 to $3\sigma$ or less, but used only the most significant TTV frequency, while KOI 841.02 exhibits a few more near-threshold frequencies. 
\begin{figure}
	\includegraphics[width=0.5\textwidth]{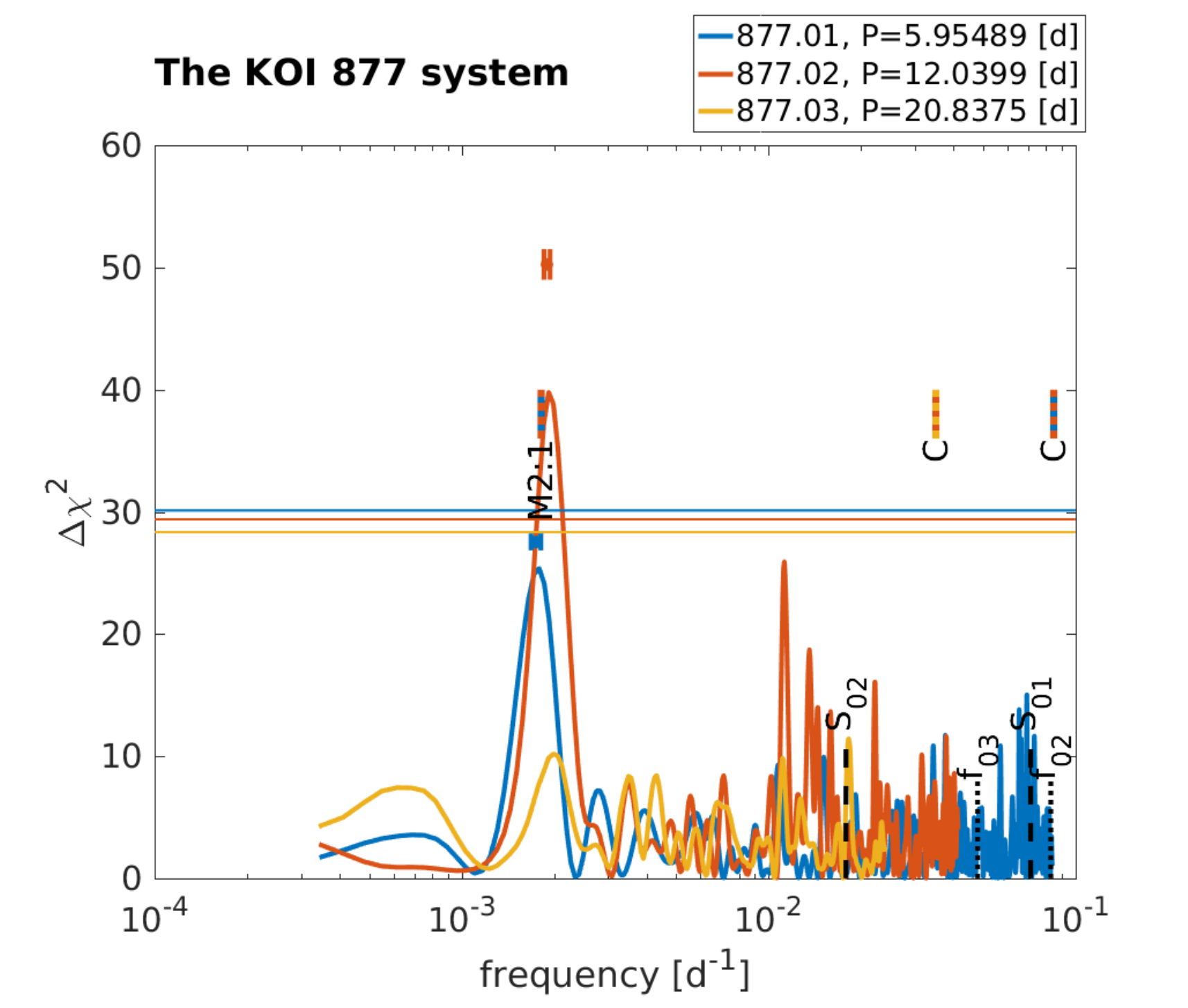}
	\caption{Similar to Figure~\ref{KOI_209_system}, for the KOI 877 system.}
	\label{KOI_877_system}
\end{figure}

\item \textbf{KOI 880 / Kepler-82:} (Figure \ref{KOI_880_system}): There are prominent and well-known TTVs (e.g. HL14, H16) on both KOI~880.01 and KOI~880.02, where the primary peak of KOI~880.01 at the super-frequency corresponding to 2:1 MMR between them. Here we detect one additional significant PA-spectral peak for KOI~880.01 and three additional peaks in KOI~880.02, and note that the most prominent TTV frequency of the KOI 880.02 at $f_{\mathrm{TTV}}=(8.15\pm0.12) \cdot 10^{-4} d^{-1}$ is offset from both the predicted super-frequency and from the observed TTV frequency of KOI~880.01 by a significant margin. Finally, we find that the most significant TTV frequency of KOI 880.04 is just below the threshold in PA (confidence=0.996) but its $\Delta \chi^2$ increases to above-threshold in the full fit. This observed TTV frequency does not correspond to any expected of the ones.
\begin{figure}
	\includegraphics[width=0.5\textwidth]{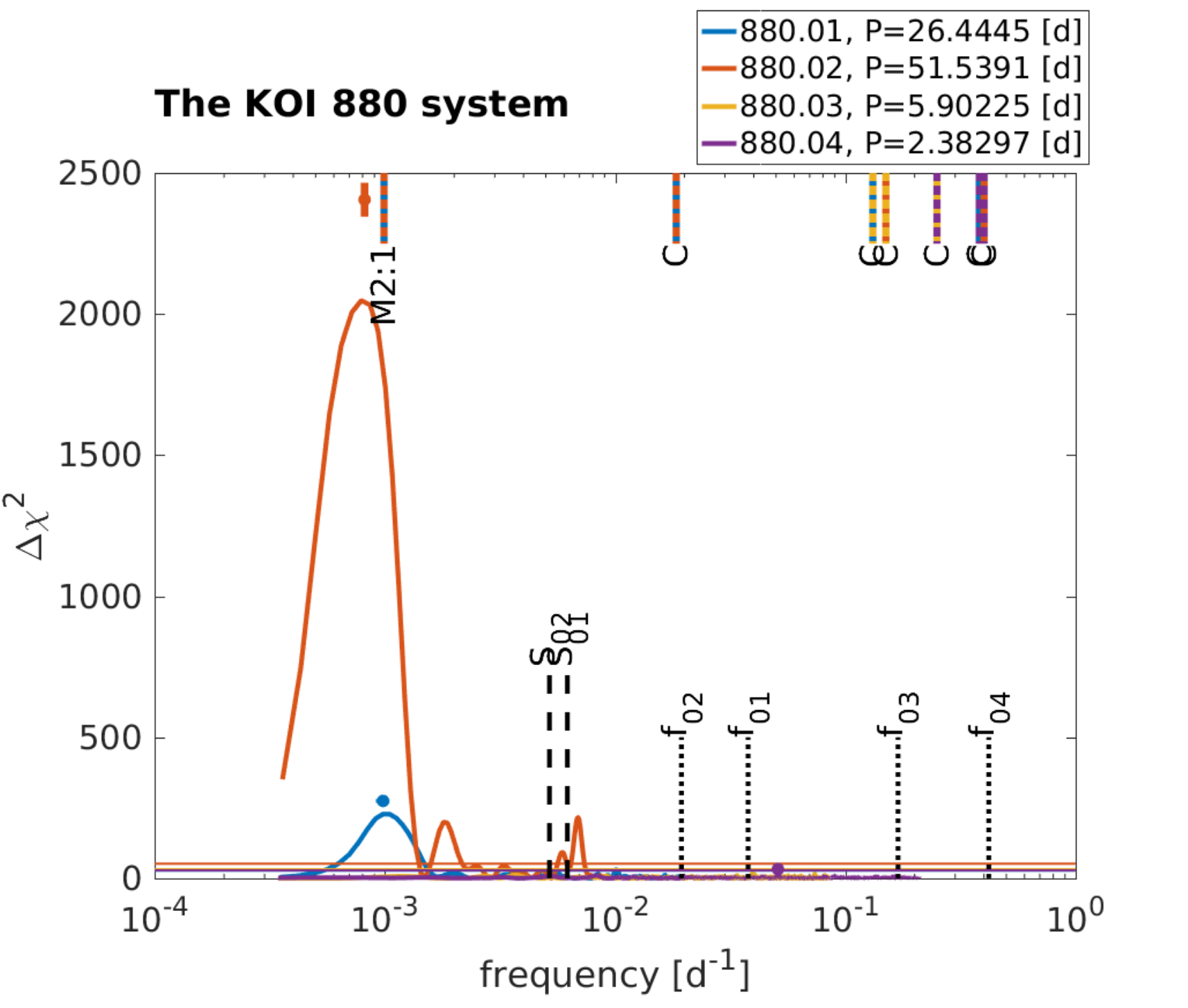}
	\caption{Similar to Figure~\ref{KOI_209_system}, for the KOI 880 system.}
	\label{KOI_880_system}
\end{figure}


\item \textbf{KOI 886 / Kepler-54:} (Figure \ref{KOI_886_system}): TTV most significant TTV frequency on both KOI 886.01 and KOI 886.02 is well known (e.g. H16). Here we find additional peaks in the PA spectrum of KOI 886.01 at frequencies which are harmonics of the main peak. In addition, the primary peaks of both KOI~886.01 and KOI~886.02 are shifted by $4-6\sigma$ from the expected super-frequency of the 3:2 MMR between them.

\begin{figure}
	\includegraphics[width=0.5\textwidth]{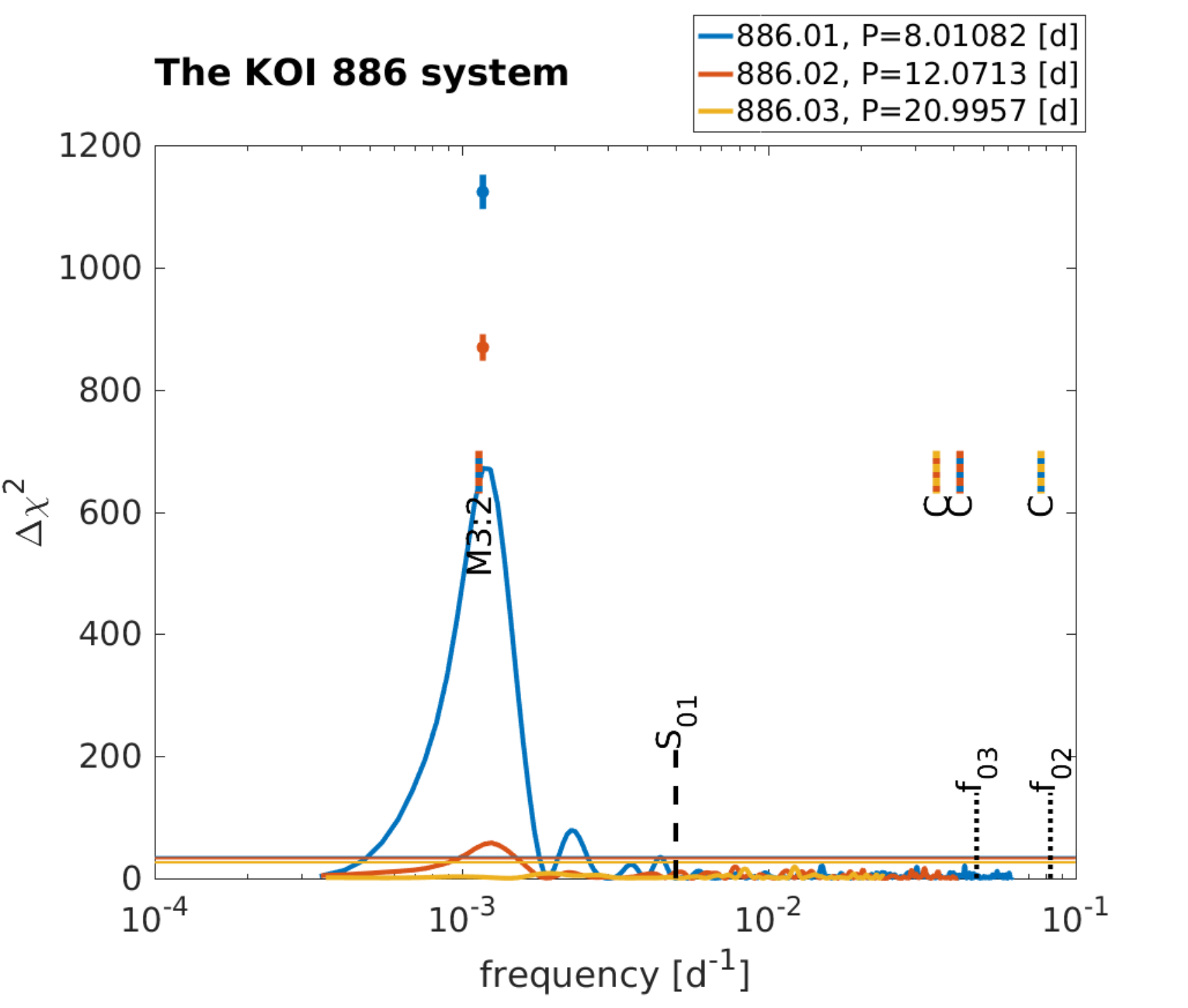}
	\caption{Similar to Figure~\ref{KOI_209_system}, for the KOI 886 system.}
	\label{KOI_886_system}
\end{figure}


\item \textbf{KOI 935 / Kepler-31:} (Figure \ref{KOI_935_system}): HL14 identified the main TTV frequency of KOIs 935.01 and 935.02 as a 2:1 MMR at the expected $f_{\mathrm{Sup}}=10.27 \cdot 10^{-4} d^{-1}$ and provided a determination of the mass of KOI 935.02. Here we find the 935.01 also has a second high-significance peak close to the expected $f_{\mathrm{Sup}}=23.01 \cdot 10^{-4} d^{-1}$ of the 4:1 MMR with 935.03. Also the PA peak of KOI 935.02 is wider than expected (and wider than  KOI 935.01's peak) - close to the expected super-frequency of the 2:1 MMR between the 935.02 and 935.03. We therefore suspect that the PA spectrum of KOI 935.02 includes the sum of two blended and similar peaks tying the three outer planets in a 1:2:4 resonance chain. This may allow determination of other masses in the system.

\begin{figure}
	\includegraphics[width=0.5\textwidth]{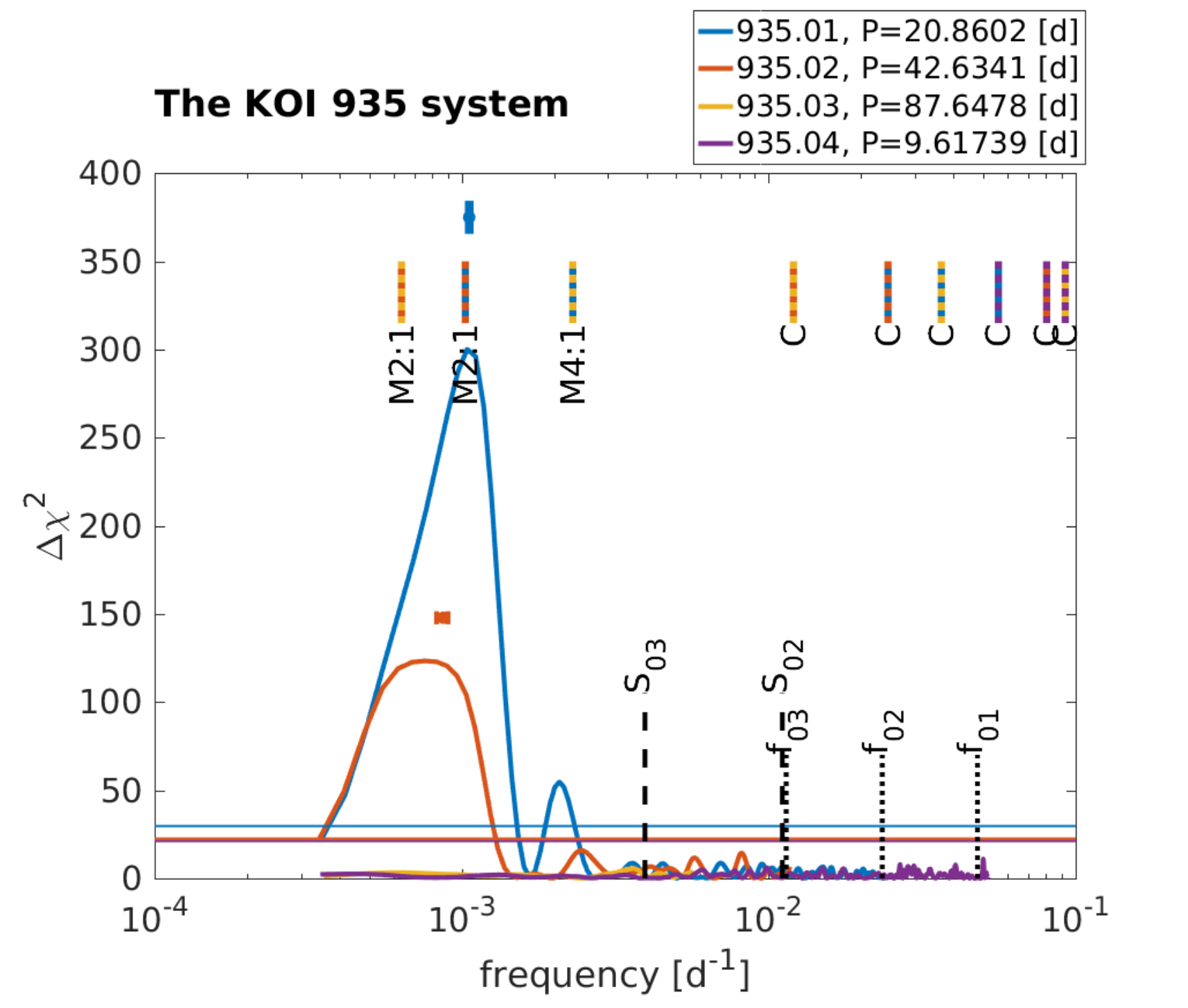}
	\caption{Similar to Figure~\ref{KOI_209_system}, for the KOI 935 system.}
	\label{KOI_935_system}
\end{figure}

\item \textbf{KOI 952 / Kepler-32:} (Figure \ref{KOI_952_system}): We find a significant peak in the PA spectrum of KOI 952.02 at a low frequency, in addition to the known peak near the 3:2 MMR super-frequency ($f_{\mathrm{Sup}}=38.649 \cdot 10^{-4} d^{-1}$) between it and 952.01, previously identified by HL14.
\begin{figure}
	\includegraphics[width=0.5\textwidth]{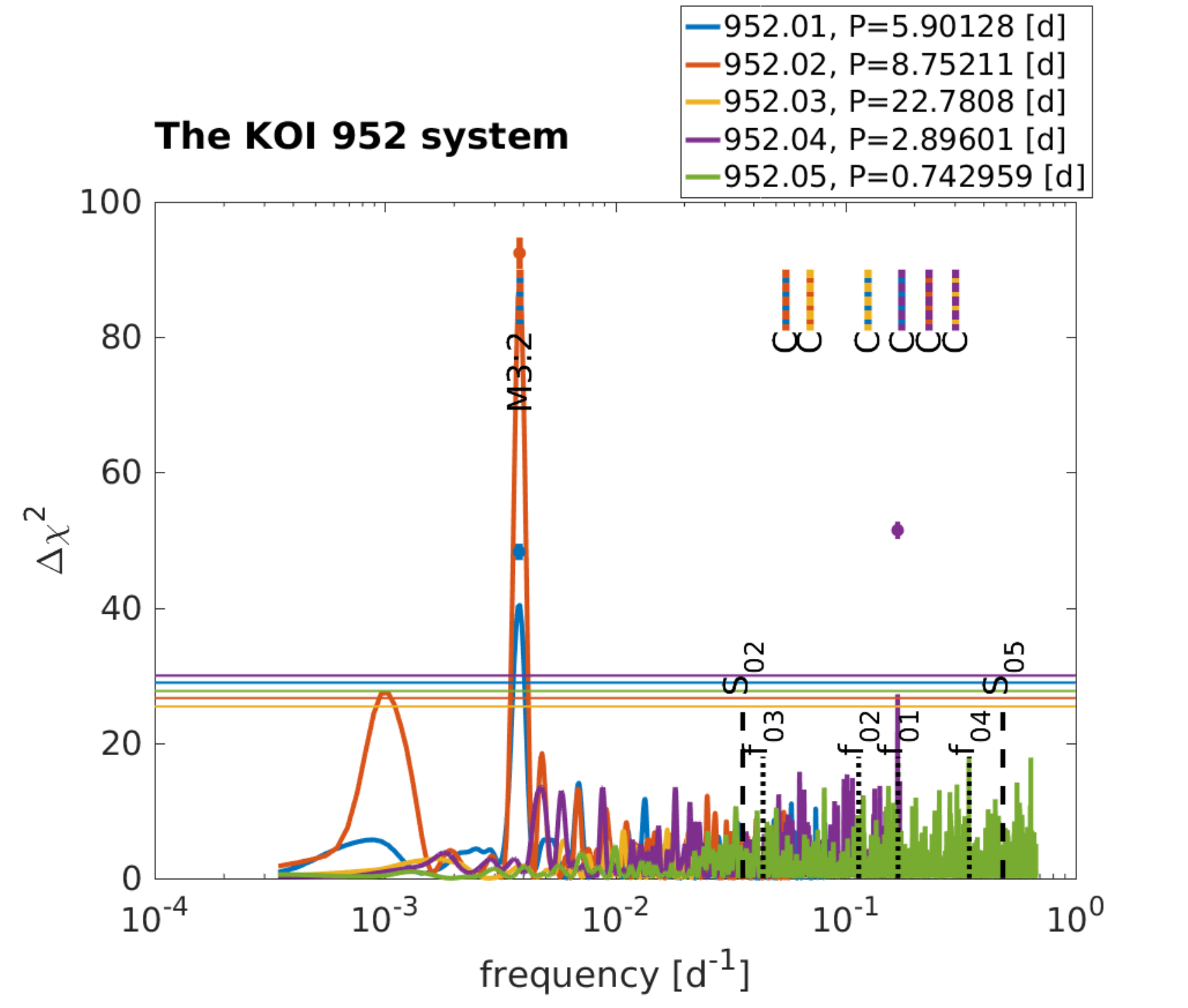}
	\caption{Similar to Figure~\ref{KOI_209_system}, for the KOI 952 system.}
	\label{KOI_952_system}
\end{figure}


\item \textbf{KOI 1102 / Kepler-24:} (Figure \ref{KOI_1102_system}): KOI 1102.01 and KOI 1102.02 have both well known TTVs (e.g. HL14, H16) with TTV frequency consistent with the 3:2 MMR between them. Here we identify another significant TTV frequency for KOI 1102.01 - so a more precise determination of the masses seems possible.

\begin{figure}
	\includegraphics[width=0.5\textwidth]{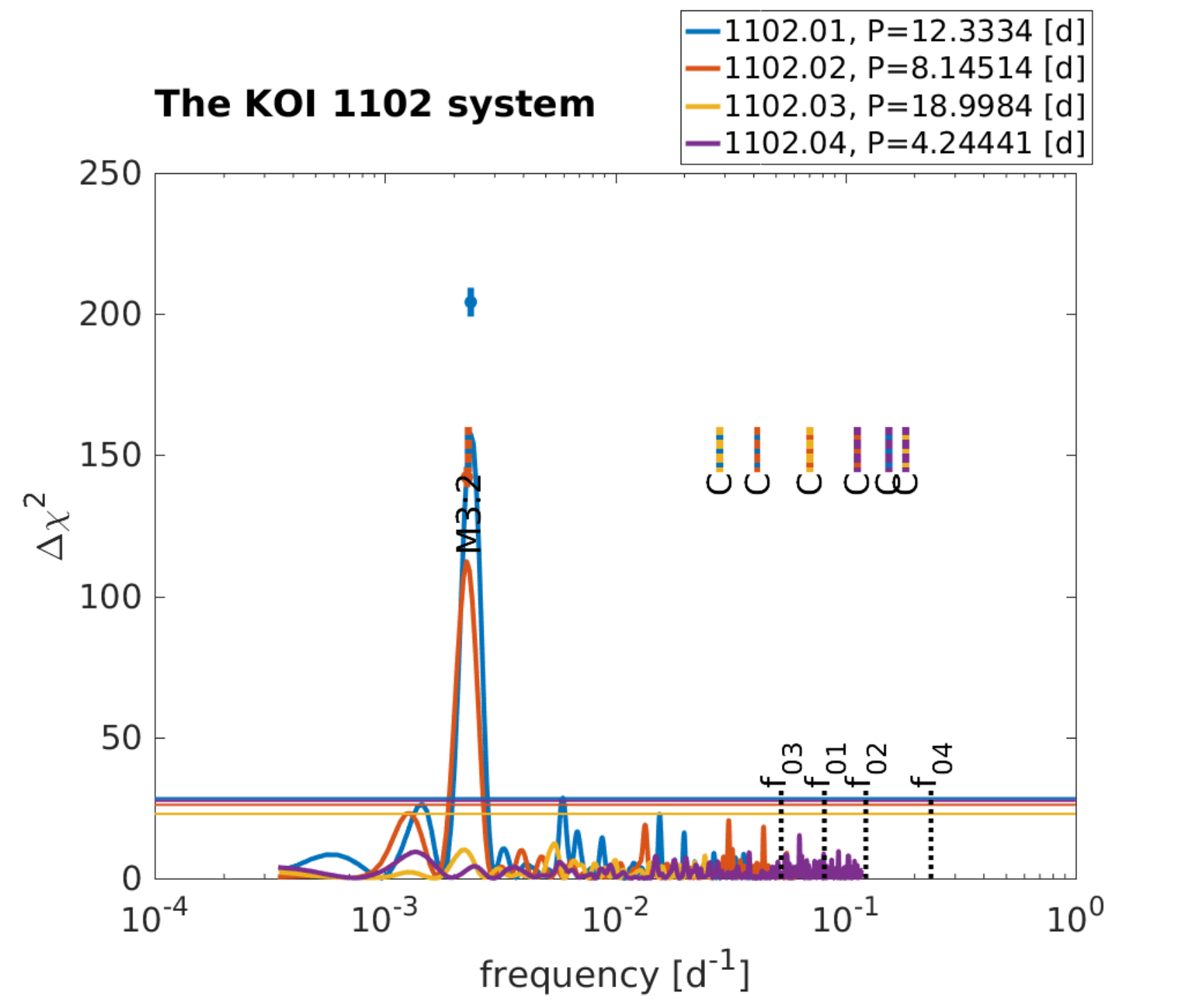}
	\caption{Similar to Figure~\ref{KOI_209_system}, for the KOI 1102 system.}
	\label{KOI_1102_system}
\end{figure}

\item \textbf{KOI 1236 / Kepler-279:} (Figure \ref{KOI_1236_system}): 
Xie 2014 identified the main TTV frequency of KOIs 1236.01 and 1236.03 as a 3:2 MMR at the expected $f_{\mathrm{Sup}}=8.4773\cdot 10^{-4} d^{-1}$ and provided a determination of their mass ($49.4^{+7.2}_{-5.9} m_\oplus$ and $37.5^{+5.5}_{-4.5} m_\oplus$ for KOI 1102.01 and KOI 1102.03, respectively). Xie 2014 did not use, however, the fact that there are more significant frequencies for both planets - including one peak for KOI 1236.03 near the expected 1:4 MMR super-frequency with KOI 1236.02, which has thus far no mass constraints, and a peaks for KOI 1236.01 near the expected "chopping" frequency with KOI 1236.03. A more precise determination of the masses seems possible.
\begin{figure}
	\includegraphics[width=0.5\textwidth]{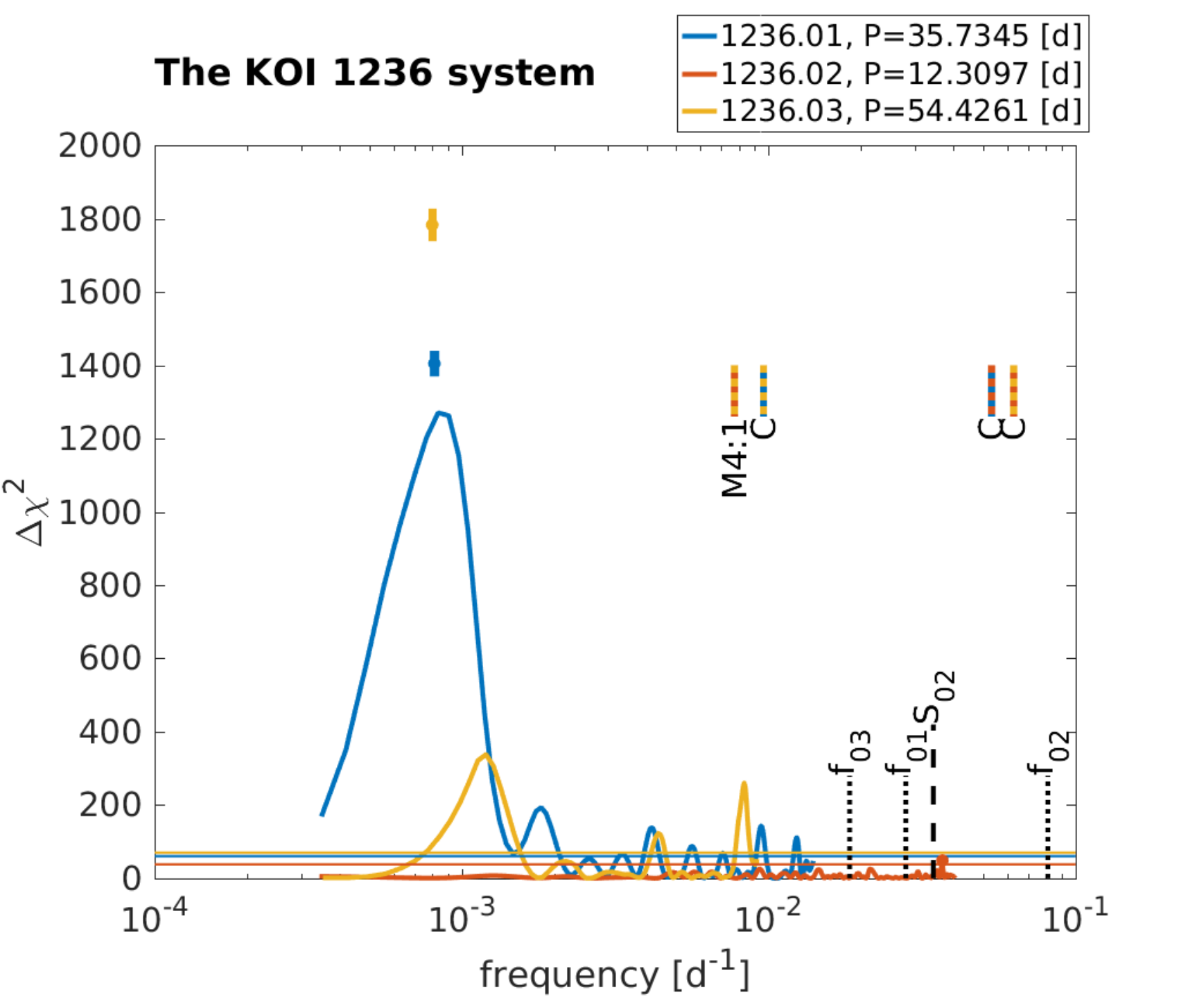}
	\caption{Similar to Figure~\ref{KOI_209_system}, for the KOI 1236 system.}
	\label{KOI_1236_system}
\end{figure}

\item \textbf{KOI 1258 / Kepler-281:} (Figure \ref{KOI_1258_system}): 
Here we detect a PA spectral peak for 1258.01 that is not consistent with any expected super-frequency of the previously known members of the system.  The transit signals of KOIs 1258.01 and 1258.02 were statistically validated (Morton \etal 2016). One of the secondary (low confidence) peaks of PA spectrum is close to the orbital frequency of KOI 1258.03, possibly confirming the latter (currently classified as a candidate). 
\begin{figure}
	\includegraphics[width=0.5\textwidth]{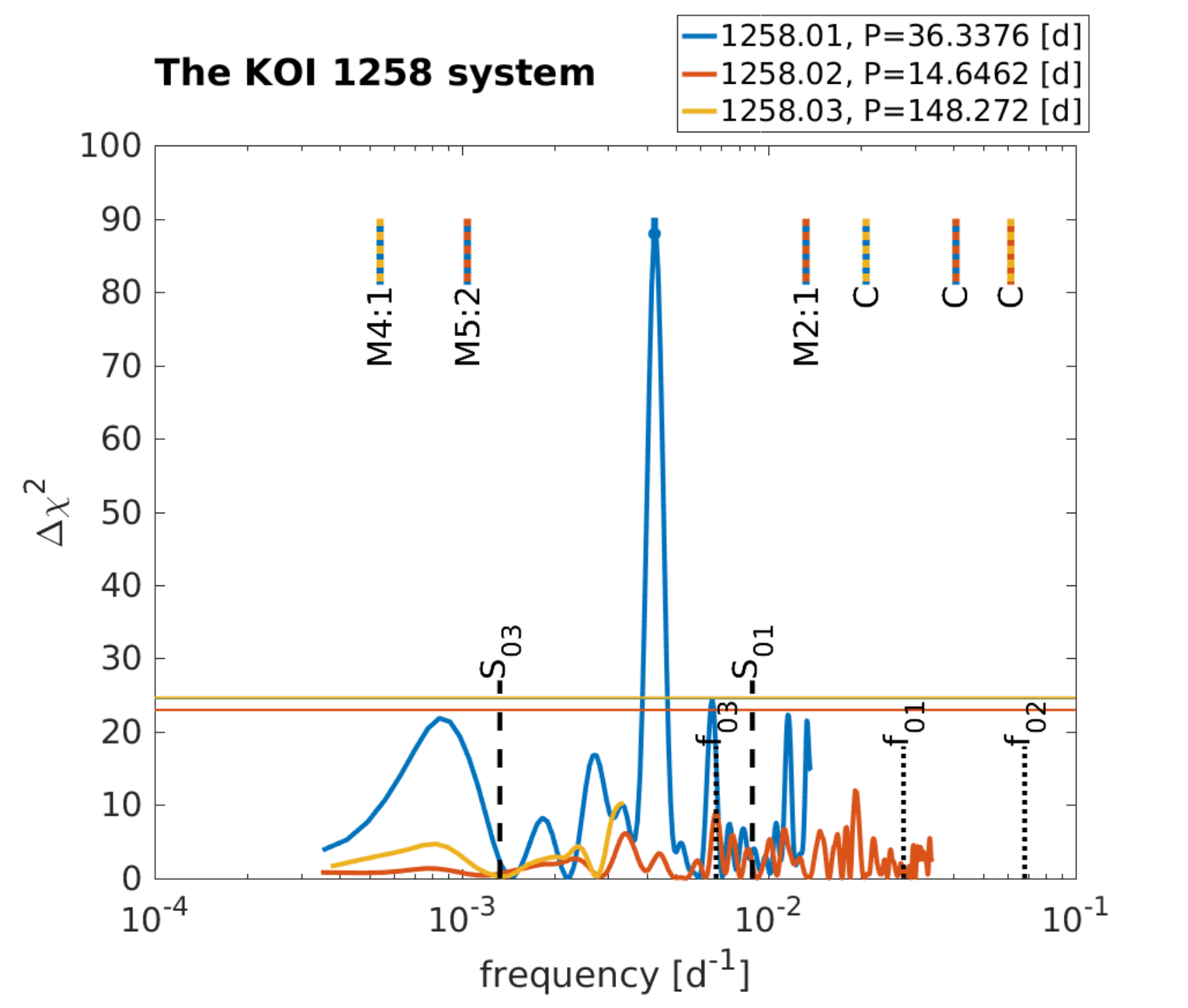}
	\caption{Similar to Figure~\ref{KOI_209_system}, for the KOI 1258 system.}
	\label{KOI_1258_system}
\end{figure}


\item \textbf{KOI 1366 / Kepler 293:} (Figure \ref{KOI_1366_system}) We find a significant TTV signal for 1366.01 at $f_{TTV}=(33.43_{-0.63}^{+0.65})\cdot 10^{-4} d^{-1}$, consistent with the expected frequency for second-order 3:1 MMR with KOI 1366.02 at $f_{\mathrm{Sup}}=34.58 \cdot 10^{-4} d^{-1}$. This fequency also appears as a smaller peak in KOI's 1366.02 PA spectrum. This confirms both planets (hitherto only having only statistical validation).
\begin{figure}
	\includegraphics[width=0.5\textwidth]{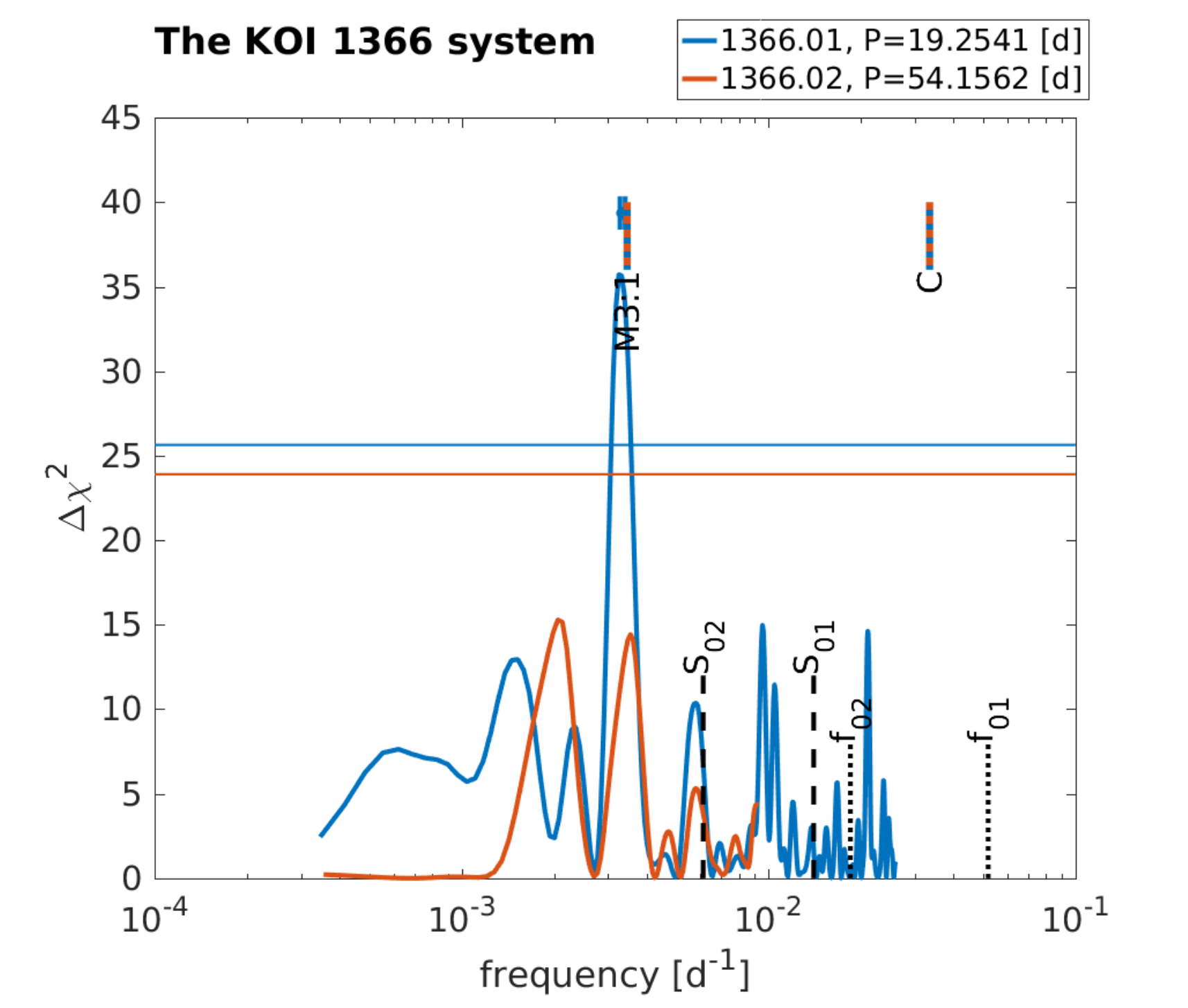}
	\caption{Similar to Figure~\ref{KOI_209_system}, for the KOI 1366 system.}
	\label{KOI_1366_system}
\end{figure}

 \item \textbf{KOI 1426 / Kepler 297:} (Figure \ref{KOI_1426_system}) The system is close to 1:2:4 MMR chain with normalized distance from resonances of $\Delta_{01,02}=-0.03748$ and $\Delta_{02,03}=0.001097$. As previously noted (HL14), the PA spectrum of 1426.02 exhibits evidence for the interaction between it and 1426.01 at the expected super-frequency and its first harmonic. The additional peaks of 1426.02 may be attributed to the orbital period of 1426.03 or the "chopping" frequency between 1426.02 and 1426.03.  Moreover, the observed TTV frequency of 1426.03 and 1426.01 at $f_{TTV}=(8.91_{-0.28}^{+0.33})\cdot 10^{-4} d^{-1}$ is consistent with the 4:1 MMR between these objects.
 We note 1426.03 is currently a still a candidate, that no RV variation was detected in the system (Santerne \etal 2016). The system has a high SNR, so it is attractive for further analysis (previous analysis by Diamond-Lowe \etal 2015 is not publicly available). 
\begin{figure}
	\includegraphics[width=0.5\textwidth]{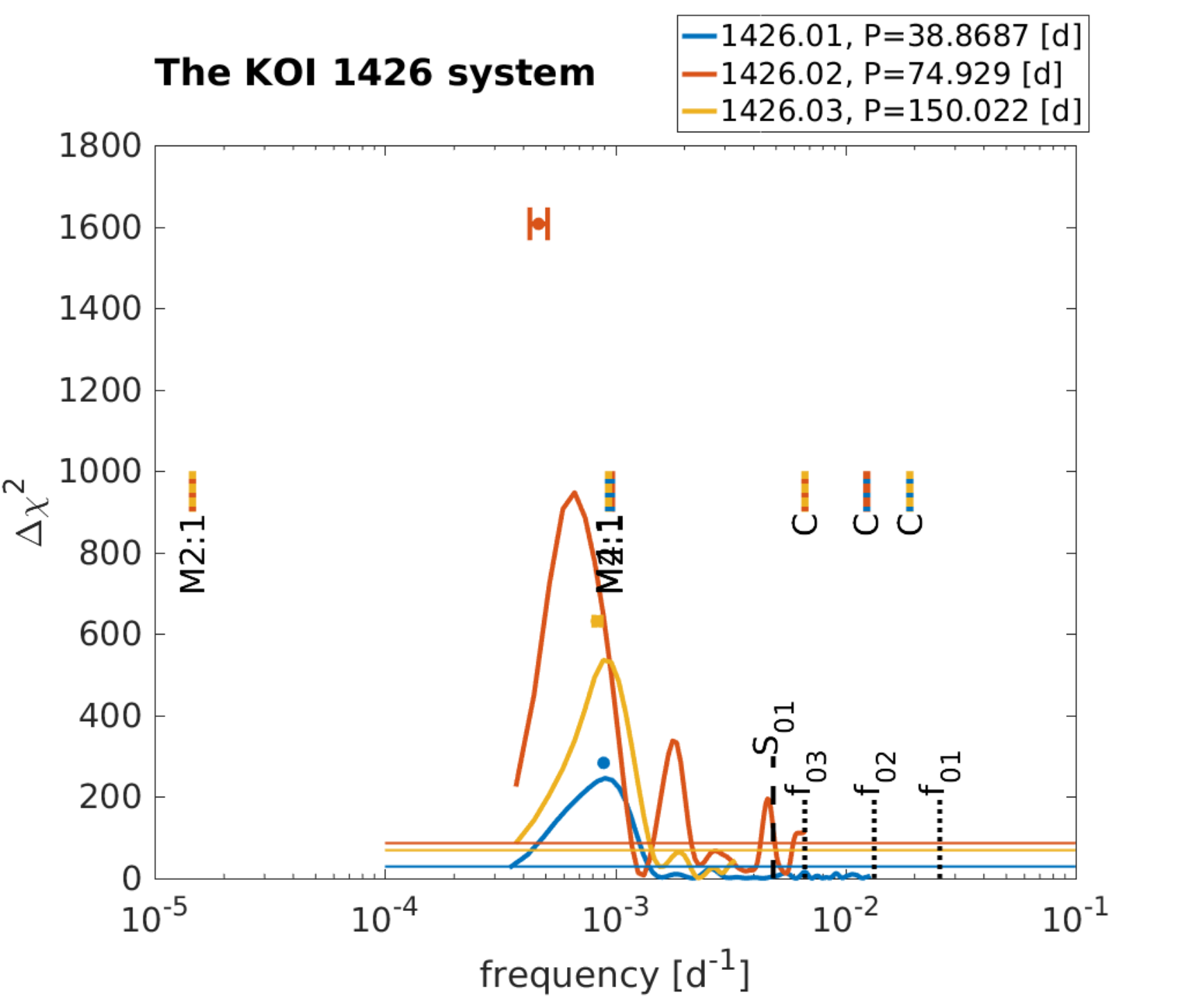}
	\caption{Similar to Figure~\ref{KOI_209_system}, for the KOI 1426 system.}
	\label{KOI_1426_system}
\end{figure}

\item \textbf{KOI 1529 / Kepler-59:} (Figure \ref{KOI_1599_system}) We detect a significant peak in 1529.02 at the same frequency as the known peak in 1529.01, both consistent with the 3:2 MMR super-frequency.
Moreover, there are peaks in the PA spectrum of KOI 1529.01 that appear to be just below the adopted significance threshold - possibly enabling absolute mass determinations for these small planets (both have radii$<2R_\oplus$)
\begin{figure}
	\includegraphics[width=0.5\textwidth]{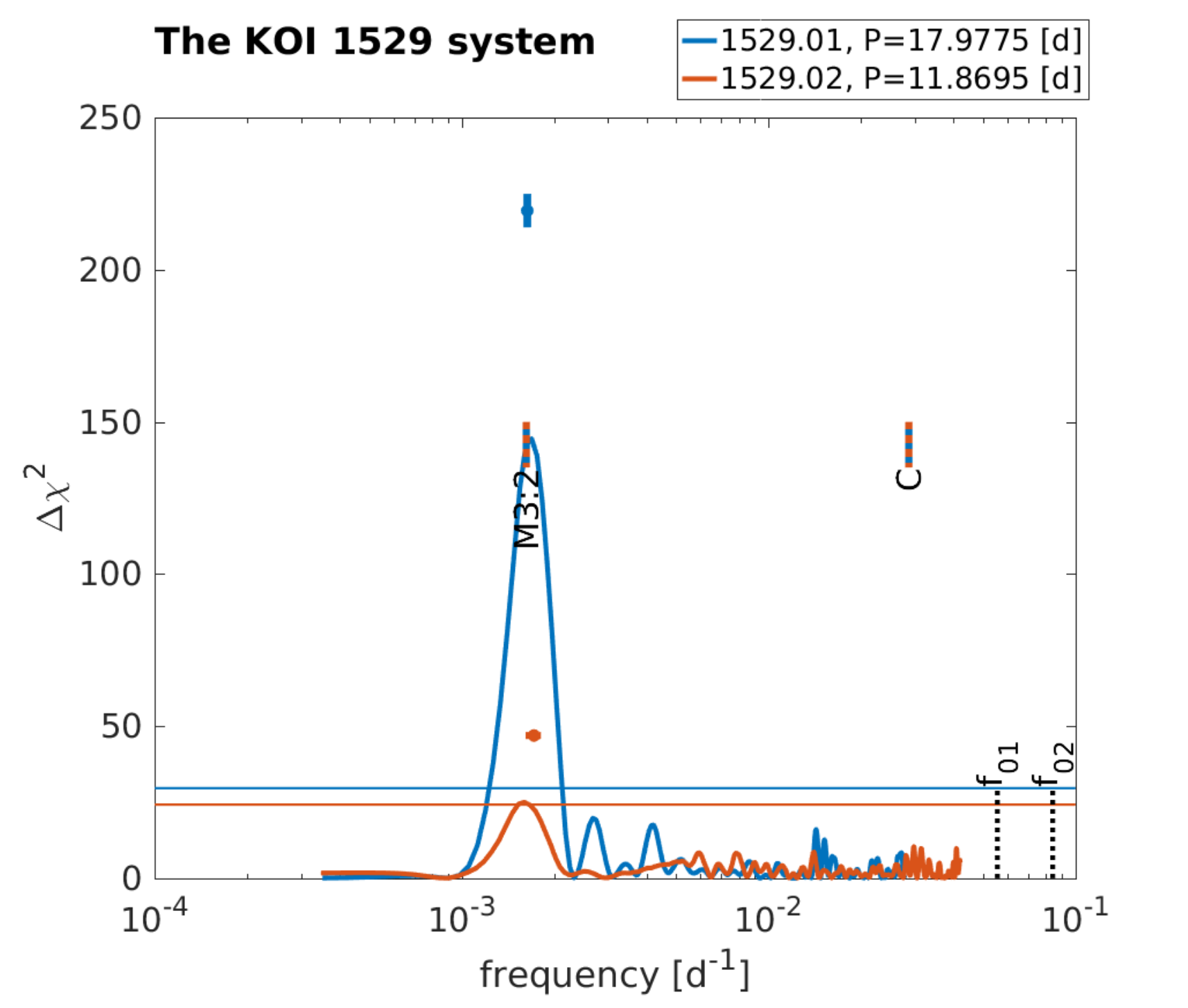}
	\caption{Similar to Figure~\ref{KOI_209_system}, for the KOI 1529 system.}
	\label{KOI_1529_system}
\end{figure}


\item \textbf{KOI 1599:} (Figure \ref{KOI_1599_system}) KOI-1599.02 is found to have TTVs with $f_{TTV}=(6.00\pm0.67)\cdot 10^{-4} d^{-1}$ which is approximately consistent with that of KOI-1599.01, but both are not consistent with the expected  3:2 MMR super-frequency. This, together with the strong ($\Delta=-0.00081$) resonance suggests the system is in resonance (and not just near resonance), rendering the usual expression for the super-frequency irrelevant. In such a case the TTV frequency, in addition to the amplitude, can constrain the planetary masses.  It is noteworthy that there are a few more significant frequencies in the PA spectrum of KOI-1599.01, but the signal is high-amplitude. Both KOIs are currently still neither confirmed or validates.
\begin{figure}
	\includegraphics[width=0.5\textwidth]{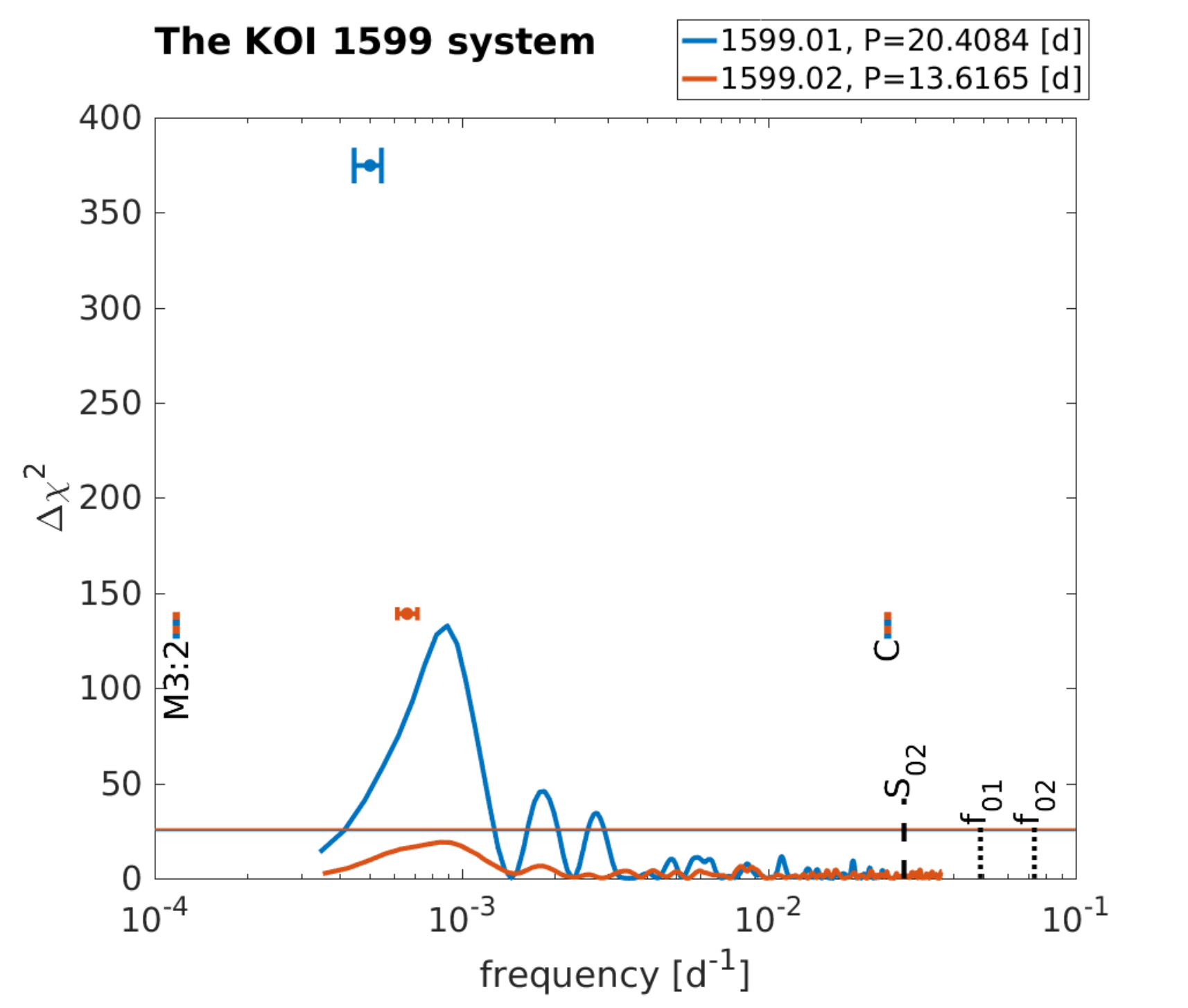}
	\caption{Similar to Figure~\ref{KOI_209_system}, for the KOI 1599 system. Note KOI-1599.01 has high-amplitude TTVs.}
	\label{KOI_1599_system}
\end{figure}



\item \textbf{KOI 1783:} (Figure \ref{KOI_1783_system})  TTVs of both candidates were detected by H16: a low TTV frequency of $f_{H16}=7.19 \pm 0.67 10^{-4} d^{-1}$ and long-term ("polynomial") TTVs for KOIs 1783.01 and 1783.02 respectively. We find (and indeed see clearly in H16's figures) that the most significant TTV frequency for KOI 1783.01 is much higher at $f_{TTV}=(35.95_{1.02}^{+0.64})\cdot 10^{-4} d^{-1}$ - which is consistent with all three of: the orbital frequency of 1783.02, the "chopping" frequency between the two candidates, and with the maximal TTV frequency of KOI 1783.01 itself. Also, we constrain the TTV frequency of KOI 1783.02 to be $f_{TTV}=(10.6_{0.77}^{+1.53})\cdot 10^{-4} d^{-1}$, and the second-most significant TTV frequency for KOI 1783.01 (at low confidence) is virtually identical to this frequency - but not close to any expected super-frequency. The two candidates therefore appear to be interacting - but further study is needed.
\begin{figure}
	\includegraphics[width=0.5\textwidth]{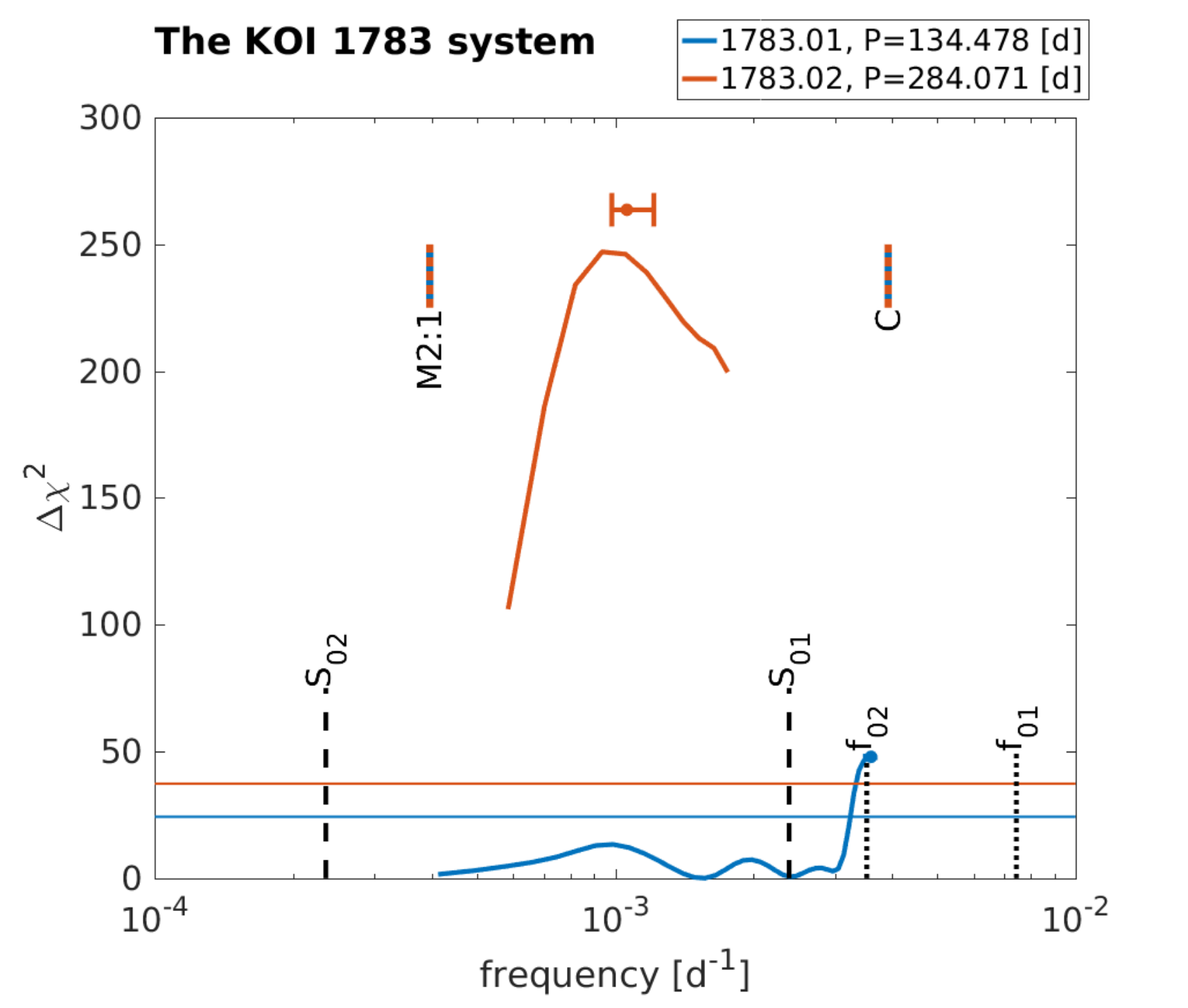}
	\caption{Similar to Figure~\ref{KOI_209_system}, for the KOI 1783 system.}
	\label{KOI_1783_system}
\end{figure}

\item \textbf{KOI 1831 / Kepler 324:} (Figure \ref{KOI_1831_system}) Known TTVs on KOIs 1831.01 and 1831.03 are anti-correlated but only polynomial [H16], and only the former was statistically validated while the latter is still a candidate. Here we detect the TTV frequency of KOI 1831.01 at $f_{TTV}=(4.24_{-0.47}^{+1.38})\cdot 10^{-4} d^{-1}$ and find it to be in agreement with the predicted 3:2 MMR super-frequency with KOI 1831.03 - but we do not detect the very significant TTVs on KOI 1831.03, as expected, since it is high-amplitude. Additionally, there are a few more significant frequencies in the PA spectrum of KOI 1831.01. This dynamically confirms KOI 1831.03, hitherto just a candidate, and possibly allows probing the absolute masses of the planets.
\begin{figure}
	\includegraphics[width=0.5\textwidth]{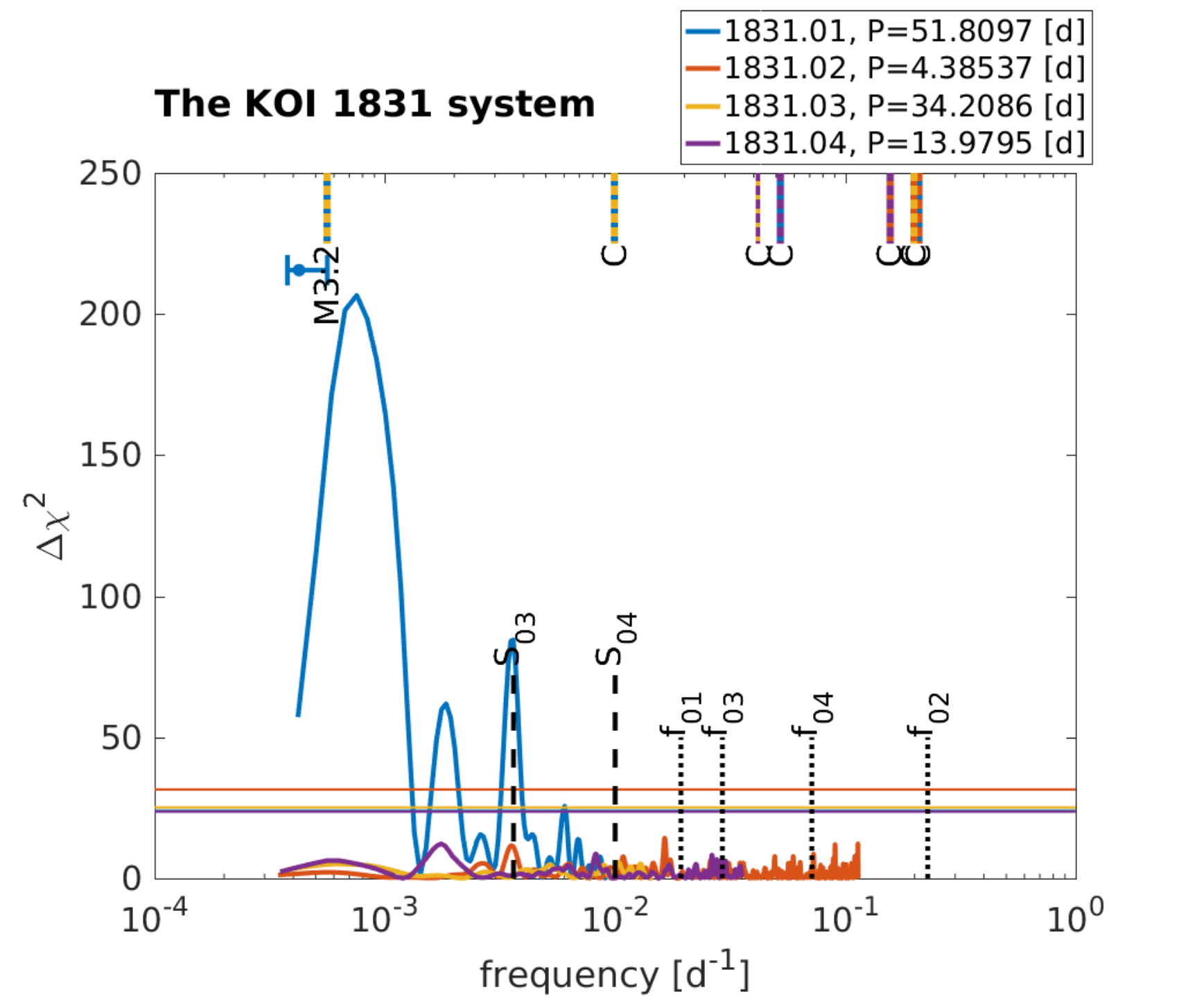}
	\caption{Similar to Figure~\ref{KOI_209_system}, for the KOI 1831 system.}
	\label{KOI_1831_system}
\end{figure}

\item \textbf{KOI 1955 / Kepler-342:} (Figure \ref{KOI_1955_system}) We find two significant TTVs in the PA spectrum of 1955.02, also present in 1955.04 (partly also seen by H16). The frequencies are away from the expected 3:2 MMR super-frequency, possibly due to the system being deep in resonance ($\Delta=0.0027$).

\vspace{1cm}
\begin{figure}
	\includegraphics[width=0.5\textwidth]{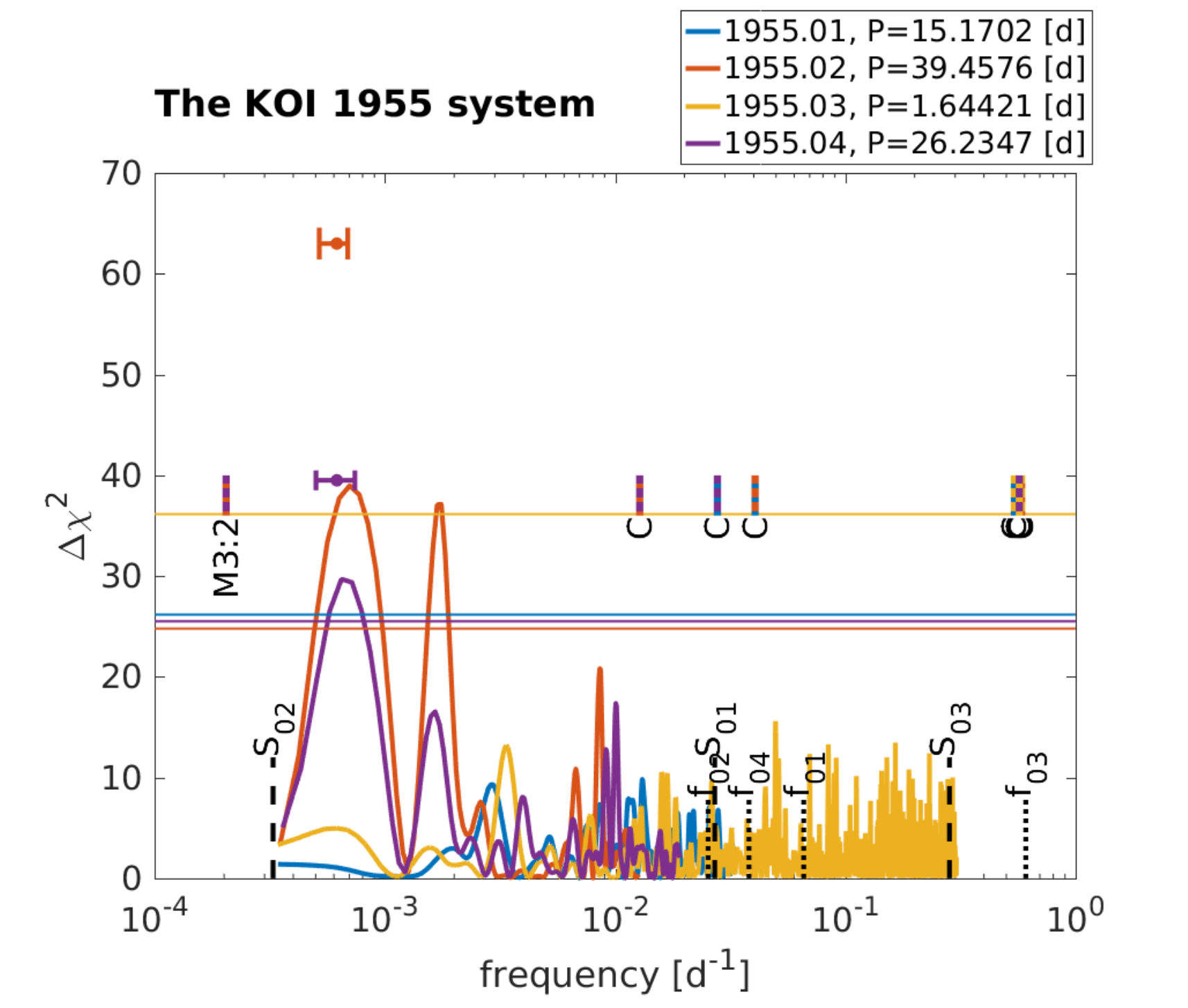}
	\caption{Similar to Figure~\ref{KOI_209_system}, for the KOI 1955 system.}
	\label{KOI_1955_system}
\end{figure}


\item \textbf{KOI 2038 / Kepler-85:} (Figure \ref{KOI_2038_system}) This well studied system (Xie 2013, H16, Hadden \& Lithwick 2016) shows significant TTV frequencies on KOIs 2038.01 and 2038.02 that are consistent with the expected 3:2 MMR between them. Here we find the system may be more interconnected, with a possible blended peak on on the 2038.02 PA spectrum related to a 2:1 MMR with KOI 2038.04, and two additional low significance peaks in the PA spectrum of 2038.01 - one occurring at the expected super-frequencies of the 2:1 MMR with 2038.04 (at $f_{\mathrm{Sup}}=410\cdot 10^{-4} d^{-1}$, and another near the unusual 9:4 MMR with 2038.03.

\begin{figure}
	\includegraphics[width=0.5\textwidth]{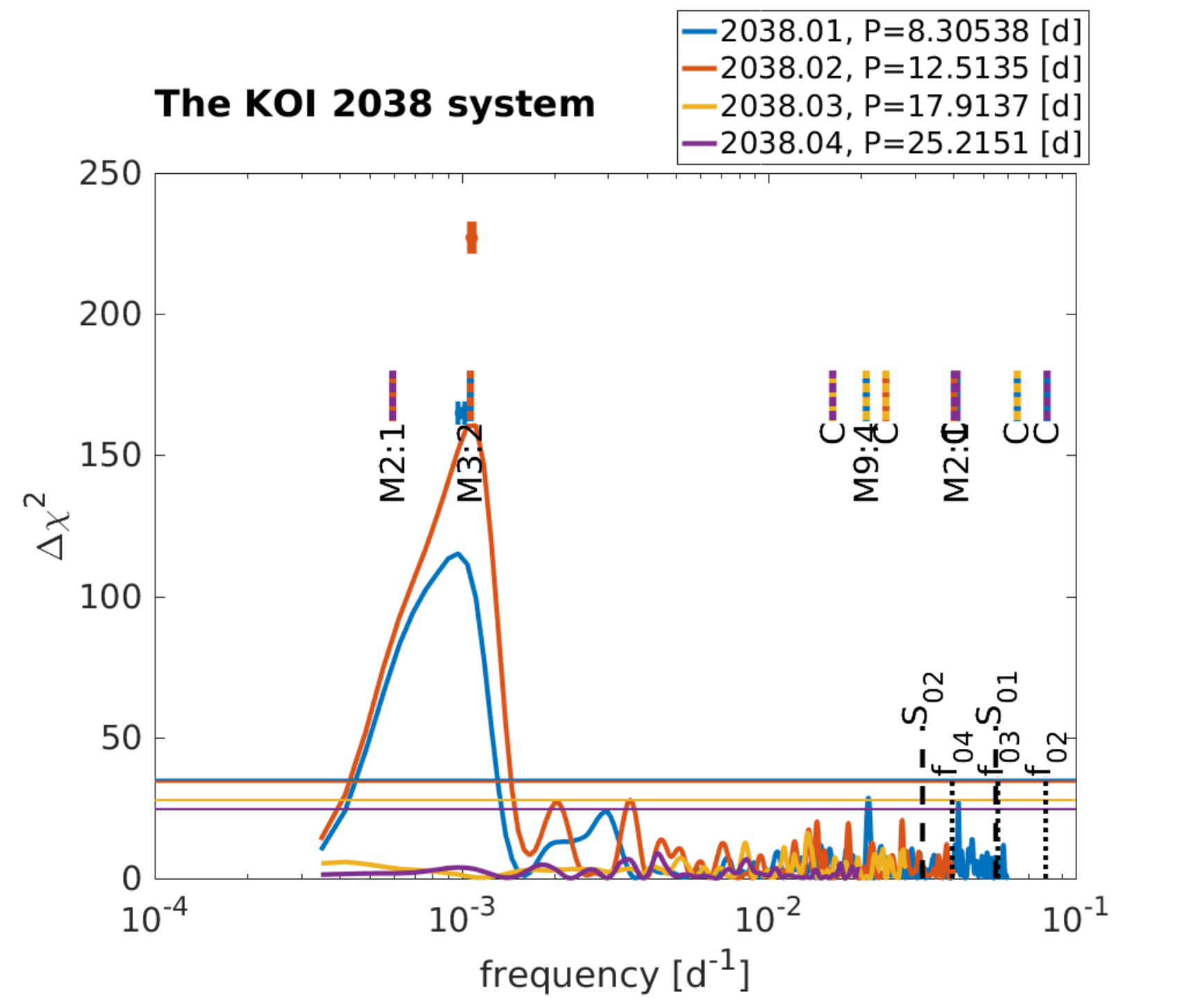}
	\caption{Similar to Figure~\ref{KOI_209_system}, for the KOI 2038 system.}
	\label{KOI_2038_system}
\end{figure}

\item \textbf{KOI 2092 / Kepler-359:} (Figure \ref{KOI_2092_system}) All three planets in the system were only statistically validated and weak ($<2\sigma$) mass limits were subsequently given by Hadden \& Lithwick (2016).
Here we find significant low frequency TTVs for all three planets which may arise either from the system residing near or within resonance. 

\begin{figure}
	\includegraphics[width=0.5\textwidth]{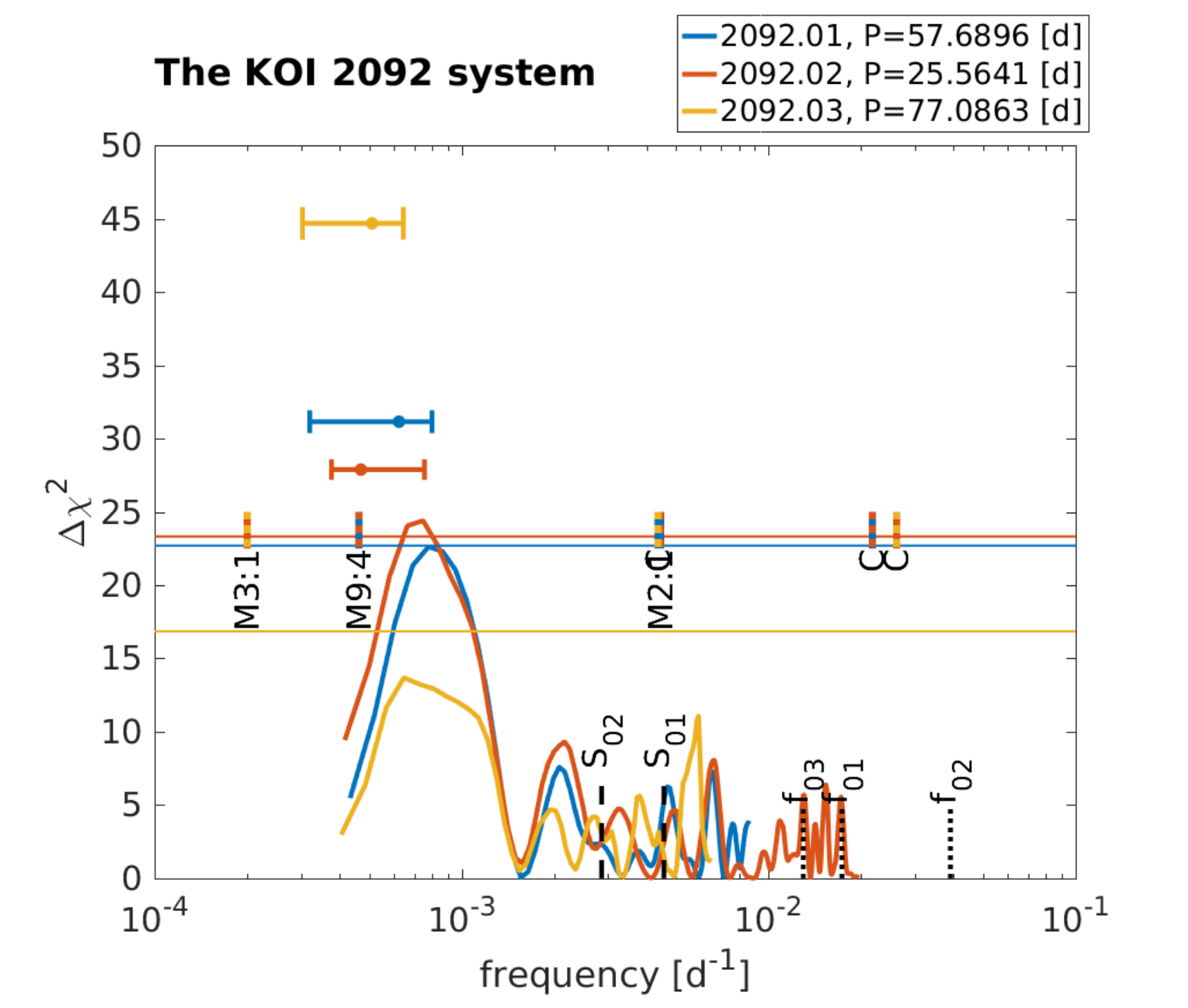}
	\caption{Similar to Figure~\ref{KOI_209_system}, for the KOI 2092 system.}
	\label{KOI_2092_system}
\end{figure}




\end{itemize}


\begin{table*}
	\caption{Results of the combined PA and the full MCMC spectral approach fit. Shown here are 449 objects which are of some interest (Confidence$\geq0.99$) and that are not EBs or likely false positives as defined in \S \ref{GenStat} ({\it i.e.} including objects for which the Confidence in their TTV is still lower than the adopted threshold for new detections of Confidence$\geq0.999$, but excluding high-amplitude objects). In all cases results are from the full non-linear spectral approach fit, except for the count of significant frequencies which is based on the PA spectrum. Please refer to \S \ref{GenStat} for general description and possible caveats. The columns are: KOI numbers (new detections are marked with an asterisk, new determination of the TTV period to a previously-polynomial TTV is marked by two asterisks); frequency of TTV signal - the fitted parameter; period of the TTV signal ($=\mathrm{f^{-1}_{TTV}}$, rounded value); $\Delta\chi^2$ of the TTV signal over the linear model (rounded value); The four cumulative $\Delta\chi^2$ tests (\S \ref{CumChi2}); the TTV signal amplitude;  the TTV signal reference time; Bootstrap test confidence estimation from 1000 runs; ratio of the scatter around the linear model to the median error; other PA-detected independent TTV frequencies $\Delta\chi^2>=0.999$ threshold (or ">5" if so many exist); the expected stroboscopic frequency; Previous references indicated by 1) H16, 2) Xie13+Xie14, 3) HL14 (high significance), 4) HL16, 5) JH16, 6) Van Eylen \& Albrecht 2015 (see \S \ref{GenStat} for details and references); Comments.  The table is available in its entirety in a machine readable format.  A portion is shown here for guidance regarding its form and content.}
    \label{ResultsTable} 
	\tiny
	\begin{tabular}{|lcccccccccllp{1.3cm}clp{1.1cm}|}
    	\hline
        KOI & $\mathrm{f_{TTV}}$     & $\mathrm{P_{TTV}}$ & $\Delta \chi^2$ & \multicolumn{4}{c}{Cumulative $\Delta \chi^2$ tests} & $A$    & $T_0$  & Conf.      & STD/Med        & Frequencies                      & Strob. freq.  & Previous      & Comments \\ 
            & $\cdot10^{-4} [\mathrm{d}^{-1}]$ & $[\mathrm{d}]$                &                 & Area & Single & RMS & Corr.     & [min]  & [KBJD] &            & (linear model) & $\cdot10^{-4} [\mathrm{d}^{-1}]$ & $\cdot10^{-4} [\mathrm{d}^{-1}]$  & references &          \\ 
        \hline

12.01    & $7.29^{+0.21}_{-0.26}$           &  1371 &   1932                     & 0.338           & -121.947           & 249.643           & 0.885          & $1.162^{+0.050}_{-0.042}$          & $1068.5^{+5.9}_{-6.3}$                   & 1                 &  4.02        & [16.1, 24.5, 200.7] $\pm3.1$          & 459.2       & 1        & Kepler-448 b \\ 
13.01    & $1746.12^{+0.25}_{-0.18}$           &     6 &  12659                     & 0.312           & -15.347           & 46.904           & 0.961          & $0.1070^{+0.0075}_{-0.0057}$          & $134.143^{+0.129}_{-0.094}$                   & 1                 &  2.27        & [1718.7, 1771.5] $\pm2.7$          & 1746.5       & 1        & Kepler-13 b \\ 
41.01    & $115.93^{+0.60}_{-0.34}$           &    86 &     48                     & 0.215           & -7.584           & 5.721           & 0.940          & $5.28^{+0.48}_{-0.69}$          & $190.6^{+1.5}_{-1.7}$                   &  0.994            &  1.62        &  --           & 152.2       &         & Kepler-100 c \\ 
42.01    & $9.913^{+0.072}_{-0.067}$           &  1009 &   3246                     & 0.165           & -83.669           & 272.222           & 0.987          & $14.95^{+0.25}_{-0.30}$          & $523.9^{+3.6}_{-3.6}$                   & 1                 &  2.46        & [19.4, 33.9, 143.2] $\pm3.1$          & 428.4       & 1        & Kepler-410 A b \\ 
46.01 *    & $699.69^{+0.75}_{-0.71}$           &    14 &     45                     & 0.258           & -2.094           & 3.791           & 0.973          & $1.35^{+0.19}_{-0.20}$          & $134.70^{+0.75}_{-0.72}$                   & 1                 &  1.15        &  --           & 1960.3       &         & Kepler-101 b \\ 
49.01    & $244.98^{+0.66}_{-0.71}$           &    41 &     42                     & 0.217           & -2.765           & 3.514           & 0.950          & $2.37^{+0.35}_{-0.38}$          & $170.8^{+2.3}_{-2.0}$                   &  0.997            &  1.33        &  --           & 1042.8       &         &  \\ 
63.01 *    & $265.79^{+0.31}_{-0.29}$           &    38 &    116                     & 0.244           & -16.469           & 14.878           & 0.959          & $0.441^{+0.032}_{-0.033}$          & $160.24^{+0.96}_{-0.91}$                   & 1                 &  2.27        & 306.0 $\pm3.1$          & 738.1       &         & Kepler-63 b, Active star \\ 
64.01 *    & $3.2^{+1.2}_{-1.3}$           &  3087 &    240                     & 0.151           & -4.605           & 14.783           & 0.985          & $4.4^{+7.1}_{-1.7}$          & $120^{+172}_{-168}$                   & 1                 &  1.26        &  --           & 2478.0       &         &  \\ 
70.01    & $335.07^{+0.58}_{-0.53}$           &    30 &     46                     & 0.539           & -3.314           & 6.149           & 0.749          & $1.44^{+0.20}_{-0.24}$          & $157.3^{+1.8}_{-1.8}$                   &  0.995            &  1.28        &  --           & 174.1       &         & Kepler-20 c \\ 
70.02    & $725.66^{+0.59}_{-0.91}$           &    14 &     14                     & 0.456           & -2.296           & 3.983           & 0.880          & $2.23^{+0.39}_{-0.39}$          & $141.9^{+0.8}_{-1.2}$                   &  0.993            &  1.22        &  --           & 2391.1       &         & Kepler-20 b \\ 
72.01    & $667.46^{+0.52}_{-0.58}$           &    15 &     23                     & 0.241           & -1.439           & 5.141           & 0.916          & $1.10^{+0.20}_{-0.16}$          & $145.12^{+0.88}_{-0.77}$                   &  0.994            &  1.32        &  --           & 11773.0       &         & Kepler-10 b \\ 
75.01 **    & $3.62^{+0.40}_{-0.29}$           &  2759 &   2308                     & 0.106           & -192.179           & 124.444           & 0.975          & $37.2^{+5.4}_{-5.7}$          & $116^{+91}_{-56}$                   & 1                 &  3.16        & [36.9, 47.1] $\pm2.5$          & 65.7       & 1        &  \\ 
82.01 *    & $214.94^{+0.46}_{-0.72}$           &    47 &     47                     & 0.426           & -4.175           & 4.717           & 0.948          & $1.61^{+0.19}_{-0.24}$          & $174.9^{+1.4}_{-1.6}$                   & 1                 &  1.49        &  --           & 94.2       &         & Kepler-102 e \\ 
84.01    & $31.80^{+0.18}_{-0.18}$           &   314 &    317                     & 0.187           & -10.790           & 13.228           & 0.990          & $5.40^{+0.26}_{-0.27}$          & $291.3^{+4.2}_{-4.2}$                   & 1                 &  1.20        &  --           & 532.6       & 1        & Kepler-19 b \\ 
89.02    & $5.39^{+0.67}_{-0.74}$           &  1854 &    182                     & 0.483           & -22.188           & 13.741           & 0.954          & $86^{+40}_{-20}$          & $10^{+104}_{-146}$                   & 1                 &  1.36        & 24.1 $\pm2.8$          & 43.5       & 1        &  \\ 
94.01 *    & $80.06^{+0.49}_{-0.52}$           &   125 &     72                     & 0.252           & -9.305           & 8.650           & 0.937          & $0.817^{+0.073}_{-0.086}$          & $198.8^{+5.8}_{-5.9}$                   & 1                 &  1.54        & [65.0, 117.4, 161.4] $\pm2.8$          & 180.1       &         & Kepler-89 d, A mutual event system \\ 
94.02 *    & $64.13^{+0.28}_{-0.27}$           &   156 &    228                     & 0.145           & -12.213           & 7.622           & 0.957          & $5.86^{+0.51}_{-0.52}$          & $149.8^{+3.3}_{-3.4}$                   & 1                 &  1.67        & (>5)          & 100.3       &         & Kepler-89 c, A mutual event system \\ 
94.03    & $14.64^{+0.27}_{-0.31}$           &   683 &    299                     & 0.061           & -27.076           & 9.659           & 0.995          & $7.35^{+0.61}_{-0.51}$          & $723^{+12}_{-17}$                   & 1                 &  1.80        & 42.4 $\pm3.5$          & 50.5       & 1        & Kepler-89 e \\ 
100.01 *    & $339.31^{+0.31}_{-0.35}$           &    29 &    108                     & 0.396           & -6.342           & 17.335           & 0.656          & $2.39^{+0.24}_{-0.22}$          & $153.99^{+0.79}_{-0.78}$                   & 1                 &  1.61        &  --           & 744.5       &         &  \\ 
103.01    & $38.111^{+0.076}_{-0.079}$           &   262 &   2419                     & 0.194           & -37.914           & 140.433           & 0.999          & $25.56^{+0.61}_{-0.46}$          & $329.6^{+1.5}_{-1.4}$                   & 1                 &  1.54        & [28.3, 48.4, 72.6] $\pm2.8$          & 487.3       & 1        &  \\ 
105.01    & $281.23^{+0.61}_{-0.65}$           &    36 &     20                     & 0.376           & -2.752           & 4.218           & 0.735          & $1.20^{+0.24}_{-0.25}$          & $150.7^{+1.8}_{-1.8}$                   &  0.990            &  1.08        &  --           & 578.0       &         &  \\ 
108.02    & $10.60^{+0.33}_{-0.34}$           &   943 &    115                     & 0.398           & -16.683           & 18.669           & 0.863          & $20.3^{+1.8}_{-1.9}$          & $767.2^{+10.2}_{-9.3}$                   & 1                 &  1.42        & 22.2 $\pm2.4$          & 39.8       & 1, 6        & Kepler-103 c \\

    \hline
    \end{tabular}
    \normalsize{}
    \vspace{0.3cm}
\end{table*}


\section{Conclusions} 
\label{Conclusions}


We introduced the technique of Spectral Approach to TTVs for the detection of transit timing variations. The Spectral Approach is: more sensitive due to the reduced number of free parameters in its model; not limited by short or low-SNR single events because it uses one global fit and not multiple event-by-event fits; unbiased since only the improvement over the linear model matters, and not the properties of linear model itself. New TTV candidates were found, and the overall set has no significant period or depth biases relative to the general \textit{Kepler} candidates population - unlike catalogs resulting from the classical approach to TTV detection. Consequently the Spectral Approach is more sensitive to TTVs of lower amplitude, around smaller-  and shorter-period planets, than the classical TTV measurement technique. We also presented the Perturbative Approximation (PA) to the Spectral Approach, a linear approximation which is much faster than the full model, albeit less sensitive to higher-amplitude TTVs, allowing to quickly identify candidates for more computationally-intensive full Spectral Approach fit. PA can also be used to other types of variations, such as impact parameter variations, and this will be further explored in future work.

Applying these techniques to \textit{Kepler} data we were able to detect 131 new TTV-bearing stars. The fact that so many new TTVs were detected is interpreted as stemming from the high planetary multiplicity uncovered by \textit{Kepler}: TTVs are not the exception but rather the rule. Of particular importance are: (a) Stars that exhibit multiple sets of transits, which sometimes allow us to link the observed TTVs to a specific planet-planet interaction, and to place constraints on the masses of the planets; (b) Stars that exhibit multiple significant TTV frequencies. (c) Planets that have TTVs that cannot be linked to other planets in the system: these planets are likely affected by other yet-unknown objects in the system. We note that the use of the full PA spectrum, and not just the most significant frequency was found to be useful.

The sensitivity and generality of PA for all targets that exhibit three or more transits, paired with its short execution time, make it highly suitable for current and future large-scale surveys ({\it e.g.} space-based K2, TESS, and PLATO as well as ground-based SuperWASP and HATNet). Moreover, the execution time of PA is so short that applying it to all ~150,000 \textit{Kepler} stars is possible - even those with currently no threshold crossing event at all. The speed will make searching for TTVs on \textit{all} stars feasible. This may be useful since small, low-mass planets and easier to deflect by gravitational interaction with other bodies in the system, and thus planets may (and probably do) exist that have escaped detection because their transit signals have been blurred by TTVs.  Such planets may become detectable once TTVs are accounted for using PA. 

\section*{Acknowledgements}

This project was supported by the Helen Kimmel Center for Planetary Science, the Minerva Center for Life Under Extreme Planetary Conditions and by the I-CORE Program of the PBC and ISF (Center No. 1829/12). AO acknowledges the support of the Koshland Foundation and McDonald-Leapman grant 
J.-W.X. acknowledges support from the NSFC Grants (11403012, 11333002,11661161014) and a Foundation for the Author of National Excellent Doctoral Dissertation of People's Republic of China. RS acknowledges the support of the ISF.
This paper includes data collected by the Kepler mission. Funding for the Kepler mission is provided by the NASA Science Mission directorate. Some/all of the data presented in this paper were obtained from the Mikulski Archive for Space Telescopes (MAST). STScI is operated by the Association of Universities for Research in Astronomy, Inc., under NASA contract NAS5-26555. Support for MAST for non-HST data is provided by the NASA Office of Space Science via grant NNX09AF08G and by other grants and contracts. This work has made use of services produced by the NASA Exoplanet Science Institute at the California Institute of Technology.

\end{document}